\documentclass[preprint]{elsarticle}
\biboptions{sort&compress}
\usepackage{lipsum}
\usepackage{lineno,hyperref}
\usepackage{amssymb}
\usepackage{amsmath}
\usepackage{multirow}
\usepackage{bigdelim}
\usepackage[T1]{fontenc}
\usepackage{color}
\usepackage{xcolor}
\usepackage{hyperref}
    \hypersetup{
        colorlinks   = true,
        citecolor    = blue
    }
\usepackage{graphicx}
\usepackage{subcaption}
\usepackage{mwe}
\usepackage{ulem}
\usepackage{soul}
\newcommand{\newtxt}[1]{#1}
\renewcommand{\sout}[1]{}
\begin{document}
%%%%%%%%%%%%%%%%%%%%%
\title{Simulation of potential and species distribution in a Li$\|$Bi liquid metal battery using coupled meshes}

\author[1]{Carolina Duczek\corref{cor1}}%
\ead{c.duczek@hzdr.de}
\author[1]{Norbert Weber}
\author[2]{Omar E. Godinez-Brizuela}
\author[1]{Tom Weier}

\cortext[cor1]{Corresponding author}

\address[1]{Helmholtz-Zentrum Dresden-Rossendorf,
Bautzner Landstr. 400,
01328 Dresden,
Germany}

\address[2]{Norwegian University of Science and Technology (NTNU),
7030 Trondheim,
Norway}

%\tnoteref{<label(s)>}
%\corref{<label(s)>}
%\fnref{<label(s)>}
%\tnotetext[<label>]{<title note text>}
%\cortext[<label>]{<corresponding author note text>}
%\fntext[<label>]{<author footnote text>}
%%%%%%%%%%%%%%%%%%%%%
\begin{abstract}
In this work a 1D finite volume based model using coupled meshes is introduced to capture potential and species distribution throughout the discharge process in a lithium-bismuth liquid metal battery while neglecting hydrodynamic effects, focusing on the electrochemical properties of the cell and the mass transport in electrolyte and cathode. Interface reactions in the electrical double layer are considered through the introduction of a discrete jump \newtxt{of the potential} modelled as periodic boundary condition to resolve interfacial discontinuities in the cell potential.
A balanced-force like approach is implemented to ensure consistent calculation at the interface level.
It is found that mass transport and concentration gradients have a significant effect on the cell overpotentials and thus on cell performance and cell voltage. By quantifying overvoltages in the Li$\|$Bi cell with a mixed cation electrolyte, it is possible to show \newtxt{that diffusion and migration current density could have counteractive effects on the cell voltage.} \sout{diffusion is lossless in the electrical sense and ohmic losses in the cell are based on migration processes only.} Furthermore, the \newtxt{simulated} limiting current density is observed to be much lower than experimentally measured, which \sout{implies that there are}\newtxt{can be attributed to} convective effects in the electrolyte that need to be addressed in future simulations.\\
The solver is based on the open source library OpenFOAM and thoroughly \sout{validated} \newtxt{verified} against the equivalent system COMSOL multiphysics \newtxt{and further validated with experimental results}. It is openly available at
\url{https://doi.org/10.14278/rodare.2313}.

\end{abstract}
%%%%%%%%%%%%%%%%%%%%%
\begin{keyword}
OpenFOAM \sep liquid metal battery \sep molten salt battery \sep species transfer
\end{keyword}
\maketitle
%%%%%%%%%%%%%%%%%%%%%
%\newpageafter{title}
%\newpageafter{author}
%\newpageafter{abstract}
%%%%%%%%%%%%%%%%%%%%%
\section{Introduction}
\label{sec:introduction}
%-----------------------------------------------------------------%
\subsection{Liquid metal battery and its application}
\label{sec:introLMB}
\begin{figure}[h!]
  \centering
  \includegraphics[width=0.25\linewidth]{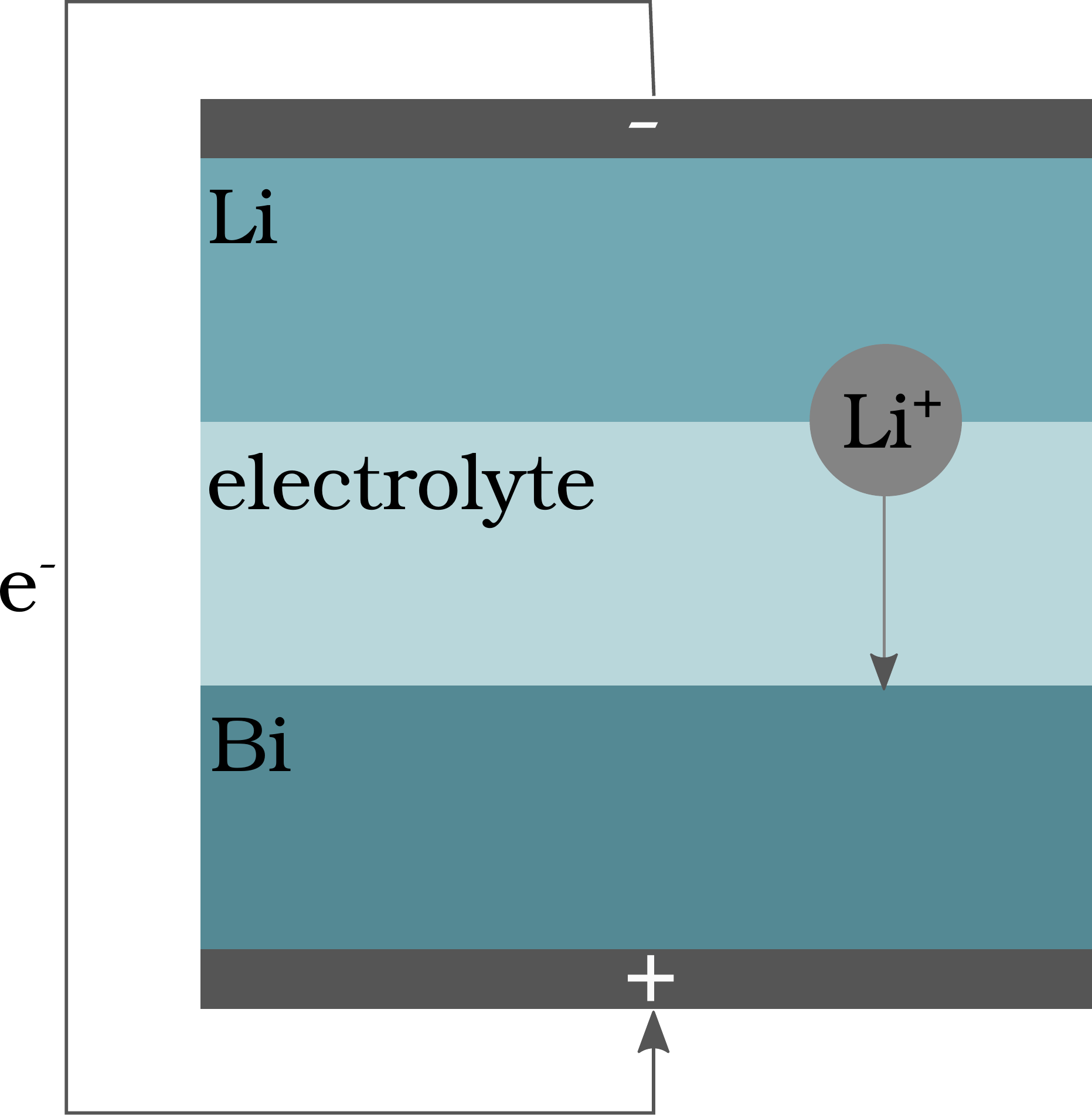}
  \caption{Schematic of a Li$\|$Bi LMB upon discharge.}
  \label{fig:schemDischarge}
\end{figure}

%-----------------------------------------------------------------%
Due to the transition from fossil fuel and nuclear energies to
renewables, the demand for stationary grid scale energy storage is
increasing. Grid fluctuations need to be balanced out and peak demands
must be supplied. \sout{Durability, high power, energy efficiency and low
cost play a major role for a suitable storage technology.}\\
The liquid metal battery (LMB) has the potential to meet \sout{all}the\sout{se} requirements of \newtxt{durability, high power, energy efficiency and low
cost} and\sout{is} therefore is a highly promising \newtxt{storage technology.}\sout{Such batteries offer high efficiency at high current densities, a long life time
and competitive prices.} Basically, LMBs consist of two liquid
electrodes and a molten salt electrolyte. The positive electrode
-- or cathode -- is a heavier metal, while the negative electrode --
or anode -- is a lighter one\footnote{In the present paper,
``cathode'' will always be used for the positive electrode and
``anode'' for the negative one since the discharge behaviour is
investigated only.}. The electrode materials \newtxt{-- mainly being earth-abundant and inexpensive --} are selected in a manner such that
they are liquid at the operating temperature while having a
sufficiently large density difference. When the system is heated, the components
self segregate into three immiscible horizontal layers.\sout{As many of the possible electrode
materials are earth-abundant and inexpensive, LMBs are expected to
provide low cost energy storage.}
Moreover, the liquid state and the high temperatures
enable superior kinetics and transport properties. At the same time,
dendrite growth and electrolyte fracture -- which are challenging in
solid-state batteries -- are avoided. On the other hand, the high
operating temperatures are disadvantageous and increase risk and
speed of corrosion. Sensitivity against motion, which can result in
short circuits, makes the technology unsuitable for portable
applications and restricts LMBs to stationary storage \cite{Kim2013,Newhouse2014}.\\
There are many possible material pairings available, of which the
lithium-bismuth cell has been investigated most intensively and
therefore, most material properties are readily available
\cite{Weber2022}.\sout{The latter is
important for developing a detailed model to investigate a battery
system numerically.} The basic working principle at discharge of a
Li$\|$Bi LMB is shown in Fig.~\ref{fig:schemDischarge}.
First, the anode metal lithium is oxidised at the anode-electrolyte
interface and then the Li$^+$ ion is transported through the
electrolyte. Thereafter, the lithium ion is reduced at the
electrolyte-cathode interface, alloys with the solvent metal bismuth
and diffuses into the cathode. The process is reversed when the cell
is charged. 

% short circuits
Motivated by the imminent risk of short circuits -- caused by strong
fluid flow in the completely liquid cell -- much theoretical and
numerical work has been devoted to LMB fluid dynamics in the
past. Especially, coupling the different regions with the fundamental equations of
electrochemistry, fluid dynamics, electromagnetics and heat transport
is a challenging task\sout{, which has been addressed by several authors}. \newtxt{In LMBs, different phenomena resulting from mass transport, heat transfer and (magnetohydrodynamically) induced flow are occurring simultaneously and heavily influence each other as well as the battery performance} \cite{Agarwal2022}. As described in the review by Kelley and Weier \cite{Kelley2018}, the investigated fluid flow phenomena include thermal convection
\cite{Shen2015,Beltran2016,Ashour2018,Kollner2017,Personnettaz2018,Ashour2019,Keogh2021},
Marangoni convection \cite{Kollner2017}, interfacial instabilities
\cite{Zikanov2015,Weber2017,Horstmann2018,Molokov2018,Tucs2018,Zikanov2018,Herreman2019},
the Tayler instability
\cite{Stefani2011,Weber2013,Weber2014,Weber2015,Herreman2015} and
electro-vortex flow
\cite{Ashour2018,Weber2018,Weber2018a,Herreman2019,Keogh2021}.

% efficiency, mass transfer
Cell efficiency is, in addition to \sout{safety}\newtxt{short-circuits}, the second most important
research topic. Generally, the cell voltage of LMBs is determined by
the open-circuit potential and various overpotentials, which are
typically divided into three groups: (i) ohmic losses, (ii) mass
transfer overpotentials caused by concentration gradients and (iii)
charge transfer losses. Already 60 years ago, it was predicted that
the latter will be negligible at liquid-liquid interfaces
\cite{Agruss1962a} -- a fact, which is still valid today
\cite{Newhouse2017}. While during the first years of LMB research it
was thought that ohmic losses in the electrolyte represent the only
\cite{Agruss1962a,Shimotake1967}, or at least major overpotential
\cite{Vogel1968b,Vogel1969}, it was found soon that concentration
polarisation can as well be important
\cite{Agruss1967,Heredy1967,Foster1967b} -- especially when applying high current
densities \cite{Agruss1963,Agruss1962a}. The maximum cell current is
known to be limited by mass transfer
\cite{Vogel1965a,Vogel1967}. While these experimentally observed
polarisation effects were attributed to the cathode, the existence of
concentration gradients within the electrolyte have at least been
subject of theoretical considerations. Especially the risk of
electrolyte solidification due to compositional changes
\cite{Blanchard2013,Ouchi2016,Percival2018,Gong2020,Gross2020} has
been discussed in this context.

Although \sout{seldom}\newtxt{barely studied} in LMBs, it is well known from literature that 
composition gradients play an important role in molten salt
electrolytes. Such phenomena have been widely investigated by
Braunstein and Vallet in several publications \cite{Vallet1978,
  Braunstein1979, Vallet1980, Vallet1982, Vallet1982a, Vallet1983} who
show that concentration gradients in molten salts can be
significant. The authors predicted concentration gradients in mixed cation
molten salts based on equations that did not include convection
terms \cite{Vallet1978,Braunstein1979}. Later, these predictions were
confirmed in several experiments with
electrolytes contained in porous bodies that completely suppress or at
least strongly damped any convective effects
\cite{Vallet1982,Vallet1982a}. These experiments were performed not only
with low-melting model systems (such as NaCl-KCl-AlCl$_3$ or
AgNO$_3$-NaNO$_3$ \cite{Vallet1982,Vallet1982a}), but even with
LMB-relevant LiCl-KCl \cite{Vallet1983}.
The observed concentration gradients are caused by the electrode
reactions, i.e.~the electric current, as well as migrational and
diffusive transport processes. This finally results in a depletion or
enrichment of the electroactive species at the interfaces between
electrodes and electrolyte
\cite{Vallet1980,Braunstein1979,Masset2007}. The same authors show 
that current density, initial composition and differences in the
mobility of the ions determine the final composition profiles
\cite{Vallet1978}.  

Overall, the literature suggests that concentration gradients
with possibly significant effects will appear not only in the
cathode, but also in the electrolyte of LMBs. These will not only lead
to concentration overpotentials \cite{Braunstein1979}, but might cause
solidification
\cite{Braunstein1979,Vallet1983,Blanchard2013,Ouchi2016,Percival2018,Gong2020,Gross2020},
changes of the electric conductance, of the kinetics at the
interfaces, increased corrosion \cite{Braunstein1979} or a reduced
energy utilisation \cite{Vallet1983}. Finally, if one interface 
completely depletes of the active species, undesired reactions might
take place, such as the deposition of a previously passive species of the
electrolyte \cite{Vallet1978}. 

Mass transport effects have, to a certain extend, been studied by
numerical simulation -- but almost exclusively within the
cathode. Investigated phenomena include pure diffusion
\cite{Weber2019,Weber2022}, thermal convection
\cite{Kelley2014,Ashour2018}, solutal convection
\cite{Personnettaz2019,Personnettaz2020,Herreman2020a,Personnettaz2022} and
electro-vortex flow
\cite{Ashour2019,Weber2020,Herreman2021,Herreman2020a,Herreman2020}. With
the exception of one model \cite{Weber2019,Weber2020}, the influence of
the flow on the cell voltage has always been strongly simplified. 

To the best knowledge of the authors, only Newhouse addressed mass transport in an
LMB electrolyte theoretically \cite{Newhouse2014}. Employing a one-dimensional
model, she assumed a small concentration overpotential over the
boundary layer in a very much simplified manner \cite{Newhouse2014}.
None of the previous studies regarded species transport or
diffusive currents inside the electrolyte. Likewise, mass transport
was up to now only simulated in the cathode. Hence, the present paper
focuses on species transport in the electrolyte and investigates
composition gradients which lead to concentration overpotentials in the
electrolyte.
\subsection{Modelling of liquid metal batteries}
\label{sec:introModelling}
Many numerical studies regarding liquid metal batteries were
performed by various authors. Thereby, the finite difference method (FDM)
\cite{Ashour2019, Ashour2018a, Weber2022}, the finite element method
(FEM) \cite{Herreman2020, Herreman2019, Herreman2021, Aguilar2021} and the finite
volume method (FVM) \cite{Ashour2018, Personnettaz2018,
  Keogh2021, Personnettaz2019, Personnettaz2020,
  Horstmann2018, Weber2017, Weber2013, Weber2015, Weber2015a,
  Weber2018, Weber2018a, Weber2020, Weber2019} were used. This
indicates that the FVM is the most used method to investigate LMBs
numerically. In general, FVM is the method of choice in
computational fluid dynamics (CFD) whereas FEM is widely used in
structural mechanics \cite{Hirsch2007}. However, this does not mean
that FDM and FEM are not suitable for such problems. Idelsohn and
O$\tilde{\mathrm{n}}$ate show that in particular cases and with wisely
chosen discretisation procedures, the FVM and FEM can yield coincident
results \cite{Idelsohn1994}.
However, one of the main advantages of the FVM is that the
conservative discretisation is automatically satisfied due to the
mathematical formulation of the method, which makes it very attractive
for the solution of convective dominant equations \cite{Hirsch2007},
like the Navier-Stokes equations. Contrarily, the FEM guarantees only
global conservation and stabilising the discretisation for
convection-dominated flow is not straightforward. So, additional
computational effort is necessary and leads to higher computational
cost for the fluid flow. For the FVM, computational times are lower
\cite{Fontes2018}. A major drawback of the FVM becomes obvious, when it
comes to the consideration of internal boundary values. Due to the
cell-centred formulation, a solution on the boundary is not well
defined. Boundary values need to be interpolated, which could potentially lead to
problems in the present study, since the electrode-electrolyte interfaces need to be
described at the boundaries of specific cells. This needs to be kept in
mind for the development of a suitable numerical solver.

As the performance of LMBs is highly dependent on the fluid flow of the
metals, the FVM is the method of choice. However, convection
is not the only transport phenomenon that needs to be accounted for.\sout{To
the best knowledge of the authors, potential and species distribution
in a whole cell were not investigated so far using the finite volume
method.} If \sout{those}\newtxt{it is possible to} \sout{can be}simulate \newtxt{potential and species distribution in the whole cell} accurately using the FVM, such
a solver would be an ideal basis to investigate the impact of various flow 
phenomena on the cell performance, which is the future goal for modelling LMBs. 

For the development of a finite volume based solver for species
transport and potential distribution in the Li$\|$Bi cell, the
single-potential approach that was introduced by Beale et
al. \cite{Beale2013,Beale2016,Weber2018} is used.
Within this approach, one potential field is used for the whole
battery. This couples the regions for anode, electrolyte and cathode
using one matrix to obtain the electric potential in the
whole domain. The remaining \newtxt{region-specific} transport equations are solved \newtxt{on local meshes} in a
segregated manner and are iteratively coupled\sout{Moreover, a multi-mesh
approach, where certain variables are solved on a global mesh,
while others are only solved in local meshes, is used}
\cite{Beale2013,Beale2016}.

\begin{figure}[h!]
  \centering
  \includegraphics[width=0.95\linewidth]{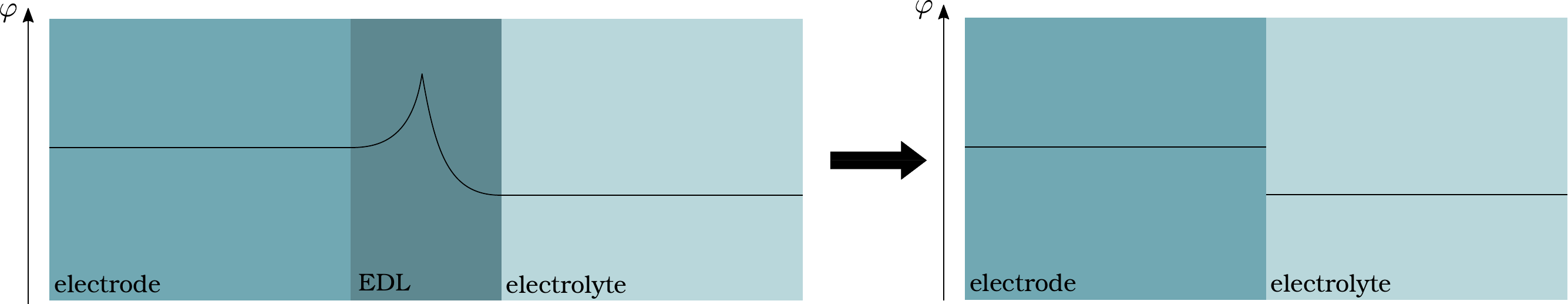}
  \caption{Schematic of an electrical double layer (left) and the simplification using a potential jump (right). Simplified adaption from \cite{Luck2016}.}
  \label{fig:schemEDL}
\end{figure}

When solving the electric potential, special care must be taken in
describing the electrode-electrolyte interface, i.e.~the
electrical double layer (EDL). While microscopic models allow to
completely resolve the EDL, certain macroscopic models simply replace
it by a discrete ``jump'' in potential. The two jumps at the
anode-electrolyte and electrolyte-cathode interfaces are then defined
by the standard potential and represent the ``driving force'' 
of the battery current. In Fig.\,\ref{fig:schemEDL}, a schematic
illustration of the micro- and macroscopic approach, based on a model
from Lueck and Latz \cite{Luck2016}, is shown.

Modelling a potential jump at an internal boundary within the FVM
framework is generally possible by two different approaches. Firstly, it might be implemented as a volumetric source term in
the Laplacian and gradient operator as done by Weber et
al. \cite{Weber2019}.\sout{Both operators are necessary when solving the
governing equations in the battery.} Thus, an additional term appears,
when grid cells that are touching the interface are involved.\sout{Within
this approach, care must be taken with non-orthogonal meshes,
discretisation and parallelisation.}

An alternative approach for implementing the potential jump is the
usage of cyclic (jump) boundary conditions \cite{Patankar1977,Beale1990,Beale1998,Moreira2017}.
Thereby, additional internal boundary conditions (BCs) are
applied to infinitely thin interfaces between the different regions
for electrodes and electrolyte. These BCs are periodic and hold an
offset that is corresponding to the potential jump\footnote{In
OpenFOAM this BC is available as cyclic patch with a fixed jump
\cite{Greenshields2015}.}. The main advantage of this approach is that
it largely relies on the standard OpenFOAM code and only requires slight changes to an existing boundary condition. For this reason, the present study employs this approach and considers interfacial jumps at anode-electrolyte and electrolyte-cathode interfaces without referring to measured or fitted values of the potential. No manipulation of the Laplacian and gradient operator is necessary and it is relatively easy to specify the regions and to identify and update the potential jump. Moreover, this is -- to the best knowledge of the authors -- the first study that investigates potential distributions in an LMB
without resorting to fitted experimental data.
%-----------------------------------------------------------------%

%%% Local Variables:
%%% mode: latex
%%% TeX-parse-self: t
%%% TeX-auto-save: t
%%% TeX-PDF-mode: t
%%% eval: (auto-fill-mode 1)
%%% eval: (flyspell-mode 1)
%%% eval: (reftex-mode 1)
%%% ispell-dictionary: "british"
%%% TeX-master: "../paper"
%%% End:

\section{Mathematical and numerical model}
\label{mathNumModel}
%-----------------------------------------------------------------%
\subsection{Potential difference and overpotentials}
\label{potDiffOver}

The difference between both electrode potentials is the open circuit voltage (OCV) of the LMB \cite{Kelley2018}
\begin{equation}
\label{eq:2-1}
E_{\mathrm{OC}} = - \frac{R T}{z F} \ln a_{\mathrm{Li(Bi)}}
\end{equation}
with the universal gas constant $R$, the temperature $T$, $z$ being
the number of exchanged electrons, $F$ the Faraday constant and $a$
the activity of the active species (Li).
When current flow is present, the available terminal voltage
\begin{equation}
\label{eq:2-2}
E = E_{\mathrm{OC}} - I \sum R - \eta_{\mathrm{c,a-e}} - \eta_{\mathrm{c,e-c}} - \eta_{\mathrm{c,c}} - \eta_{\mathrm{a,a-e}} - \eta_{\mathrm{a,e-c}}
\end{equation}
corresponds to the OCV being reduced by several voltage losses
\cite{Swinkels1971}. These overpotentials are outlined in
Tab.\,\ref{tab:overpot} and are considered to have positive values.
\begin{table}[h]
  \begin{center}
  \caption{Different types of overpotentials influencing the cell voltage of a battery.}
  \begin{tabular}{ l p{5cm} c c }
  \hline
  symbol & description & & group \\ 
  \hline
  $I \sum R = \eta_{\Omega}$ & \sout{ohmic}\newtxt{resistance} overpotential & & \\  
  $\eta_{\mathrm{c,a-e}}$ & concentration overpotential \newline at anode-electrolyte interface & \rdelim\}{6}{11pt} & \multirow[b]{3}{*}{$\eta_{\mathrm{mt}}$} \\
  $\eta_{\mathrm{c,e-c}}$ & concentration overpotential \newline at electrolyte-cathode interface & & \\ 
  $\eta_{\mathrm{c,c}}$ & concentration overpotential \newline in the cathode & & \\
  $\eta_{\mathrm{a,a-e}}$ & activation overpotential \newline at anode-electrolyte interface & \rdelim\}{4}{11pt} & \multirow[b]{2}{*}{$\eta_{\mathrm{ct}}$}\\
  $\eta_{\mathrm{a,e-c}}$ & activation overpotential \newline at electrolyte-cathode interface & &  \\
  \hline   
  \end{tabular}
  
  \label{tab:overpot}
  \end{center}
\end{table}
%
%In Eq.\,\eqref{eq:2-2} $I\sum R$ is the \textit{ohmic overpotential} $\eta_{\Omega}$ with the current $I$ and the resistance $R$, $\eta_{\mathrm{c,a-e}}$ and $\eta_{\mathrm{c,e-c}}$ are \textit{concentration overpotentials} in the electrolyte at anode and cathode interfaces and $\eta_{\mathrm{a,a-e}}$ and $\eta_{\mathrm{a,e-c}}$ are the corresponding \textit{activation overpotentials}. Moreover, there is a concentration overpotential in the cathode due to the inhomogeneous lithium distribution. This is denoted with $\eta_{\mathrm{c,c}}$. $\eta_{\mathrm{c,a-e}}$, $\eta_{\mathrm{c,e-c}}$ and $\eta_{\mathrm{c,c}}$ can be summarized as \textit{mass transfer overpotential} $\eta_{\mathrm{mt}}$, $\eta_{\mathrm{a,a-e}}$ and $\eta_{\mathrm{a,e-c}}$ as \textit{charge transfer overpotential} $\eta_{\mathrm{ct}}$.All overvoltage are considered to have positive values.
%
Mass transfer overpotentials $\eta_{\mathrm{mt}}$ result from inhomogeneous lithium distribution, whereas \sout{charge transfer overpotentials $\eta_{\mathrm{ct}}$ result from chemical reactions. The }resistance overpotential is mainly determined by the electrolytic resistance and can easily be calculated using Ohm's law. The \sout{activation}{charge transfer} overpotentials are primarily related to the electrode-electrolyte electron transfer\newtxt{, but their influence can be neglected since the exchange current densities for Li$\|$Bi liquid metal batteries are high due to the high temperatures as well as rapid and facile charge-transfer reactions}\sout{and may be calculated using the Butler-Volmer equation . Newhouse and Sadoway found that the exchange current densities for Li$\|$Bi liquid metal batteries are high due to the high temperatures and rapid and facile charge-transfer reactions} \cite{Newhouse2017}. \sout{This implies that the current is not limited by electron transfer and the influence of the activation overpotential can be neglected.}\newtxt{Ohmic losses are considered as the most dominant overpotentials in LMBs} \cite{Kelley2018}, nevertheless, the present study aims to quantify the influence of concentration overpotentials in cathode and electrolyte on the cell performance. \\
The mass transfer overpotential in the whole cell can be obtained as \cite{Vetter1967}
\begin{equation}
\label{eq:2-2_1}
\eta_{\mathrm{mt}} = \frac{RT}{zF} \sum \nu_i \ln \frac{a_i}{\overline{a}_i}
\end{equation}
\newtxt{being the sum of all mass transfer overpotentials} with \newtxt{$\nu_i$ denoting the number of proton charges carried by an ion $i$,} the \sout{activity in the absence of current. Mathematically, this is transferable to the activity in the bulk of the region or the}volume-averaged activity $\overline{a}_i$ \newtxt{and the activity of a species at the related interface $a_i$}. Generally, such mass transfer overpotentials are a result of the current flow and the corresponding composition changes in the electrolyte \cite{Newman2004}.
Following Eq.\,\eqref{eq:2-2_1}, the concentration overpotential in the cathode can simply be calculated as \cite{Personnettaz2019,Vetter1967}
\begin{equation}
\label{eq:2-3}
\eta_{\mathrm{c,c}} = \left| \frac{RT}{zF} \left(\ln \frac{a_{\mathrm{Li}}^{\mathrm{c-e}}}{\overline{a}_{\mathrm{Li(Bi)}}} \right) \right|.
\end{equation}
At both electrolyte-electrode interfaces the following is deduced:
\begin{equation}
\label{eq:2-4}
\begin{split}
\eta_{\mathrm{c, a-e}} &= \left| \frac{RT}{zF} \left( \nu_{\mathrm{Li^+}} \ln \frac{a_{\mathrm{Li^+}}^{\mathrm{a-e}}}{\overline{a}_{\mathrm{Li^+}}} \right) \right| \\\
\eta_{\mathrm{c, e-c}} &= \left| \frac{RT}{zF} \left( \nu_{\mathrm{Li^+}} \ln \frac{a_{\mathrm{Li^+}}^{\mathrm{e-c}}}{\overline{a}_{\mathrm{Li^+}}} \right) \right|.
\end{split}
\end{equation}
\sout{For $\overline{a}_i$, volume averaged values of the activity are used.}
%-----------------------------------------------------------------%
\subsection{Problem description and simplifications}
\label{problem}
\begin{figure*}[h]
    \centering
    \begin{subfigure}[b]{0.475\textwidth}
        \centering
        \includegraphics[height=5cm]{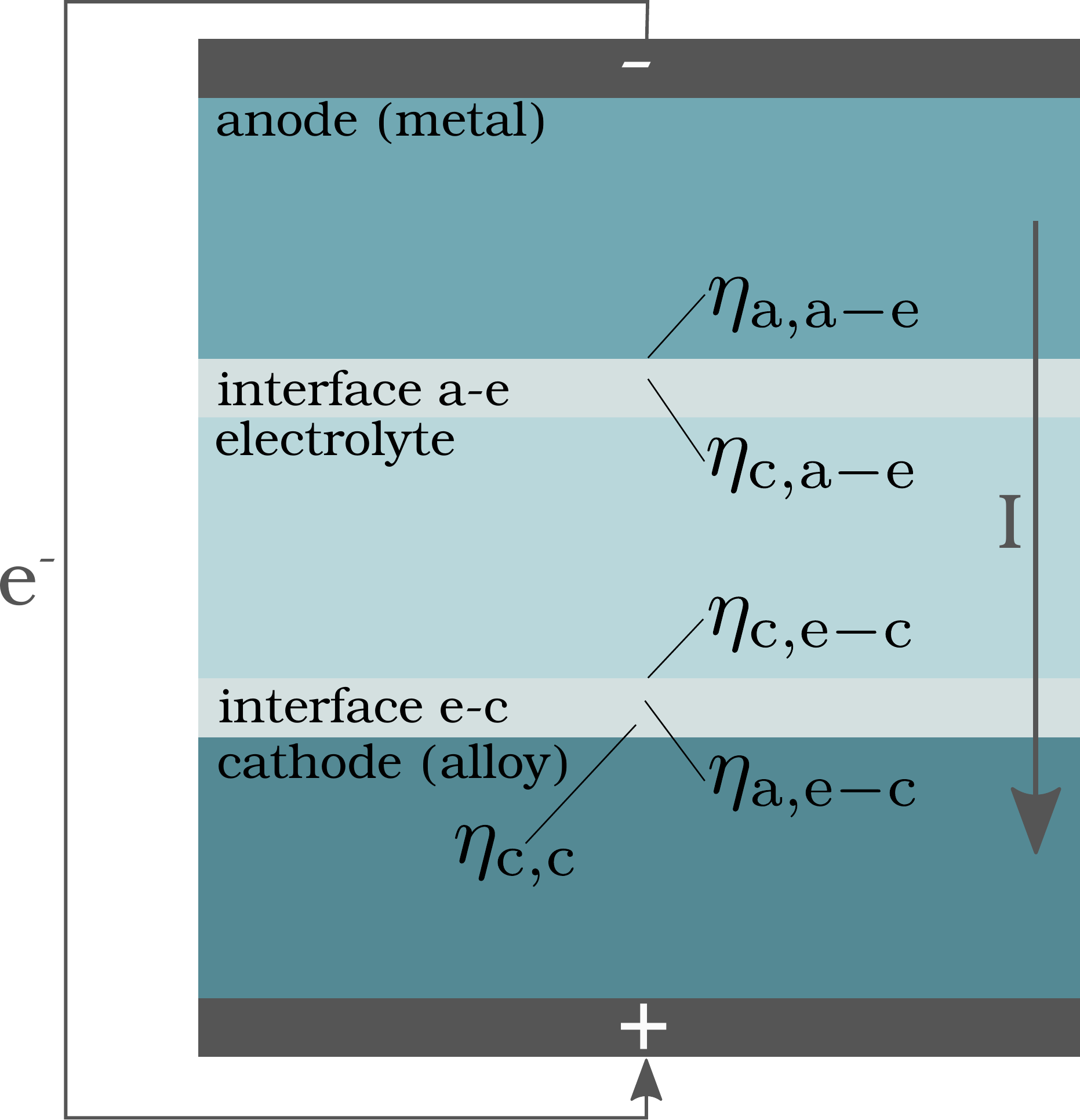}
        \caption[]%
        {{\small}}  
        \label{fig:simulationSetup}
    \end{subfigure}
    \hfill
    \begin{subfigure}[b]{0.475\textwidth}  
        \centering 
        \includegraphics[height=5cm]{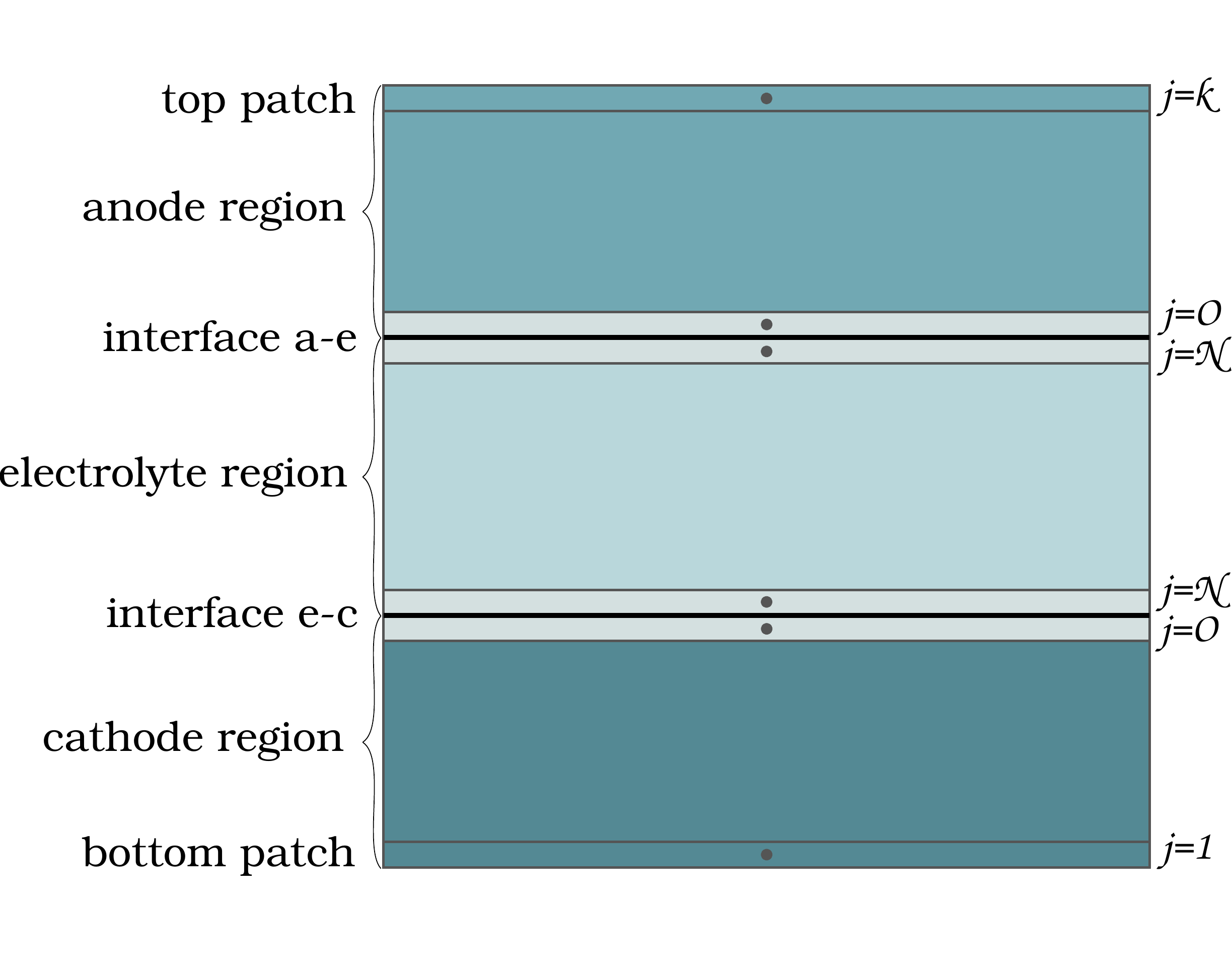}
        \caption[]%
        {{\small}}  
        \label{fig:simulationDomain}
    \end{subfigure}
    \caption[]
    {\small Simulation (a) setup and (b) domain.} 
    \label{fig:simulationSetupDomain}
\end{figure*}

The present study investigates a simplified model of an LMB as it can be seen in Fig.\,\ref{fig:simulationSetup}. Interfaces between the regions for anode, cathode and electrolyte are infinitesimally thin.\sout{ Current collectors are assumed to be perfect conductors and the whole domain is indicated as global region.} Further, the figure shows the location where the overpotentials presented in Tab.\,\ref{tab:overpot} are located.
For each region, different constants, parameters, initial fields, boundary conditions and physical properties are applied.\sout{Simulations are done only for discharge of the battery. Investigations of species distribution and overpotentials are done by solving the potential distribution in the cell during the discharge process.}\\
Figure \ref{fig:simulationDomain} shows the simulation domain with
corresponding patch names. The computational domain contains $j$
cells. For better visualisation only a few specific cells are
illustrated. Here, the bottom patch belongs to $j=1$ and the top
patch to $j=k$. The BC for the interfaces apply for both
electrode-electrolyte interfaces; the owner cell of the jump BC is always located in the electrode, while the neighbour cell is located in the electrolyte.
%
%\begin{figure}[h]
%  \includegraphics[width=\linewidth]{chapters/domain.png}
%  \caption{Simulation domain}
%  \label{fig:simulationDomain}
%\end{figure}
%
Simulations are done ``quasi'' one-dimensional using the FVM
\newtxt{with the model being implemented in the open-source library OpenFOAM} \cite{Weller1998}. Moreover the following simplifications are made:
\begin{itemize}
\item All convective and thermal effects\footnote{Fluid flow, heat transfer, Joule heating and volume change.} are neglected,
%\item the temperature is constant in the whole domain and does not change during the simulation, 
\item activation overpotentials are not taken into account,
\item the electrolyte is assumed to have ideal ionic behaviour,
%\item magnetohydrodynamic effects are not considered,
\item the current collectors are not modelled, 
\item and diffusion coefficients as well as activity coefficients are calculated based on additional \sout{simplifications}\newtxt{assumptions} due to the lack of exact values. 
\end{itemize}
\newtxt{These simplifications allow to investigate mass transport in the electrolyte fundamentally. By neglecting fluid flow, the maximal possible diffusion loss may be quantified and be compared with ohmic losses. While the results are directly valid for those batteries, where the electrolyte has been soaked into an inert filler material} \cite{Shimotake1969,Swinkels1971}, \newtxt{convection might be important for fully liquid cells} \cite{Kelley2018}. \newtxt{The latter aspect is briefly discussed at the end of section} \ref{applicationDiscussion}.
\sout{Applying the simplifications above, the model is implemented in the open-source library OpenFOAM \cite{Weller1998}.}
%
%-------------------------------------------------------------
\subsection{Transport equations and boundary conditions}
\label{eqnBC}
In this section, the electrochemical equations for molten electrodes
and electrolytes are presented. They follow Newman \cite{Newman2004},
if nothing else is mentioned. There are two transport equations that
will be discussed\sout{  in the following section}: first, the transport of
charge and second, the transport of species. Equations in section
\ref{eqnGlobal} are relevant for the whole computational domain, while
those in section \ref{eqnElec} and \ref{eqnCath} are solved on the electrolyte or cathode sub-grid. 
%
%-----------------------------------------------------------------%
%\subsubsection{\sout{Global/General}\colorbox{yellow}{Entire battery}}
\subsubsection{Entire battery}
\label{eqnGlobal}
 When a potential difference is applied to an electric conductor, electrons start to flow due to the gradient in potential $\varphi$. The current density $\mathbf{j}$ is related to the flow of electrons and can be calculated using Ohm's law
\begin{equation}
\label{eq:2-5}
\mathbf{j} = - \sigma \nabla \varphi
\end{equation}
with the electric conductivity $\sigma$. Taking the conservation of charge
\begin{equation}
\label{eq:2-6}
\nabla \cdot \mathbf{j} = 0
\end{equation}
into account, this leads to the Laplace equation for the potential distribution
\begin{equation}
\label{eq:2-7}
\nabla \cdot \left( \sigma \nabla \varphi \right) = 0.
\end{equation}
For calculating the potential, the global mesh is used. Under galvanostatic control, a Neumann BC \\
\begin{equation}
\label{eq:2-8}
\sigma_k \left. \frac{\partial \varphi_k}{\partial \mathbf{n}} \right|_{f} = \frac{I}{A_{f,k}}
\end{equation}
for the top patch is used, where $n=k$. Here, $\mathbf{n}$ denotes the normal face vector at the face $f$ and the associated area $A$.\sout{ In OpenFOAM this is predefined as fixedGradient BC.} At the bottom patch, a Dirichlet BC
\begin{equation}
\label{eq:2-9}
\left. \varphi_{\mathrm{1}} \right|_{f} = \mathrm{const.} = 0
\end{equation}
with $k=1$ applies.\sout{ In the present case, the liquid metal battery is grounded at the bottom, so the value at the bottom patch is zero.}
%
%-----------------------------------------------------------------%
\subsubsection{Electrolyte}
\label{eqnElec}
When solving species conservation within the electrolyte layer, the Laplace equation needs to be substituted by a Poisson equation in the following manner. In the electrolyte the current represents the net flux of the charged species and the current density is defined as 
\begin{equation}
\label{eq:2-10}
\mathbf{j} = F\sum_{i=1}^n \nu_i \mathbf{N}_i,
\end{equation}
with \sout{$\nu_i$ denoting the number of proton charges carried by an ion $i$ and }the flux density of each species $\mathbf{N}_i$. 
The latter is expressed by the Nernst-Planck equation
\begin{equation}
\label{eq:2-11}
\mathbf{N}_i = \underbrace{-\nu_i \mu_i F c_i\nabla\varphi}_{\mathrm{migration}}-\underbrace{D_i\nabla c_i}_{\mathrm{diffusion}} + \underbrace{c_i \mathbf{u}}_{\mathrm{convection}}
\end{equation}
with the concentration $c_i$, the mobility $\mu_i$ and the diffusion coefficient $D_i$ of a species as well as the velocity $\mathbf{u}$. The Nernst-Einstein relation 
\begin{equation}
\label{eq:2-12}
\mu_i = \frac{D_i}{RT}
\end{equation}
is used to calculate the mobility of a species. % Strictly speaking,
%equation \eqref{eq:2-12} holds only for infinitely dilute
%solutions. However, for simplicity it is  used for the molten salt.% Elsewhere, friction and interaction between the ions needs to be considered \cite[p.299ff.]{Newman2004}. 
%The species flux is caused by migration, diffusion and convection. Migration is the movement of charged species due to an electric field, diffusion is caused by concentration gradients $\nabla c_i$ of the species and convection addresses the bulk motion of the fluid. 
Further, a material balance for each species 
\begin{equation}
\label{eq:2-13}
\underbrace{\frac{\partial c_i}{\partial t}}_{\mathrm{accumulation}} = \underbrace{-\nabla\cdot\mathbf{N}_i }_{\mathrm{net \, input}} + \underbrace{R_i}_{\mathrm{production}}
\end{equation}
with the production $R_i$ can be derived.
Finally, the condition of electroneutrality
\begin{equation}
\label{eq:2-14}
\sum_i \nu_i c_i = 0
\end{equation}
applies for electrolytic solutions. Equations \eqref{eq:2-10}, \eqref{eq:2-11}, \eqref{eq:2-13} and \eqref{eq:2-14} describe the transport processes in electrolytic solutions.
In the present study\sout{, fluid flow and species production are neglected (} $R_i=0$ and $\mathbf{u}=0$. When combining Eq.\,\eqref{eq:2-10} and Eq.\,\eqref{eq:2-11}, it can be seen that the current density 
\begin{equation}
\label{eq:2-15}
\mathbf{j} = - F^2 \nabla \varphi \sum_i \nu_i^2 \mu_i c_i - F \sum_i \nu_i D_i \nabla c_i
\end{equation}
can be divided into two parts as well.\sout{, since the last term on the right-hand side (RHS) is zero due to electroneutrality.} The first term describes the migrational current density $\mathbf{j}_\mathrm{m}$ and the second the diffusive current density $\mathbf{j}_\mathrm{d}$. Since the current due to migration is closely related to the flow of electrons, the electric conductivity in the salt can be written as
\begin{equation}
\label{eq:2-16}
\sigma_{\mathrm{salt}} = F^2 \sum_i \nu_i^2 \mu_i c_i,
\end{equation}
which finally yields
\begin{equation}
\label{eq:2-17}
\begin{split}
\mathbf{j} &= - \sigma \nabla \varphi - F \sum_i \nu_i D_i \nabla c_i \\\
	   &= \mathbf{j}_\mathrm{m} + \mathbf{j}_\mathrm{d}.
\end{split}
\end{equation}
\sout{In case of the electrolyte containing only two ions, the diffusive current density is zero due to electroneutrality.} Taking conservation of charge into account leads to the Poisson equation 
\begin{equation}
\label{eq:2-18}
\nabla \cdot \left( \sigma \nabla \varphi \right) = \nabla \cdot \left(- F \sum_i \nu_i D_i \nabla c_i \right)
\end{equation}
for calculating the potential distribution in an electrolyte. 
\sout{In the absence of concentration gradients in the electrolyte, the current can be determined by migration only and Eq.\,\eqref{eq:2-17} reduces to Eq.\,\eqref{eq:2-5} which describes the current density in the electrodes.}\\
From the equations in section \ref{eqnElec}, it can be concluded that the potential distribution inside the whole liquid metal battery can be calculated using only the Poisson equation \eqref{eq:2-18}. The source term on the right-hand side (RHS) of Eq.\,\eqref{eq:2-18} is zero in the electrodes. Since the present study solves the potential jump at the interfaces between electrode and electrolyte as well, the meshes of electrolyte and electrode can be coupled. Solving the potential distribution on a global mesh respects both, the potential jumps as internal jumps and the different regions in the battery. The potential jumps are addressed in section \ref{potJump} in more detail. \\
Further, the species -- or mass -- transport can be derived by combining Eq.\,\eqref{eq:2-11} and Eq.\,\eqref{eq:2-13}. This leads to
\begin{equation}
\label{eq:2-19}
\frac{\partial c_i}{\partial t} = \nu_i \mu_i F \nabla \cdot \left( c_i \nabla \varphi \right) + \nabla \cdot \left(D_i \nabla c_i \right)
\end{equation}
which describes the transport of each species.\sout{As fluid flow is neglected, the last term on the RHS is zero.} At all boundaries not belonging to an electrolyte-electrode interface, zero flux boundary conditions apply.\\
For the electrolyte, a general case with more than two species and one active species is regarded. Therefore, each species must be classified as active or passive. Only active species -- here lithium cations -- take part in the electrode-reactions. To \sout{start}\newtxt{derive their BCs}, Eq.\,\eqref{eq:2-11} is reordered leading to an expression for the potential gradient. Multiplying with $\nu_i / D_i$, summing up over all species and using the Nernst-Einstein relation (Eq.\,\eqref{eq:2-12}) gives
\begin{equation}
\label{eq:2-20}
\nabla \varphi = - \frac{R T \sum_i \mathbf{N}_i \frac{\nu_i}{D_i}}{F \sum_i \nu_i^2 c_i}.
\end{equation}
Together with Eq.\,\eqref{eq:2-15}, the following can be derived for an arbitrary species $j$:
\begin{equation}
\label{eq:2-21}
\nabla c_j = \frac{- \mathbf{N}_j + \nu_j D_j c_j \frac{\sum_i \mathbf{N}_i \frac{\nu_i}{D_i}}{\sum_i \nu_i^2 c_i}}{D_j}.
\end{equation}
For a species $j$, the current density can be calculated as $\mathbf{j} = F \nu_j \mathbf{N}_j$ using Eq.\,\eqref{eq:2-10}. The term $\sum_i \mathbf{N}_i \frac{\nu_i}{D_i}$ must include only active species, since no passive species flows across the boundaries. 
To differentiate between active (index $1$) and passive (index $p$) species, two boundary conditions are needed. Firstly, for the active species
\begin{equation}
\label{eq:2-22}
\nabla c_1 = \frac{\mathbf{j}}{F \nu_1 D_1} \left(\frac{\nu_1^2 c_1}{\sum_i \nu_i^2 c_i} - 1 \right)
\end{equation}
can be found. Secondly, for the passive species
\begin{equation}
\label{eq:2-23}
\nabla c_p = \frac{\mathbf{j}}{F \nu_1 D_1} \frac{\nu_1 \nu_p c_p}{\sum_i \nu_i^2 c_i} 
\end{equation}
is valid.
Since the condition of electroneutrality (Eq.\,\eqref{eq:2-14}) must be fulfilled, the species transport needs to be solved for $n-1$ species only. The concentration distribution for species $n$ is calculated from electroneutrality.\\
The species transport is solved on the local mesh for the electrolyte region, using the mapped current density. Here, both interfaces are treated with Neumann BCs according to the derived conditions for active (Eq.\,\eqref{eq:2-22}) and passive species (Eq.\,\eqref{eq:2-23}).\sout{ Index $i$ refers to the species in the electrolyte.}\\
%
%-----------------------------------------------------------------%
\subsubsection{Cathode}
\label{eqnCath}
At discharge, lithium dissolves into the bismuth cathode. There, no migration occurs since no charged species is present. Therefore, the transport equation for the molar concentration respectively lithium distribution \sout{(Eq.\,\eqref{eq:2-19}) simplifies to}\newtxt{reads}
\begin{equation}
\label{eq:2-24}
\frac{\partial c}{\partial t} = \nabla \cdot \left(D \nabla c \right).
\end{equation}
For the \newtxt{boundary condition of the} lithium concentration at the cathode interface, the derivation is quite straightforward. Equation \eqref{eq:2-10} reduces to $\mathbf{j} = Fz\mathbf{N}$, since there is only a flux of lithium. Together with Eq.\,\eqref{eq:2-24} the boundary condition is
\begin{equation}
\label{eq:2-25}
\nabla c = - \frac{\mathbf{j}}{F z D}.
\end{equation}
To solve the lithium distribution in the cathode, a zero gradient Neumann BC 
%%
%\begin{equation}
%\label{eq:2-26}
%\left. \frac{\partial c_1}{\partial \mathbf{n}} \right|_{f} = 0
%\end{equation}
%%
is applied at the bottom patch.\sout{ while the BC used for the electrolyte-cathode interface follows Eq.\,\eqref{eq:2-25}}
%
%-----------------------------------------------------------------%
\subsection{Potential jump}
\label{potJump}
\sout{At the phase boundary between two substances, particularly metal and electrolyte, an electrical potential difference forms. One phase must carry a positive charge and the other phase a negative charge at the phase boundary. Together they form the electrical double layer. In section ref{sec:introLMB} it was explained that the present study does not resolve the EDL numerically.} Instead \newtxt{of resolving the EDL numerically,} the electrode potential of each half-cell can be expressed -- on a macroscopic scale -- as a potential jump between electrode and electrolyte. In the following it is shown how this is done in particular.\\
The Li$\|$Bi liquid metal battery is a concentration cell and the electrode reaction at the anode
\begin{equation}
\label{eq:2-28}
\mathrm{Li} \longrightarrow \mathrm{Li}^+ + e^-,
\end{equation}
and at the cathode
\begin{equation}
\label{eq:2-29}
\mathrm{Li}^{+} + e^- \longrightarrow \mathrm{Li(Bi)}
\end{equation}
occur due to the material paring Li$\|$Bi.
So, the potential jumps at the anode-electrolyte interface and the
electrolyte-cathode interface are in accordance to the Nernst-Equation
\begin{equation}
\label{eq:2-30}
\varphi = \varphi_0 + \frac{R T}{z F} \ln \frac{a_{\mathrm{ox}}}{a_{\mathrm{red}}}
\end{equation}
with the standard potential $\varphi_0$ and the activity of the reduced and oxidised species $a$. For a Li$\|$Bi LMB, the number of exchanged electrons is $z=1$. \\
Since concentration cells are made out of two half cells, the potential jumps at the anode and cathode are the following:
\begin{equation}
\label{eq:2-31}
\Delta \varphi_{\mathrm{a-e}} = \varphi_0 + \frac{R T}{z F} \ln \frac{a_{\mathrm{Li^+}}}{a_{\mathrm{Li}}},
\end{equation}
\begin{equation}
\label{eq:2-32}
\Delta \varphi_{\mathrm{e-c}} = \varphi_0 + \frac{R T}{z F} \ln \frac{a_{\mathrm{Li^+}}}{a_{\mathrm{Li(Bi)}}}.
\end{equation}
The activity of pure lithium is $a_{\mathrm{Li}}=1$; for the other
values, see section \ref{material}.

\begin{figure}[h]
  \includegraphics[width=\linewidth]{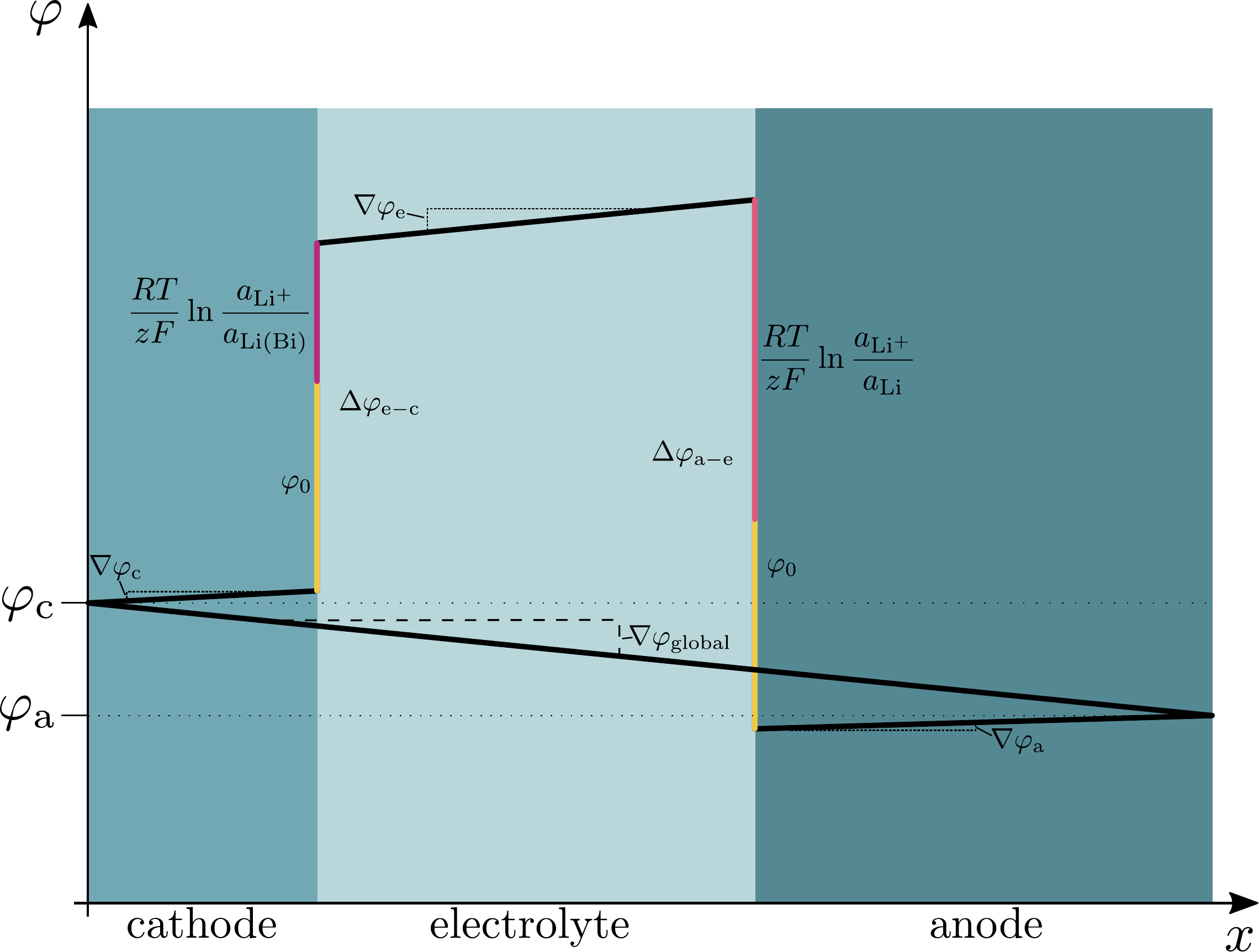}
  \caption{Relation between open circuit voltage and potential jumps without consideration of the overpotentials.}
  \label{fig:sketchOCV}
\end{figure}

In OpenFOAM, the potential jump can be modelled using a jump cyclic
boundary condition \cite{Takayama2011,Moreira2017}. The basis for this
is the predefined fixedJump BC which is slightly modified by the
authors of the present study, so that the jump can be updated in every
time step. Mathematically the potential jump over an interface $f$ is expressed as
\begin{equation}
\label{eq:2-33}
\left. \varphi_{\mathrm{N}} \right|_{f} = \left. \varphi_{\mathrm{O}} \right|_{f} - \Delta \varphi,
\end{equation}
where $\mathrm{O}$ refers the owner and $\mathrm{N}$ to the neighbour cell as it was shown in Fig.\,\ref{fig:simulationSetup}. The OCV of the battery (see Eq.\,\eqref{eq:2-1}) is the difference between the two potentials of the electrodes $E_{\mathrm{OC}} = \varphi_{\mathrm{c}} - \varphi_{\mathrm{a}}$. Figure \ref{fig:sketchOCV} shows a schematic sketch of the potential profile in the battery. Here, the potential jump $\Delta \varphi$ at both interfaces indicated by Eq.\,\eqref{eq:2-33} can be seen as well.
In a global manner, the potential gradient in the cell needs to be negative since for discharge, the potential of the anode must be smaller than the one of the cathode. For the specific regions, the potential gradient is positive as it is indicated by Eq.\,\eqref{eq:2-5}.
%
%-----------------------------------------------------------------%
\subsection{Material properties}
\label{material}
To compute the initial Li concentration in Bi, the density of the
Li(Bi) alloy in the cathode is required. The latter is obtained (in
kg/m$^3$) as an empirical formula which is based on values from Wax et
al. \cite{Wax2011} and Steinleitner et al. \cite{Steinleitner1975} as
\begin{equation}
\label{eq:2-34}
\rho = -7357 x^2 - 2045 x + 9863
\end{equation}
with $x$ denoting the molar fraction of Li in Bi. For alternative
density functions, see \cite{Weber2022}.
Subsequently, the concentration of lithium can be computed as
\begin{equation}
\label{eq:2-37}
c = \frac{x \rho}{x M_{\mathrm{Li}} + \left(1 - x\right) M_{\mathrm{Bi}}}. 
\end{equation}
The molar concentrations of the different species in the electrolyte can be calculated as
\begin{equation}
\label{eq:2-38}
c_i = \chi_i c_{\mathrm{salt}} 
\end{equation}
with $\chi$ being the molar fraction of the species in the
electrolyte. To calculate the total concentration of the electrolyte
\begin{equation}
\label{eq:2-39}
c_{\mathrm{salt}} = \frac{\rho_{\mathrm{salt}}}{M_{\mathrm{salt}}}
\end{equation}
the molar mass 
\begin{equation}
\label{eq:2-40}
M_{\mathrm{salt}} = \sum \chi_i M_i
\end{equation}
needs to be calculated using the electrolyte composition, where the
subscript $i$ refers to the salt components and not to the single species. Here, eutectic LiCl-KCl
(58.8:41.2 mol\%) is considered. \\
The electric conductivity of the electrolyte
%%
%\begin{equation}
%\label{eq:2-41}
%\sigma_{\mathrm{salt}} = \frac{F^2}{R T} \sum{\nu_i^2 D_i c_i},
%\end{equation}
%%
can be used to estimate the diffusivity of the species. \sout{In molten
salts, ions behave very different than in dilute solutions. Further,
$D_i$ may be highly dependent on the composition of the salt. }Morgan and Madden \cite{Morgan2004} investigated the diffusion coefficients of the species in the here concerned molten salt for some temperatures and in dependence of the molar fraction of the potassium ion $\chi_{\mathrm{K}^+}$. For the present $\chi_{\mathrm{K}^+}=0.206$, it is found that $D_{\mathrm{Li}^+} > D_{\mathrm{K}^+} > D_{\mathrm{Cl}^-}$. To adapt the diffusion coefficients to a specific temperature, ratios between the $D_i$ obtained by Morgan and Madden \cite{Morgan2004} at $T=900$K, were derived. It is found that
\begin{equation}
\label{eq:2-42}
\begin{split}
A &= \frac{D_{\mathrm{Cl}^-}}{D_{\mathrm{Li}^+}} = 0.807 \\\
B &= \frac{D_{\mathrm{K}^+}}{D_{\mathrm{Li}^+}} = 0.895.
\end{split}
\end{equation}
Applying the relations above to \newtxt{Eq}.\,\eqref{eq:2-16} \newtxt{with using Eq}.\,\eqref{eq:2-12}, the diffusion coefficient for lithium in LiCl-KCl (eutectic) 
\begin{equation}
\label{eq:2-43}
D_{\mathrm{Li}^+} = \frac{\sigma_{\mathrm{salt}} R T}{F^2 \left( c_{\mathrm{Li}^+} + A c_{\mathrm{Cl}^-} + B c_{\mathrm{K}^+} \right)} 
\end{equation}
can be derived. Ionic diffusion coefficients for the other species
follow from the relations in Eq.\,\eqref{eq:2-42}.

Another important property is the activity of Li in Bi and the
activity of the various ions in the molten salt. The latter are of
interest for the calculation of the Nernst equation \eqref{eq:2-30}
and the potential jump at the interfaces between electrolyte and
electrodes. As the present study aims to develop and demonstrate the
electrochemical model in its simplest form, the activity of Li in Bi
is assumed not to deviate from ideality and is described as
\begin{equation}
\label{eq:2-45}
a_\mathrm{Li(Bi)} = x_\mathrm{Li(Bi)}.
\end{equation}
For an overview on realistic activities of Li in Bi, see \cite{Weber2022}.

Unfortunately, values for activities of molten salts are often only
approximately known and determinable
\cite{Burgot2017, Salanne2008}. Therefore, the ionic activities of
Li$^+$, K$^+$ and Cl$^-$ are estimated using Temkin's model as
\cite{Temkin1945,Fellner1983} 
\begin{equation}
\label{eq:2-44_1}
\quad a_i = \chi_i
\end{equation} 
again assuming the activity coefficient to be unity. Please note that $\chi_i$ describes here the mole fraction of an ion related to all ions of the same charge. So, in terms of determining the activity of a salt, separate solutions of cations and anions are considered.

The values used for the simulation will be explicitly given in section \ref{valSimPar}.
%
%-----------------------------------------------------------------%

%%% Local Variables:
%%% mode: latex
%%% TeX-master: "../paper"
%%% TeX-parse-self: t
%%% TeX-auto-save: t
%%% TeX-PDF-mode: t
%%% eval: (auto-fill-mode 1)
%%% eval: (flyspell-mode 1)
%%% eval: (reftex-mode 1)
%%% ispell-dictionary: "british"
%%% End:

\section{Numerical implementation}
\label{numImplement}
%-----------------------------------------------------------------%
\subsection{Modified species transport equation}
\label{eqnTrans}
\sout{Besides identifying equations and defining BCs, the numerical
implementation of those is important and sometimes challenging.}In our
model, special care needs to be taken in expressing the migrational flux when solving the species transport equation \eqref{eq:2-19}. 
As explained in detail in \ref{chapter:appSpeciesEqn}, limitations in
the gradient calculation of the potential require rewriting the
migrational term.
Thus, Eq.\,\eqref{eq:2-19} is reformulated as 
\begin{equation}
\label{eq:3-1}
\frac{\partial c_i}{\partial t} = \nabla \cdot \left(\Psi c_i \right) + \nabla \cdot \left(D_i \nabla c_i \right)
\end{equation}
with
\begin{equation}
\label{eq:3-2}
\Psi = \nu_i \mu_i F \nabla \varphi.
\end{equation}
Reordering Eq.~\eqref{eq:2-17} gives
\begin{equation}
\label{eq:3-6}
\Psi = - \nu_i \mu_i F \left(\frac{\mathbf{j} + F \sum_i \nu_i D_i \nabla c_i}{\sigma} \right),
\end{equation}
\newtxt{so that the electric potential is not required any more for solving the species transport.}\sout{which does not require the electric potential any more.} This
formulation is necessary to avoid the generation of divergence near
the electrode-electrolyte interface. \sout{The given boundary conditions for the concentration gradient $\nabla c_i$ are capable to handle this issue.}
%
%-----------------------------------------------------------------%
\subsection{Mesh coupling and simulation routine}
\label{meshCoupling}
The computational domain consists of three regions for anode, electrolyte and cathode. Since the electric current passes all three regions, a classical segregated solver with potential coupling only at the interfaces using boundary conditions is not suitable and shows poor convergence \cite{Weber2018}. Facing this problem, Weber et al. introduced a parent-child mesh technique for calculating the potential distribution in a liquid metal battery. For that matter, global variables are solved on a global (parent) mesh and local (child) meshes for specific regions are used to solve local variables. Before, this approach has already been used for calculations of thermal conduction \cite{Beale2013,Beale2016} and eddy-current problems \cite{Beckstein2017}.\\
\sout{According to Beale et al. \cite{Beale2016} non overlapping regions can be modelled in three ways: (i) by a single global mesh for all regions and one set of transport equations, (ii) by a set of coupled meshes with a set of transport equations each that can be coupled at their boundaries and (iii) by an integrated cell concept where the global mesh provides a framework for the solution and individual meshes support the solution with additional transport equations solved locally.}
The latter matches with the parent-child-mesh technique used in the present study. Field variables and region properties can be mapped between the different meshes and they share common values. Globally, the boundaries between the regions are internal boundaries, while locally, the boundaries are external. This allows more flexibility and better computational efficiency while more detailed simulations can be performed. \\
In the present study the electric potential and current density are
solved on the global mesh. Further, there is a mesh for each electrode
and one for the electrolyte. The anode is not of special interest, so
there are no additional transport equations solved. In the cathode,
the concentration of lithium in bismuth is calculated. Whereas, in the
electrolyte the corresponding electric conductivity, the species
distribution and the diffusive current density are computed. Here,
it is important to consider that the electric conductivity is
different in each region. This will further be discussed in section
\ref{discretisation}. 

\begin{figure}
  \includegraphics[width=\linewidth]{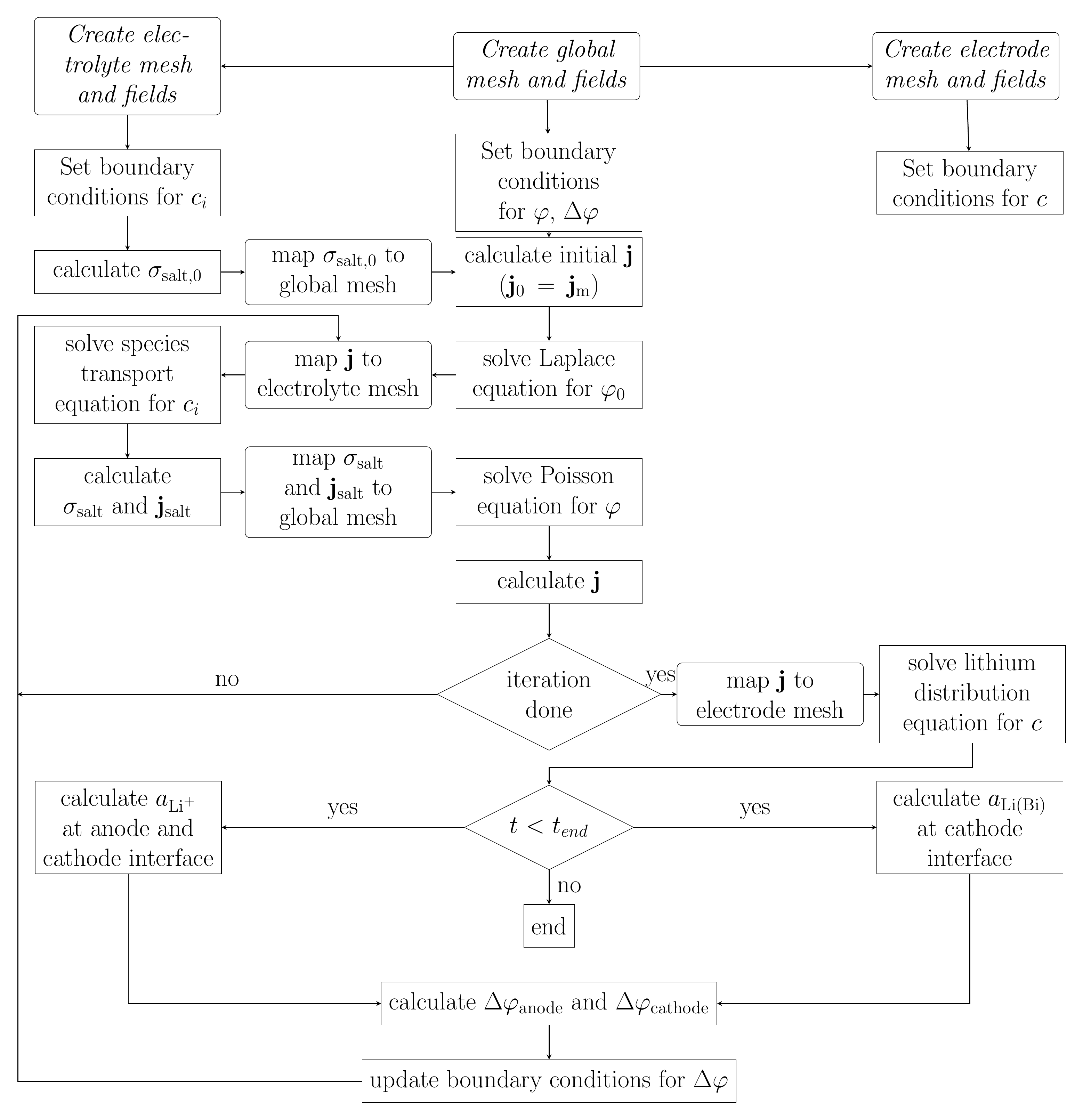}
  \caption{Flowchart of the simulation model.}
  \label{fig:simulationScheme}
\end{figure}

In Fig.\,\ref{fig:simulationScheme}, the scheme of the simulation model is shown.
It can be seen that the electric conductivity $\sigma$ and the current density $\mathbf{j}$ are the variables of interest for mapping.
As an overview, Tab.\,\ref{tab:eqnMeshes} shows the equations solved
in the different regions. All of them are solved in a segregated manner and are iteratively coupled. 
\begin{table}[h]
  \begin{center}
  \caption{Meshes and the related equations.}
  \begin{tabular}{ l l }
  \hline
  mesh & equations \\ 
  \hline
  global & Eq.\,\eqref{eq:2-18} and Eq.\,\eqref{eq:2-17}   \\  
  electrolyte & Eq.\,\eqref{eq:3-1} using Eq.\,\eqref{eq:3-6}   \\
  cathode & Eq.\,\eqref{eq:2-24}  \\ 
  \hline   
  \end{tabular}
  
  \label{tab:eqnMeshes}
  \end{center}
\end{table}

\sout{To connect the different regions, it is important to give special
attention to the electric boundary conditions. Similar to Colli and
Bisang \cite{Colli2018}, who proposed a model with region coupling to
calculate potential and current distribution, the calculations can be
done for a given local potential at any electrode, fixed cell
potential difference or galvanostatic control. To express this
numerically, Dirichlet and Neumann boundary conditions at the parent 
mesh are used.}
Further, \newtxt{continuity of the current density ensures sufficient region and mesh coupling.}\sout{current density must be continuous from cell to cell.} \newtxt{Numerically, this requires continuity at the cell faces -- especially at those cells connecting two regions.}\sout{ especially from cells being located in different regions respectively in different child meshes. This can be expressed as
\begin{equation}
\label{eq:3-7}
\sigma_{\mathrm{N}} \left. \frac{\partial \varphi_{\mathrm{N}}}{\partial n} \right|_{f} = - \sigma_{\mathrm{O}} \left. \frac{\partial \varphi_{\mathrm{O}}}{\partial n} \right|_{f} = j_n,
\end{equation}
where $n$ is the coordinate normal to the cell face $f$, pointing out of the cell and $j_n$ is the current density normal to the boundary.}
\newtxt{The potential jump has a similar importance. Both leads to the consequence that special discretisation in the interface region is required.}\sout{The interfaces between electrodes and electrolyte need special treatment due to the potential jump. They are created as internal patches with periodic boundary conditions, which physically connect two boundaries.} This will be elaborated in the following sections.
%
%-----------------------------------------------------------------%
\subsection{Discretisation}
\label{discretisation}
The potential jump and the multi-mesh approach require special attention for the discretisation of some terms in the solved equations. Especially the Poisson equation (Eq.\,\eqref{eq:2-18})\sout{ and the species transport (Eq.\,\eqref{eq:3-1})} needs to be considered.\\ 
Equation \eqref{eq:2-24} for calculating the lithium distribution in the cathode \newtxt{and }Eq.\,\eqref{eq:3-1} \newtxt{for the species transport in the electrolyte} do not need special treatment. Here, a first order Euler implicit scheme for the time discretisation, a second order linear scheme for the Laplacian \newtxt{and an upwind differencing interpolation scheme for the divergence term are used}.
%
%\subsubsection{Discretisation of the potential \sout{transport}\colorbox{yellow}{distribution}}
%\label{discretisationPot}
%
Due to the different regions with highly differing electric conductivities, sharp jumps at the interfaces are present. As \newtxt{continuity at the cell faces is required}\sout{shown in Eq.\,\eqref{eq:3-7}}, the face values for the potential and the conductivity are \sout{required for continuity}\newtxt{needed}. In the FVM, cell values are interpolated to the faces. To avoid wrong potential calculations due to inconsistency caused by numerical errors, the electric conductivity must be interpolated harmonically. This can be expressed as
\begin{equation}
\label{eq:3-8}
\sigma_f = \left( \frac{\delta_{\mathrm{O}}}{\delta \sigma_{\mathrm{O}}} + \frac{\delta_{\mathrm{N}}}{\delta \sigma_{\mathrm{N}}} \right)^{-1} 
\end{equation}
with $\delta_{\mathrm{O}}$ and $\delta_{\mathrm{N}}$ being the distance from cell centre to cell face of owner respectively neighbour cell and the distance between both cell centres $\delta$ \cite{Weber2018, Jasak1996, Beckstein2017}.\sout{This scheme, as an extension of a scheme by Jasak was also used by Beckstein et al. to evaluate the discontinuity in the potential gradient, based on the discontinuity of electric conductivity between conducting and non-conducting regions.}\\
Moreover, care must be taken when discretising Eq.\,\eqref{eq:2-17} and Eq.\,\eqref{eq:2-18} since they include $\sigma$. First, the current density calculation is described. OpenFOAM uses the Gauss theorem \cite{Greenshields2015} to compute the gradient
\begin{equation}
\label{eq:3-9}
\int_V \nabla \varphi dV = \int_S d\mathbf{S} \varphi = \sum_f \mathbf{S}_f \varphi_f
\end{equation}
with the volume $V$ and the surface area $S$ of a cell, the surface
area vector $\mathbf{S}$ and the face area vector $\mathbf{S}_f$. So,
the face values of the potential are needed.  Following this, the
potential gradient is discretised using the weightedFlux scheme, which
is derived in a similar way as harmonic interpolation as in \cite{Weber2018}
\begin{equation}
\label{eq:3-10}
\varphi_f = w \varphi_{\mathrm{O}} + \left(1 - w \right) \varphi_{\mathrm{N}}
\end{equation}
with
\begin{equation}
\label{eq:3-11}
w = \frac{\delta_{\mathrm{N}} \sigma_{\mathrm{O}}}{\delta_{\mathrm{O}} \sigma_{\mathrm{N}} + \delta_{\mathrm{N}} \sigma_{\mathrm{O}}},
\end{equation}
which gives accurate solutions for the migrational current density.

Solving Eq.~\eqref{eq:2-18} for the electric potential, a consistent
discretisation like in the balanced-force approach
\cite{Popinet2009, Thirumalaisamy2018} is employed to guarantee numerical accuracy.\sout{
and to avoid numerical smearing. Originally, this approach is often used in
multi-phase solvers to compute the pressure jump caused by surface
tension at the boundary between two phases. Both forces must be
calculated at the faces for consistency} This idea
is applied to the present model, which means that the left-hand side
(LHS) and RHS of the Poisson equation \eqref{eq:2-18} need to be in
exact balance to calculate the potential jump correctly. The LHS is
discretised using a harmonic Laplacian scheme
\begin{equation}
\label{eq:3-12}
\int_V \nabla \cdot \left( \sigma \nabla \varphi \right) dV = \int_S d\mathbf{S} \cdot \left( \sigma \nabla \varphi \right) = \sum_f \sigma_f \mathbf{S}_f \cdot \nabla \left( \varphi \right)_f.
\end{equation}
It can be seen that the potential gradient is calculated at the cell faces.\sout{Using the harmonic scheme prevents interpolational errors of $\sigma$.} The RHS\sout{ can be interpreted as the divergence of the diffusive current density and} is discretised as
\begin{equation}
\label{eq:3-13}
\int_V \nabla \cdot \mathbf{j}_{\mathrm{d}} dV = \int_S d\mathbf{S} \cdot \mathbf{j}_{\mathrm{d}} = \sum_f \mathbf{S}_f \cdot \left( \mathbf{j}_{\mathrm{d}}\right)_f.
\end{equation}
Here, the face values of the diffusive current density $\left(\mathbf{j}_{\mathrm{d}}\right)_f$ are needed. To avoid inaccurate face values, interpolation from cell to face must be avoided. Instead, the face value $\left(\mathbf{j}_{\mathrm{d}}\right)_f$ is calculated directly from the face values of the concentration gradient $\nabla (c_i)_f$:
\begin{equation}
\label{eq:3-14}
\sum_f \mathbf{S}_f \cdot \left( \mathbf{j}_{\mathrm{d}}\right)_f = \sum_f \mathbf{S}_f \cdot \left(-F \sum \nu_i D_i \left(\nabla c_i\right)_f \right).
\end{equation}
This is accurate, since $\left(\nabla c_i\right)_f$ is implemented as BC.
Using the procedure above, $\mathbf{j} = \mathbf{j}_{\mathrm{m}} + \mathbf{j}_{\mathrm{d}}$ (Eq.\,\eqref{eq:2-17}) is continuous over the computational domain. \\
%
%-----------------------------------------------------------------%
%\subsubsection{Discretisation of the species transport}
%\label{discretisationSpec}
%%
%The electric conductivity in the electrolyte, which is given by
%Eq.\,\eqref{eq:2-16}, is directly obtained from the species
%distribution. As previously discussed, it is of high importance to
%obtain accurate conductivities especially near the
%interfaces. Therefore, care must be taken when discretising the
%species transport equation, as well. The first order Euler implicit
%scheme is used for the time discretisation and a second order linear
%scheme for the Laplacian. The divergence term was derived in section
%\ref{eqnTrans} and is discretised as \cite{Greenshields2015}
%%
%\begin{equation}
%\label{eq:3-15}
%\int_V \nabla \cdot \left(\Psi c \right) dV = \int_S d\mathbf{S} \cdot \left(\Psi c \right) = \sum_f \mathbf{S}_f \cdot \Psi_f c_f.
%\end{equation}
%%
%Values for $c_f$ are calculated through the chosen upwind differencing interpolation scheme.
%
%-----------------------------------------------------------------%

%%% Local Variables:
%%% mode: latex
%%% TeX-master: "../paper"
%%% TeX-parse-self: t
%%% TeX-auto-save: t
%%% TeX-PDF-mode: t
%%% eval: (auto-fill-mode 1)
%%% eval: (flyspell-mode 1)
%%% eval: (reftex-mode 1)
%%% ispell-dictionary: "british"
%%% End:

%\section{\colorbox{yellow}{Verification and} validation}
\section{Verification and validation}
\label{validation}
To \sout{validate}\newtxt{verify} the numerical solver, a comparative study with the commercial FEM solver COMSOL Multiphysics is performed. \newtxt{Detailed results can be found in} \ref{chapter:appValidationComsol}\newtxt{. Experimental investigations of the composition gradients in LiCl-KCl from Vallet et al.} \cite{Vallet1983} \newtxt{are further taken into consideration for the validation of the presented numerical solver.}
%
%-----------------------------------------------------------------%
%
%\subsection{Simulation results \colorbox{yellow}{for comparison OpenFOAM-COMSOL}}
\subsection{Simulation results for comparison OpenFOAM-COMSOL}
\label{valSimResults}
\subsubsection{Simulation parameters}
\label{valSimPar}
In the following, a Li$|$LiCl-KCl$|$Bi liquid metal
battery at $T=450^\circ$C is considered. The LiCl-KCl electrolyte is
an eutectic mixture (58.8:41.2 mol\%) and can be classified as \sout{binary }salt system of three ions -- in the present study the electrolyte is therefore named ``ternary''\sout{ -- and two components sharing a common ion}. The geometry of the battery is simplified according to Fig.\,\ref{fig:simulationSetup}\newtxt{, but numerically considered to be one-dimensional}.\sout{ with a cross section of 1cm$^2$. Current collectors are neglected.} The height of the anode is $h_{\mathrm{a}} = 16$mm, of the electrolyte $h_{\mathrm{e}} = 5$mm and of the cathode $h_{\mathrm{c}} = 4$mm. The anode-electrolyte interface is located at $x=9$mm and the electrolyte-cathode interface at $x=4$mm. A discharge current density of $\mathbf{j} = -100$mA/cm$^2$ and a time of $t=600$s are chosen. 
Molar mass, density and electric conductivity for the electrodes and the electrolyte are shown in Tab.\,\ref{tab:matProp}.\\
\begin{table}[h]
  \begin{center}
  \caption{Material properties at 450$^\circ$C \cite{Iida1988, Iida2015, Janz1975}.}
  \begin{tabular}{ l l l l}
  \hline
  Property & Li & Bi & LiCl-KCl \\ 
  \hline
  M in g/mol & 6.94 & 208.98 & 55.64\\
  $\rho$ in kg/m$^3$ & 491 & 9839 & 1671 \\  
  $\sigma$ in S/m & 2.78$\cdot 10^6$ & 7.14$\cdot 10^5$ & 157.28 \\  
  \hline   
  \end{tabular}
  
  \label{tab:matProp}
  \end{center}
\end{table}
Estimating the initial concentration of the species and their diffusion coefficients in the electrolyte is done in accordance with the equations stated in section $\ref{material}$. The initial concentration of the electrolyte is $c_{\mathrm{salt}}=30\,028$ mol/m$^3$. Using the molar fractions of each species, \sout{resulting from the eutectic mixture,} the concentrations for each ion can be calculated as shown in Tab.\,\ref{tab:initCon}. The diffusion coefficients follow from Eq.\,\eqref{eq:2-43} and Eq.\,\eqref{eq:2-42}. The diffusivity of lithium in bismuth\sout{ $D_{\mathrm{Li(Bi)}}$} is calculated according to Newhouse \cite{Newhouse2014}:
\begin{equation}
\label{eq:4-1}
D_{\mathrm{Li(Bi)}} = \exp{\frac{-4.081 c_{\mathrm{Li}} - 0.01315}{c_{\mathrm{Li}}^2 + 0.3742 c_{\mathrm{Li}} + 0.001572}}
\end{equation}
with $c$ denoting the molar concentration of Li in Bi in mol/cm$^3$.
\begin{table}[h]
  \begin{center}
  \caption{Initial concentrations.}
  \begin{tabular}{ l l l l l}
  \hline
  Species & Li(Bi) & Li$^+$ & Cl$^-$ & K$^+$ \\ 
  \hline
  $z$ / $\nu$ & 1 & 1 & -1 & 1 \\
  $c$ in mol/m$^3$ & 13125 & 8828 & 15014 & 6186 \\  
  $D$ in m$^2$/s & 4.43$\cdot 10^{-9}$ & 3.84$\cdot 10^{-9}$ & 3.1$\cdot 10^{-9}$ & 3.43$\cdot 10^{-9}$ \\  
  \hline   
  \end{tabular}
  
  \label{tab:initCon}
  \end{center}
\end{table}
The initial concentration of lithium in bismuth can be derived from Eq.\,\eqref{eq:2-34} and \eqref{eq:2-37}. Here, an arbitrary initial state of charge with an initial molar fraction of lithium $x_0=0.236$ is chosen. 

To \sout{validate}\newtxt{verify} the presented solver\newtxt{, obtain deeper and piecewise detailed results as well as to investigate the influence of individual parameters on the cell performance}, four test cases \newtxt{of}\sout{with} different \sout{extent of }complexity are used:
\begin{enumerate}
        \item binary electrolyte (LiCl), equal diffusion coefficients, potential jump is not modelled
        \item binary electrolyte (LiCl), potential jump is not modelled
        \item ternary electrolyte (LiCl-KCl), potential jump is not modelled
        \item ternary electrolyte (LiCl-KCl), fixed potential jump is modelled  
\end{enumerate}
Here, ``binary'' refers to a salt with two ions. Although the test cases are purely fictive, the values in Tab.\,\ref{tab:matProp} are
used. For test case 1 it is assumed that $D_{\mathrm{Cl}^-} =
D_{\mathrm{Li}^+}= 3.84\cdot 10^{-9}$ m$^2$/s and $c_{\mathrm{Li}^+}=c_{\mathrm{Cl}^-}=13239$ mol/m$^3$, while for test case 2
$c_{\mathrm{Li}^+}=c_{\mathrm{Cl}^-}=14654$ mol/m$^3$ is used. The
potential jumps in test case 4 are arbitrarily chosen as $\Delta
\varphi_{\mathrm{a-e}} = 0.8$V and $\Delta \varphi_{\mathrm{e-c}} =
1$V. In all simulations, the active species lithium is calculated from
electroneutrality. An additional grid study and electroneutrality
study can be found in \ref{chapter:appGrid} and
\ref{chapter:appElectro}. 

For each simulation, the species concentration in the electrolyte and
the potential distribution in the whole cell are compared between
COMSOL multiphysics and OpenFOAM to \sout{validate the solver}\newtxt{verify the latter}.\sout{Further, these very simple test cases allow to obtain a first impression of the influence of
various parameters on the cell performance.} \newtxt{Only main findings are presented here, additional information and figures can be found in} \ref{chapter:appValidationComsol}. \newtxt{In general, the results -- even if only presented in supplementary material -- of the OpenFOAM simulations highly coincide with the results obtained using COMSOL. The former can therefore be considered to be verified.}
%
%\subsubsection{\colorbox{yellow}{Test cases}}
\subsubsection{Test cases}
\label{valCases}
%
%\subsubsection{\sout{Validation} Test case 1}
%\label{valCase1}
%
Test case 1 is the most simple one as it is highlighted in
\ref{chapter:appSpeciesEqn}. In the absence of concentration gradients in the electrolyte, the current can be determined by migration only and Eq.\,\eqref{eq:2-17} reduces to Eq.\,\eqref{eq:2-5} which describes the current density in the electrodes as well. Here, no diffusive current is flowing and only the Laplace equation for the potential distribution
Eq.\,\eqref{eq:2-7} needs to be solved.\sout{The species distribution and species concentration gradient for the Li$^+$ and Cl$^-$ ions are equal, since the electroneutrality condition Eq.\,\eqref{eq:2-14} is preserved within a tolerance of machine accuracy. The simulation results are shown in Fig.\,\ref{fig:resCase1}.
It can be seen that the solutions of OpenFOAM and COMSOL are in good
accordance and match perfectly with each other in every case except of
the species gradient at $t_0$.} Analytically calculating the diffusive
current density of the lithium ion from the concentration gradient for
the initial state using Eq.\,\eqref{eq:2-22} and Eq.\,\eqref{eq:2-17}
gives $j_{\mathrm{d, Li^+}}^{t_0} = 500$ A/m$^2$ which fits
with the results obtained by the OpenFOAM simulation.\sout{In contrast, the
COMSOL results are not reliable. For the chloride ion, the sign of the
diffusive current is the opposite. Hence, it can be concluded that the
diffusive current density is zero. As it can be seen in
Fig.\,\ref{fig:case1Spec}, the species distribution is point symmetric
to the electrolyte centre, since the concentration gradient for the
active and passive species are equal (see Eq.\,\eqref{eq:2-22} and
Eq.\,\eqref{eq:2-23}). Moreover, the potential gradient $\nabla
\varphi$ is linearly dependent on $\nabla c$ -- as shown by
Eq.\eqref{eq:2-17}. Therefore, the former needs to show a symmetric
trend in the electrolyte as well. Figure \ref{fig:case1gradPot}
confirms this.} Further, the electric losses in the electrolyte are\sout{ is
slightly increasing but} almost constant over time.\sout{, which indicates
that the changes in the electric conductivity are relatively
small in this case where the electrolyte contains two ions with equal diffusion coefficients.} \sout{Due to the symmetry of $c$ and Eq.\,\eqref{eq:2-16} it can be
deduced that the electric conductivity is symmetric, too.} 
%
%\subsubsection{\sout{Validation} Test case 2}
%\label{valCase2}
%

Not assuming equal diffusion coefficients in test case 2 makes it necessary to solve the Poisson equation for the potential distribution, Eq.\,\eqref{eq:2-18}. Thus, the diffusive part of the current density has now an impact on the solution\newtxt{: it influences the electric losses significantly, which indicates}\sout{The results can be found in Fig.\,\ref{fig:resCase2}. As in test case 1, the results of COMSOL and OpenFOAM match perfectly except for the diffusive current density in COMSOL at the initial time step.
Again, species distribution and diffusive current density of Li$^+$
and Cl$^-$ are equal and symmetric, the latter is true for the
electric conductivity as well. However, the potential gradient in the
electrolyte is no longer symmetric. Here, it can be concluded that the influence of the diffusion coefficient on the potential gradient is large.}
\sout{On a first glance, it seems that the electric conductivity in the electrolyte increases over time since the slope flattens. Due to the dependence $U = \frac{j l}{\sigma}$ with $U$ being the electric loss and $l$ being the resistor length, it arises the presumption that the electric loss in the electrolyte
decreases over time. Having a closer look at Eq.\,\eqref{eq:2-16}
reveals that this is partly misleading. Investigating the electric conductivity, which is constant on average, this indicates}that the overall ohmic loss
in the electrolyte is \sout{influenced}\newtxt{determined} not only by electric conductivity, but also by the diffusive current. The latter -- or diffusion itself
-- contributes to charge transport without causing ohmic losses. In
the present case, the electric loss decreases over time about
approx.~6\% during the whole simulation, so a better cell performance
due to increasing diffusive current could be expected. Similarly, the
diffusive current density is approx.~6\% of the current density on
average at the end of the simulation (cf.~Fig.\,\ref{fig:resCase2}). A clear coincidence can be seen
here. In contrast, unequal diffusion coefficients have no impact
on the species distribution as long as only two species are
involved\footnote{The values of the concentration gradients for test case 1
and 2 are coincident.}.
%
%\subsubsection{\sout{Validation} Test case 3}
%\label{valCase3}
%
\sout{Potential and potential gradient in test case 3 show the same behaviour as in test case 2. It can be seen in Fig.\,\ref{fig:resCase3_1} that the diffusive current leads to a decreasing electric loss over time. Compared to test case 2, $\nabla \varphi$ and $\varphi$ are slightly lower, which means that the electric loss decreases a little more than in test case 2.}

\begin{figure}[h]
  \centering
  \includegraphics[width=0.5\linewidth]{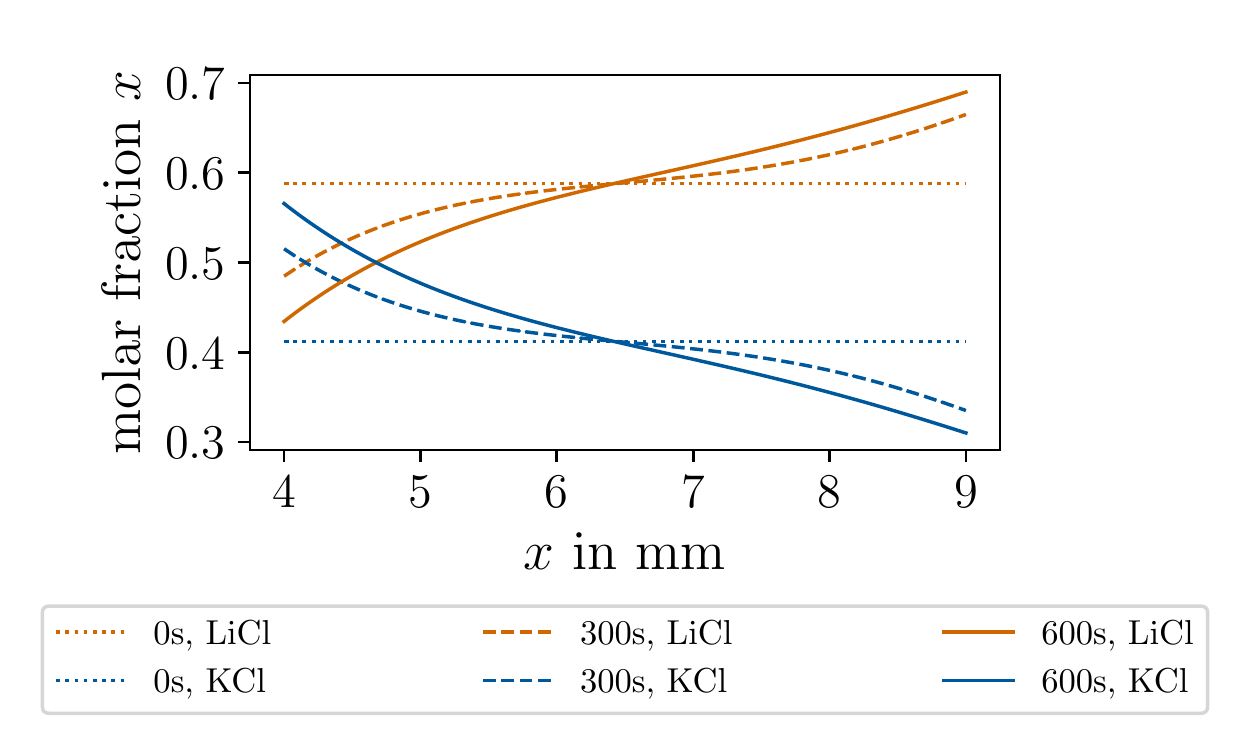}
  \caption{Test case 3; results for OpenFOAM simulation: species distribution of LiCl and KCl expressed in molar fraction.}
  \label{fig:resCase3_3}
\end{figure}

Now, a closer look to the species distribution is necessary since a ternary electrolyte is considered \newtxt{in test case 3}.\sout{It can be seen in Fig.\,\ref{fig:resCase3_2} that the Li$^+$, K$^+$ and Cl$^-$ species distributions and gradients are not equal any more. Nevertheless, they fulfil electroneutrality in terms of machine accuracy.}\sout{The electric conductivity $\sigma$ is constant on average over
time.} Calculating the initial diffusive current density for each ion
using Eq.\,\eqref{eq:2-22}, Eq.\,\eqref{eq:2-23} and
Eq.\,\eqref{eq:2-17} gives $j_{\mathrm{d, Li^+}}^{t_0} = 706$
A/m$^2$, $j_{\mathrm{d, Cl^-}}^{t_0} = 404$ A/m$^2$ and
$j_{\mathrm{d, K^+}}^{t_0} = -184$ A/m$^2$. \newtxt{In the simulations those values are obtained as well as it is shown in} Fig.\,\ref{fig:resCase3_1}.\sout{For later times, these values are obtained in the simulations only for the Cl$^-$ ion, being the common ion in both salts. Here, the distribution and gradient show symmetric behaviour, which is not the case for the Li$^+$ and K$^+$ ions when $t>t_0$. However, qualitatively the gradient of each species shows the correct sign and order, which indicates that the results are plausible.}
\sout{Moreover, }As it can be seen in Fig.\,\ref{fig:resCase3_3}, more LiCl is accumulated at the anode while more KCl is accumulated at the cathode interface and vice versa.\sout{ The deviations from symmetry at the anode are larger than at the cathode.} Compared with experimental work, the presented results are in qualitative good agreement with the results from Vallet et al.~\cite{Vallet1983} who determined composition gradients in LiCl-KCl\sout{ and observed a similar asymmetry}.
\sout{Moreover, an asymmetric potential gradient is observable
within the electrolyte -- as in test case 2. This leads to the logical
consequence that the reason for the asymmetry must lie in the
diffusive part of the material balance for the species transport
Eq.\,\eqref{eq:2-19}. Contrary to case 2, the concentration gradients
are now $\nabla c_{\mathrm{Li^+}} \ne \nabla c_{\mathrm{Cl^-}} \ne
\nabla c_{\mathrm{K^+}}$, which causes different material transport
for each species at both interfaces between electrolyte and
electrodes.} 
\sout{The fact that the deviation of the concentrations at the
cathode is larger than the one at the anode interface can be justified
as follows: the diffusion coefficient of lithium is larger than for
potassium, causing a locally varying electric conductivity that is
larger at the anode interface and smaller at the cathode
interface.} \sout{Together with Fig.\,\ref{fig:case3gradPot}} \sout{It is followed that the migrational current at the anode interface decreases less
than at the cathode. As a consequence of charge conservation, the
diffusive current at the anode interface needs to behave likewise. The
latter is only possible, if the concentration gradient of LiCl is
smaller than the one for KCl.} \\
%
%\subsubsection{\sout{Validation} Test case 4}
%\label{valCase4}
%
In order to approach the \sout{validation}\newtxt{verification} test cases closer to real conditions in a cell, two potential jumps are added at the electrolyte-electrode interfaces in test case 4. Everything else is taken from test case 3. The potential jumps $\Delta \varphi_{\mathrm{a-e}}$ and $\Delta \varphi_{\mathrm{e-c}}$ always refer to a viewpoint from electrolyte to electrode. \sout{In Fig.\,\ref{fig:resCase4_1} the potential distribution in the cell is shown. Since $\nabla \varphi_{\mathrm{case}4} \approx \nabla \varphi_{\mathrm{case}3}$, the potential gradient is not presented again and it is referred to Fig.\,\ref{fig:case3gradPot}.
As in the previous validation test cases, the results of OpenFOAM and COMSOL match very well except for the initial time step.}\sout{It can be seen that}\newtxt{Here again,} the potential at the anode decreases over time, which agrees with the previous observations: the electric resistance declines due to the diffusive current being lossless in terms of charge transfer. 
%

%\subsection{\colorbox{yellow}{Comparison with experiment}}
\subsection{Comparison with experiment}
\label{valVallet}
\newtxt{After the verification with COMSOL, the solver is validated
  using experimental data. Vallet et al.~electrolysed eutectic
  LiCl-KCl contained in porous yttria felt at 425$^\circ$C} \cite{Vallet1983}. \newtxt{The cell consisted of a 4.16\,mm thick
  electrolyte between two solid Li-Al electrodes and was operated for
  242\,s at 254\,mA/cm$^2$. After electrolysis, the sample was cooled
  down to room temperature in less than one second to analyse its
  composition. Atomic absorption spectroscopy (AAS) with a lateral
  resolution of 0.4\,mm and an accuracy of approximately 6\% was used
  together with energy-dispersive X-ray spectroscopy (EDX) to obtain
  the KCl molar fraction. Several EDX scans with an accuracy of 8\% in
  a slice of less than 0.18\,mm were averaged to obtain a single
  point. Finally, the data was fitted by an empirical formula.}

\newtxt{Modelling this experiment, the same material properties as in
  section} \ref{valSimPar} \newtxt{are used. The effective diffusion
  coefficients in the yttria matrix are obtained as}
\begin{equation}
\label{eq:4-2}
D_{\mathrm{eff},i}=\frac{D_i}{\tau^2}
\end{equation}
\newtxt{with the tortuosity $\tau$. As the latter had not been
  measured for the felt under consideration, a value of $\tau = 1.17$ for a
  similar yttria matrix is used} \cite{Vallet1982a}. \newtxt{The felt reduces
the diffusivities by almost 30\%}.
\begin{figure*}[h]
    \centering
    \includegraphics[width=0.5\textwidth]{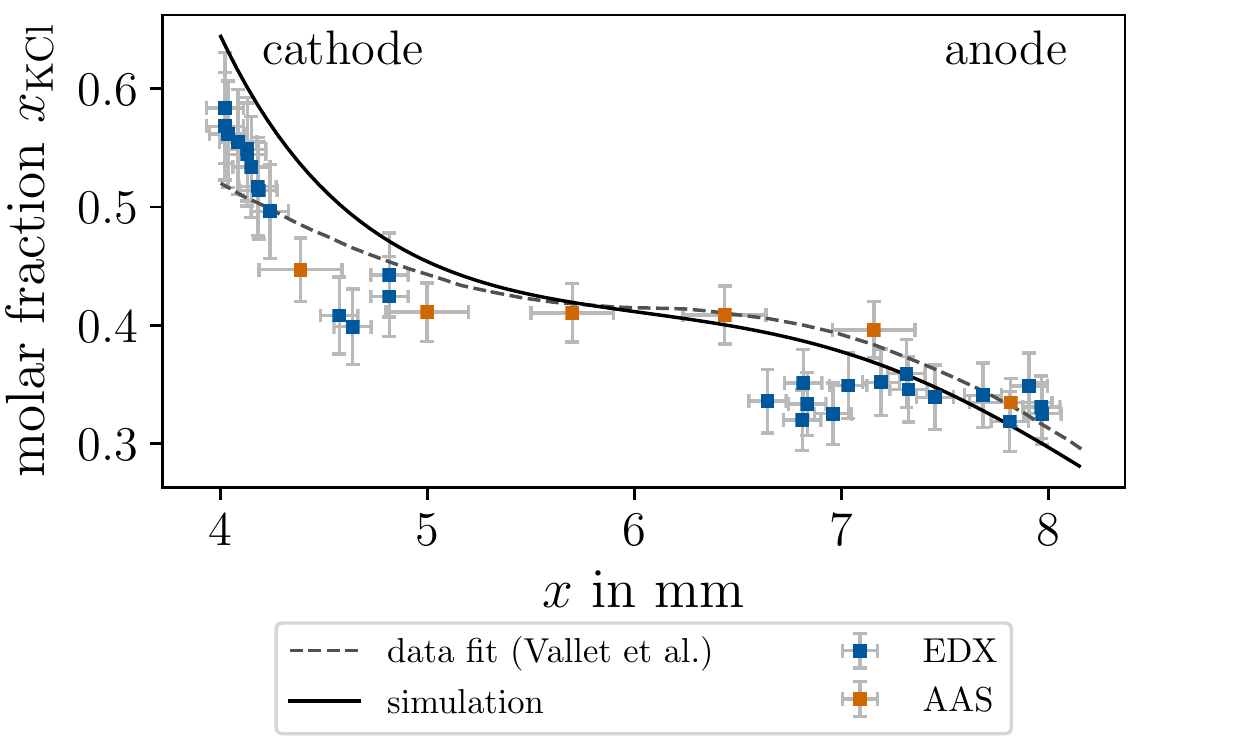}
    \caption[]
    {\small Distribution of KCl in cell No. 2 of Vallet et
      al. \cite{Vallet1983}.}
    \label{fig:vglVallet}
\end{figure*}

Fig.\,\ref{fig:vglVallet} \newtxt{shows the numerical and experimental
  results, which match fairly well. The remaining deviations can be
  attributed to a number of different sources, such as imperfect
  wetting of the yttria matrix by the salt, an inhomogeneous yttria
  felt and current density} \cite{Vallet1983}, \newtxt{measurement
  errors of EDX and AAS of up to 10\%, back-diffusion during cooling
  and the not exactly known tortuosity. As the latter was measured for
  a felt with a porosity of 90\% instead of $85\pm 4$\% as in this
  experiment, the effective diffusion coefficients might in reality be
  lower. The largest uncertainty comes, however, from the diffusion
  coefficient itself. Available measurement data is very scarce and
  might contain considerable error. Further, the diffusivity is
  expected to depend on concentration}
\cite{Newhouse2014}. Tab.\,\ref{tab:Di} \newtxt{gives an overview of
  diffusion coefficients in LiCl-KCl available in literature. Taking
  into account that even for similar temperatures the values
  vary up to a factor of six, the observed deviation
  between experiment and simulation appears to be acceptable.}
\begin{table}[h]
  \begin{center}
  \caption{Various diffusion coefficients for Li$^+$, Cl$^-$ and K$^+$ in LiCl-KCl molten salt.}
  \begin{tabular}{ l l l l l l}
  \hline
  \multicolumn{3}{l}{$D_i$ in $10^{-9}$ m$^2$/s} & $T$ in
  $^{\circ}$C & \multicolumn{2}{c}{Source} \\
  \cline{1-3} 
  Li$^+$ & Cl$^-$ & K$^+$\\
  \hline
  1.0 & 1.0 & 1.0 & 425 & estimate &
  \cite{Vallet1978,Braunstein1979}\\
  0.7 & 1.2 & 1.1 & 367 &molecular dynamics&
  \cite{Caccamo1980} \\
  2.02 & 1.75 & 2.25 & 373 &molecular dynamics& \cite{Lantelme1982} \\
  1.0 & & 1.0 & 450 &estimate& \cite{Hiroi1984}\\
  2.6 & & 2.8 & 430 &measured\footnotemark[1]& \cite{Yang1994}\\
  6.2 & 6.1 & 6.2 &757&molecular dynamics&\cite{Ribeiro2003}\\
  5.7 & 4.6 & 5.1 & 627 &molecular dynamics& \cite{Morgan2004}\\
  0.668 & & & 450 &measured& \cite{Chen2008}\\  
  %&&&molecular dynamics  & \cite{Chakraborty2013}\\
  4.37 & 3.7 & 4.89 & 527 &molecular dynamics& \cite{Bengtson2014}\\
  10.0 & & 10.0 & 827 & molecular dynamics &\cite{Wang2015}\\
  2.4 & 1.2 & 2.0 & 450 &no reference& \cite{Zhou2022}\\
  3.8 & 3.1 & 3.4 & 425 & present work\\
  %Moynihan\\
  \hline   
  \end{tabular}
  \label{tab:Di}
  \end{center}
\end{table}

\footnotetext[1]{Measured mobilities have been used to compute the
  diffusion coefficients via the Nernst-Einstein relation.}
%
%-----------------------------------------------------------------%

%%% Local Variables:
%%% mode: latex
%%% TeX-master: "../paper"
%%% TeX-parse-self: t
%%% TeX-auto-save: t
%%% TeX-PDF-mode: t
%%% eval: (auto-fill-mode 1)
%%% eval: (flyspell-mode 1)
%%% eval: (reftex-mode 1)
%%% ispell-dictionary: "british"
%%% End:

\section{Application to a Li$\Vert$Bi battery and discussion}
\label{applicationDiscussion}
In this section, the previously derived solver is used to investigate
the overpotentials introduced in section \ref{potDiffOver} in a
Li|LiCl-KCl|Bi cell. Dimensions and material properties are in
accordance to those described before in section
\ref{valSimPar}. Moreover, the potential jumps will be calculated in
every time step based on the Nernst equation (Eq.\,\eqref{eq:2-31} \&
Eq.\eqref{eq:2-32}) as well as the corresponding activities at the
electrode-electrolyte interfaces.

Investigations were done with different current densities. In order
to identify a working range of the cell, it is important to know the limiting current density $j_{\mathrm{lim}}$. \sout{The latter
corresponds to the current density at which the active species is
completely depleted at one interface. }For the LiCl-KCl electrolyte the
diffusive limiting current density reads
\begin{equation}
\label{eq:5-1}
j_{\mathrm{lim, d}} = - z F \frac{\overline{c}}{\delta} D
\end{equation}
with $\delta$ being the maximum thickness of the diffusion layer. In
steady state and neglecting convection, the latter can be approximated
as the half electrolyte thickness \cite{Vetter1967}.\sout{Please note that
not only diffusion contributes to mass transfer in the boundary
layer. The electric field and thus migrational species transport can
lead to an in- or decrease in the limiting current
\cite{Bagotskii2006}. In case of the reactant ion -- lithium in the
present case -- having a diffusion coefficient considerably
different from the other ions, the influence of migration is expected
to be large cite{Newman2004}.} In the present application case\sout{, solving
Eq.\,\eqref{eq:5-1} gives} $j_{\mathrm{lim}} \approx -130$mA/cm$^2$. Nevertheless, comparative simulations are performed with $j_1=-100$mA/cm$^2$, $j_2=-150$mA/cm$^2$, $j_3=-200$mA/cm$^2$ and $j_4=-250$mA/cm$^2$.

In the following, the potential in the entire cell and species
distribution in electrolyte and cathode are analysed. Specifically, (i) \sout{potential difference between both electrodes which corresponds
to }the terminal voltage of the battery \sout{$\varphi_{\mathrm{c}} -
\varphi_{\mathrm{a}}$, }(ii) the OCV and (iii) the different
overpotentials giving information about the composition of the cell
voltage\sout{ at discharge} will be studied. This makes it possible to compare
the simulations with the theoretical principles\sout{ (see
Eq.\,\eqref{eq:2-2})}. Further, it is analysed how the ratio of  
concentration overpotentials to ohmic overpotential for different
current densities develops.\sout{As the cell voltage is highly dependent on
the potential jumps at the interfaces, which in turn depend on the
activities, both jumps are analysed in more detail as well.} Lastly, a
polarisation curve is drawn for the applied current density range
between 0 and $-250$mA/cm$^2$.

\begin{figure*}[h]
    \centering
    \begin{subfigure}[b]{0.475\textwidth}
        \centering
        \includegraphics[width=\textwidth]{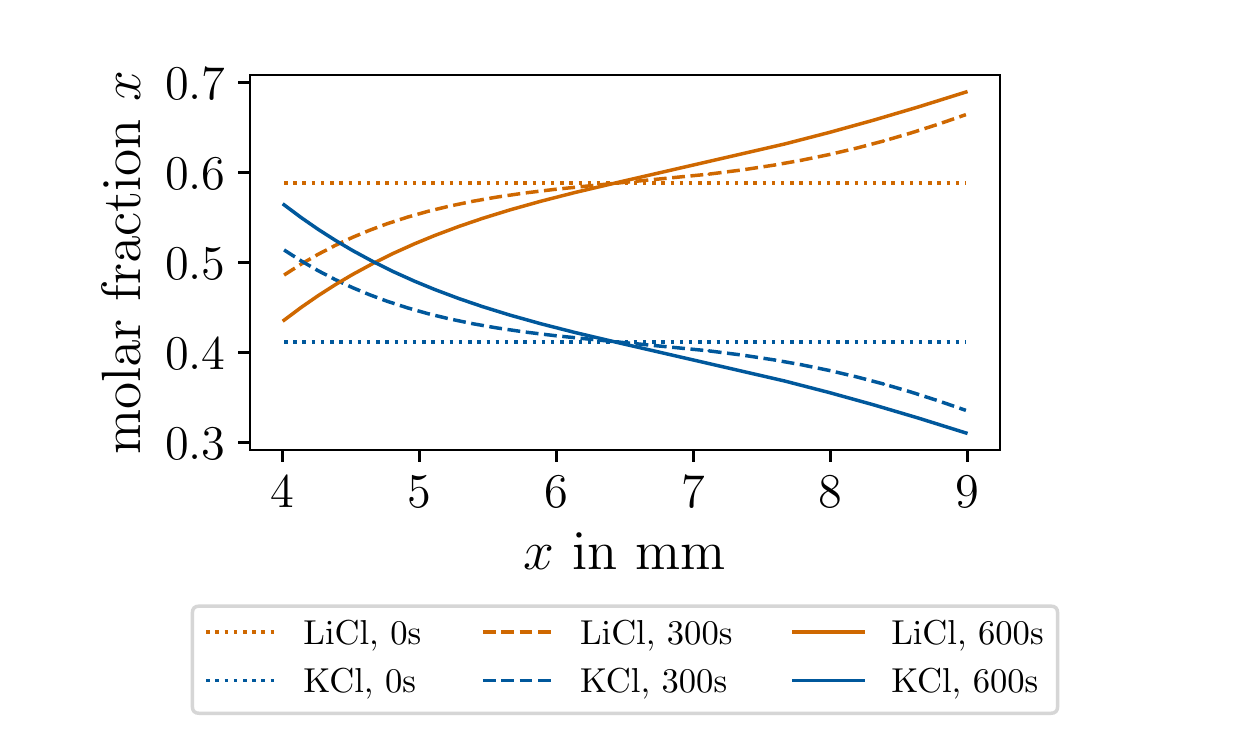}
        \caption[]%
        {{\small}}  
        \label{fig:caseAppSpecSalt}
    \end{subfigure}
    \hfill
    \begin{subfigure}[b]{0.475\textwidth}  
        \centering 
        \includegraphics[width=\textwidth]{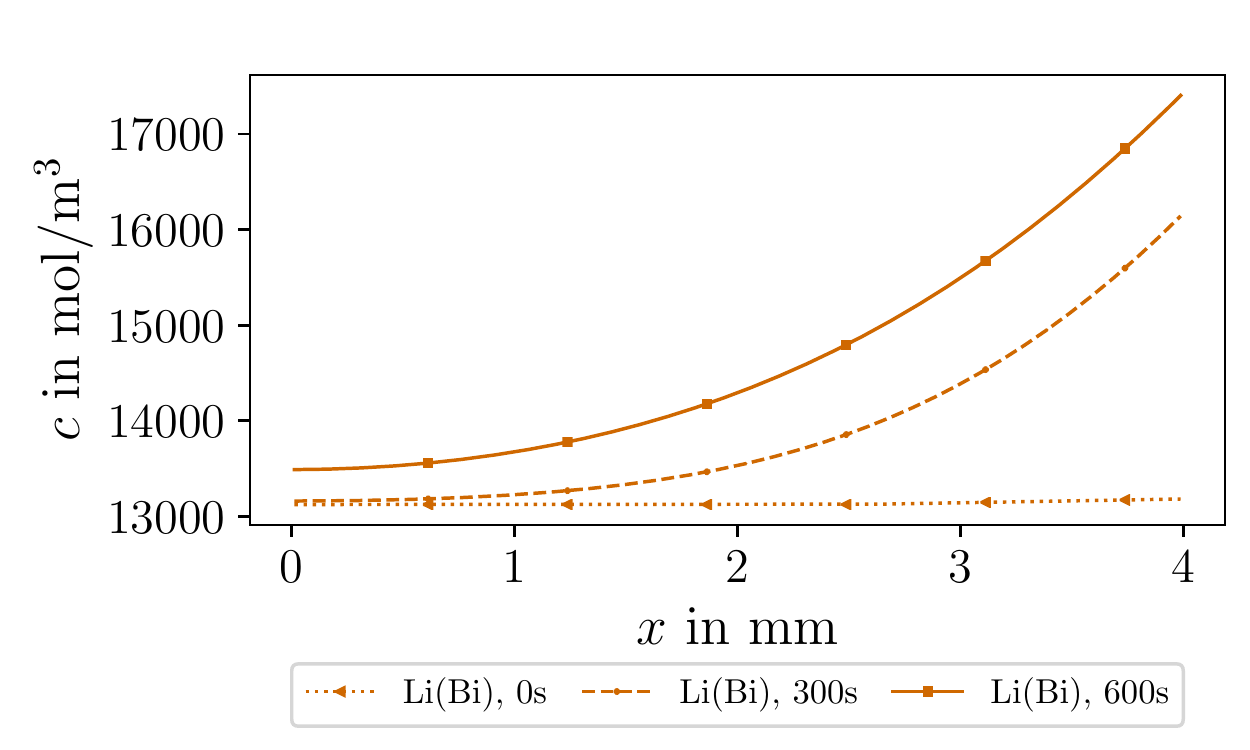}
        \caption[]%
        {{\small}}  
        \label{fig:caseAppSpecCath}
    \end{subfigure}
    \caption[]
    {\small Species distribution for discharge current density $j_1=-100$mA/cm$^2$ in (a) the Li|LiCl-KCl|Bi electrolyte and (b) Li(Bi) cathode for different discharge times.} 
    \label{fig:resApp_2}
\end{figure*}

Before showing comparative results, some outcomes of the simulation
for $j_1$ are presented.
The distribution of LiCl and KCl in the electrolyte (see
Fig.\,\ref{fig:caseAppSpecSalt}) shows the same behaviour as in the\sout{validation} \newtxt{verification} study in section \ref{valSimResults}. \sout{More LiCl is accumulated
at the anode-electrolyte interface, while more KCl is present at the
electrolyte-cathode interface.} In contrast to the previous
simulations, the lithium distribution in the bismuth cathode, as it
can be seen in Fig.\,\ref{fig:caseAppSpecCath}, is modelled and
analysed now as well. Arriving from the electrolyte, the Li diffuses
into the cathode and accumulates at the interface. For the cell
voltage, the distribution of lithium in bismuth has a decisive impact. It can be
seen in Eq.\,\eqref{eq:2-1} that the corresponding activity determines
the cell voltage and is significantly involved in the potential jump
at the cathode. 

\begin{figure*}[h]
    \centering
    \begin{subfigure}[b]{0.475\textwidth}
        \centering
        \includegraphics[width=\textwidth]{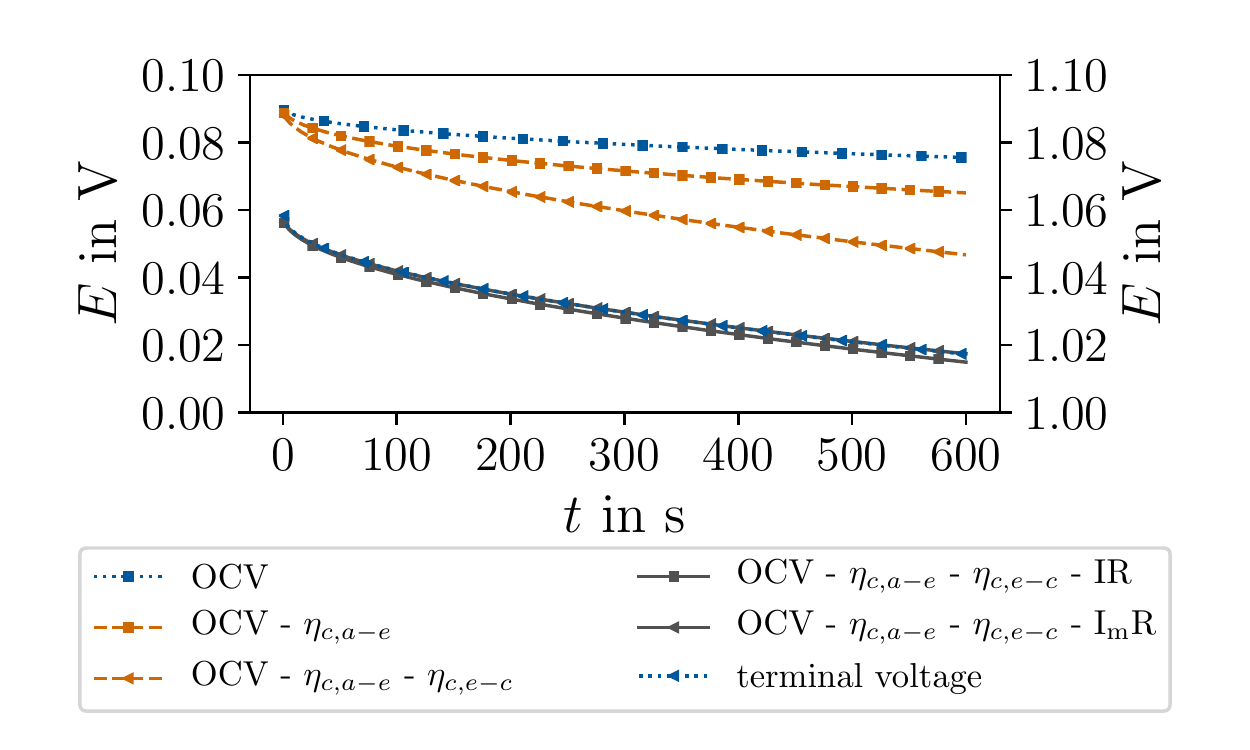}
        \caption[]%
        {{\small}}  
        \label{fig:caseAppVoltage}
    \end{subfigure}
    \hfill
    \begin{subfigure}[b]{0.475\textwidth}  
        \centering 
        \includegraphics[width=\textwidth]{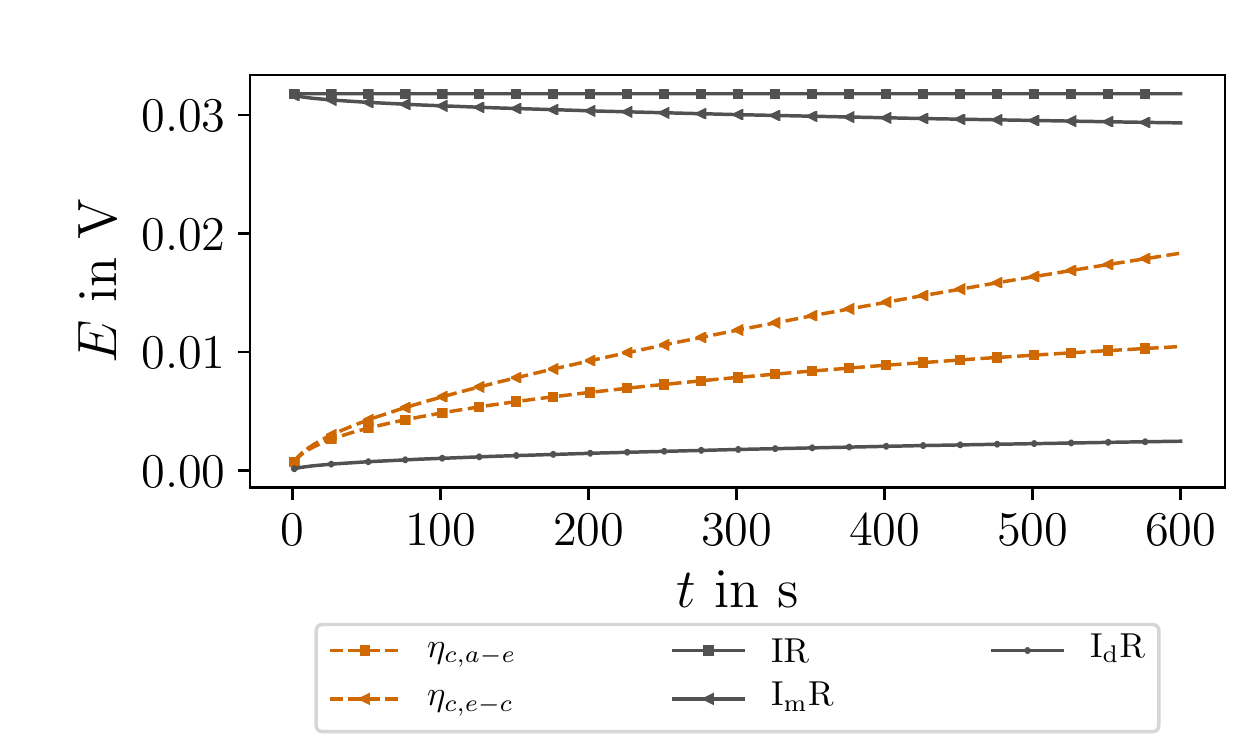}
        \caption[]%
        {{\small}}  
        \label{fig:caseAppOverpot}
    \end{subfigure}
    \caption[]
    {\small Analysis of (a) the cell voltage and (b) magnitude of overvoltages for discharge current density $j_1=-100$mA/cm$^2$. The second axis in (a) describes a potential shift of +1V for better comparability with experimental results. Note that in (b) only absolute values are shown; the term $I_\mathrm{d}R$ has in reality a negative sign.}
    \label{fig:resApp_3}
\end{figure*}

Now, a closer look to the composition of the overvoltages according to
Eq.\,\eqref{eq:2-2} is taken. \sout{Apart from that, it is assessed, if the
given formula is able to reproduce the terminal voltage of the
battery. }The results are shown in
Fig.\,\ref{fig:caseAppVoltage}. There, the theoretical open circuit
voltage is calculated from the activity of Li(Bi)
(Eq.\,\eqref{eq:2-45}), while the terminal voltage is a direct outcome
of the simulation. Both are coloured in blue. Here it is apparent that
the cell voltage of the Li$\|$Bi cell is much lower, than the value
known from pertinent literature \cite{Chum1981, Swinkels1971, Shimotake1969}. For the used Li$\|$Bi cell an open cell voltage of 0.9-1V
was observed under comparable conditions
\cite{Cairns1969, Chum1981}. The reason for this deviation is simply
that the activity coefficient was assumed to be one, when the activity
of lithium in bismuth $a_{\mathrm{Li(Bi)}}$ is calculated. To
``compensate'' for this simplification, and to make the results more
illustrative, the cell voltage is linearly shifted by +1V by
adding a second axis to	the figures.

As discussed in section \ref{potDiffOver}, concentration and ohmic
overpotential lower the cell voltage. Here it is important to mention
that the OCV is updated in each time step based on the interface 
Li-concentration in the cathode. Therefore, the OCV curve already
includes the concentration overpotential in the cathode.
The orange curves in
Fig.\,\ref{fig:caseAppVoltage} show the cell voltage including
concentration overpotentials in the electrolyte. It can be clearly
seen that the latter have a significant influence on the cell
voltage. Thus, it is not advisable to neglect them.
In their recent study, Weber et al.~\cite{Weber2022} modelled the cell voltage of a
Li$\|$Bi \newtxt{cell} without considering the concentration overpotentials in the
electrolyte. Here, deviations between model and
experiment were observed especially at high current densities (see
e.g.~Fig.\,15a in \cite{Weber2022}), which might be attributed in part to
concentration effects in the electrolyte.

Lastly, the grey curves in Fig.\,\ref{fig:caseAppVoltage} consider
ohmic losses as well. In the simplest assumption\sout{(Eq.\,\eqref{eq:2-2})}, the entire current contributes to the ohmic loss in the cell. The present study shows that this is not the case since the measured terminal voltage (dotted blue line with
triangle icon) does not match with the theoretical $E$ (following
Eq.\,\eqref{eq:2-2}, solid grey line with square icon). In contrast,
when considering only the migrational current when calculating ohmic
losses, the theoretical formula matches well with the simulation
results.
As it was already discussed previously, the diffusive current
contributes to charge transfer and therefore increases -- at least in
the present case -- the cell voltage and enhances the performance of
the entire cell. Consequently, it can be stated that only the
migrational part is important to define the ohmic loss.

In Fig.\,\ref{fig:caseAppOverpot}, the overvoltages are explicitly
shown, where $I_{\mathrm{d}}R$ represents the amount of lossless
charge transfer expressed as ``voltage gain\footnote{Undoubtedly,
there is no real voltage gain in the cell. This expression is used only
as descriptive representation here.}'' for better clarification of its
sign. It can be immediately seen that the ohmic overvoltage has the
largest influence on the cell voltage. However, concentration
overpotentials in the electrolyte are increasing with time and \sout{the sum
of both }reach\sout{es} a similar value as the former. While the concentration
overpotential at the anode-electrolyte interface is comparatively
small and will approach \sout{saturation}\newtxt{steady state}, it is larger and might
 approach infinity at the electrolyte-cathode interface.\sout{ Both can
be derived from Eq.\,\eqref{eq:2-4}.}

Vallet et al.~\cite{Vallet1982a} found experimentally that the magnitude of concentration
overpotentials in an AgNO$_3$-NaNO$_3$ electrolyte at 300$^{\circ}$C
and 150 mA/cm$^2$ varies between 5 to 15mV at each interface. Further, Braunstein et al.~\cite{Braunstein1979} predicted analytically
that the concentration overpotential for a LiCl-KCl electrolyte at
450$^{\circ}$C under ideal conditions would be about 17mV. Since the magnitude of the concentration
overpotential in the present study (see
Fig.\,\ref{fig:caseAppOverpot}) is in the same range, it is emphasised
that the concentration overpotentials in the electrolyte are indeed
important and cannot be neglected.

\begin{figure}[h]
  \centering
  \includegraphics[width=0.5\linewidth]{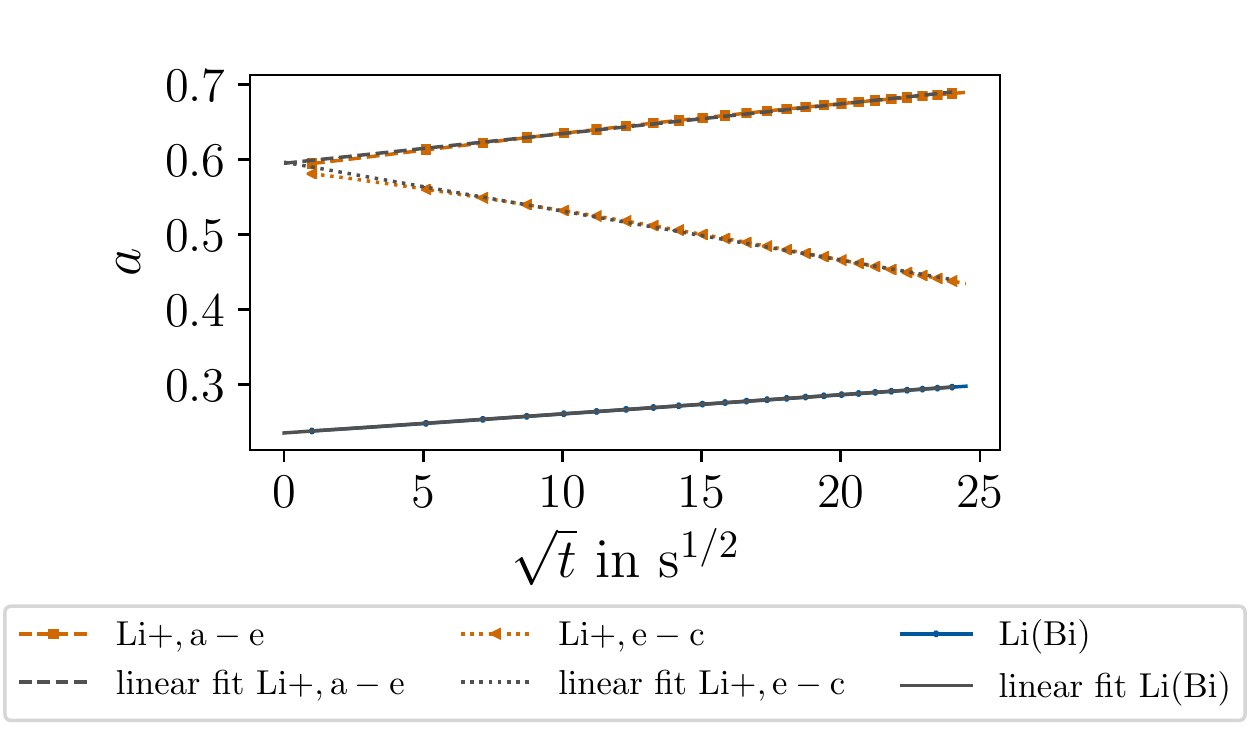}
  \caption{Parameter fit of activities for discharge current density $j_1=-100$mA/cm$^2$.}
  \label{fig:resApp_4}
\end{figure}

Concentration typically depends on the square root of time in
processes controlled by diffusion \cite{Brett1993}. In
Fig.\,\ref{fig:resApp_4} it can be seen that plotting
$a_{\mathrm{Li(Bi)}}$ versus the square root of time results in a
nearly perfect straight line, a good indication that the transport in
the cathode is -- as expected -- purely diffusive. In return, the
transport in the electrolyte is not purely diffusive, since the linear
fit of $a_{\mathrm{Li^+, a - e}}$ does not entirely cover the
simulated points. This indicates that\sout{ mass transport in the boundary
layer is not purely diffusive but that} migration \newtxt{in the boundary layer} has to be taken into
account. If fluid flow would be considered, convection would be
important as well.

\begin{figure*}[h]
    \centering
    \begin{subfigure}[b]{0.475\textwidth}
        \centering
        \includegraphics[width=\textwidth]{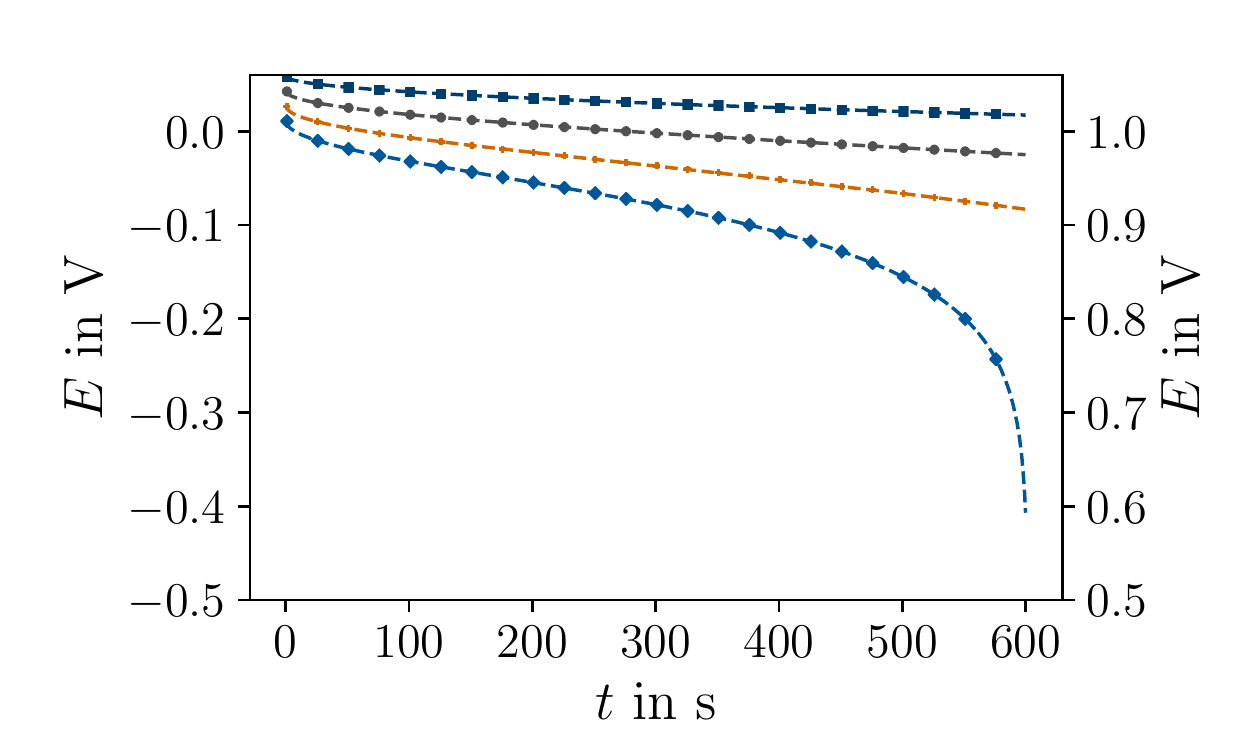}
        \caption[]%
        {{\small}}  
        \label{fig:caseAppVoltageAll}
    \end{subfigure}
    \hfill
    \begin{subfigure}[b]{0.475\textwidth}  
        \centering 
        \includegraphics[width=\textwidth]{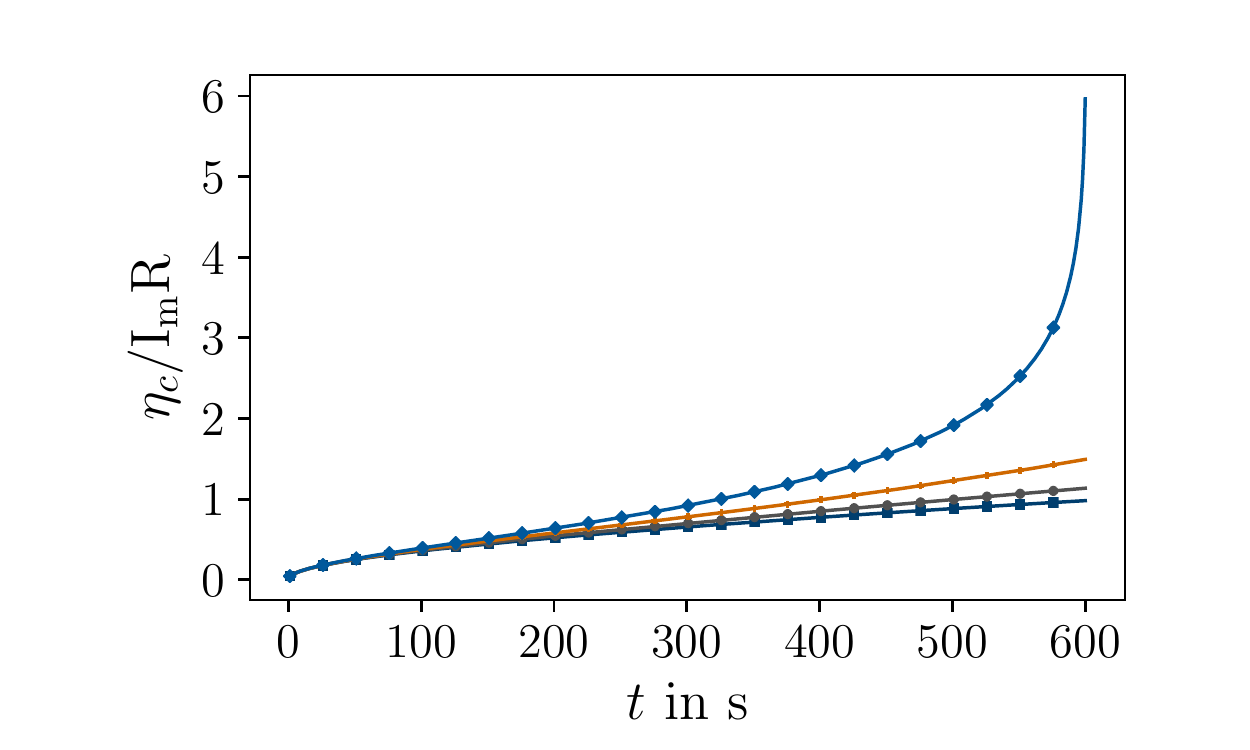}
        \caption[]%
        {{\small}}  
        \label{fig:caseAppOverpotAll}
    \end{subfigure}
    \vskip\baselineskip
    \begin{subfigure}[b]{0.85\textwidth}   
        \centering 
        \includegraphics[width=\textwidth]{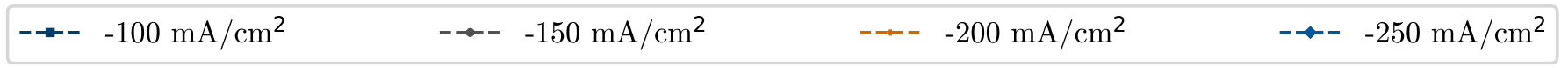}
        {{\small}}
    \end{subfigure}
    \caption[]
    {\small Comparison of (a) the cell voltage and (b) ratio of concentration overpotential to ohmic overpotential caused by migrational current for different discharge current densities. The second axis in (a) describes a potential shift of +1V for better comparability with experimental results.} 
    \label{fig:resApp_5}
\end{figure*}

Besides analysing individual overvoltages and the corresponding
phenomena, it is important to compare the impact of different current
densities. Figure \ref{fig:caseAppVoltageAll} shows, what is trivial:
the terminal voltage decreases when the current density
increases. Simultaneously the voltage losses increase. Regarding
those, in Fig.\,\ref{fig:caseAppOverpotAll} it can be observed that
the ohmic losses dominate for short \sout{discharge }time or high \sout{discharge rates}\newtxt{currents} \sout{that}\newtxt{where the concentration of the active ion in the electrolyte is far away from steady state}\sout{ are far away from steady state concentration in the electrolyte}.\sout{ Without exception, the percentage of the concentration
overpotentials increases with time and becomes dominant for large
discharge rates.} Especially, when the concentration in the electrolyte
approaches zero at the cathode interface\sout{(e.g., for $j=-250$mA/cm$^2$ at approx. 450s)}, the influence of the concentration overpotential increases
exponentially. The corresponding effects at the anode and cathode
interface are closely linked with activity, thus with concentration
and concentration gradient of the species. In general that means:
the higher the current density, the higher the concentration gradient; thus, the activity of lithium at the cathode approaches zero faster, while
it approaches \sout{saturation}\newtxt{steady state} faster at the anode interface. In
galvanostatic operation, the former might lead to chronopotentiometric
transition, i.e.~a sudden cell voltage rise, and undesired
electrochemical reactions \cite{Vallet1978}.

\sout{As a next step, the potential jumps for different current densities
are compared in Fig.\,\ref{fig:resApp_6}. A second axis for the jump
at the cathode is added again to compensate for the simplified
activity of Li in Bi. It can be seen that the absolute
values of the Nernst potential at both interfaces increase with
current density. Simultaneously, it can be observed that the jump at
the anode approaches saturation. The latter can be justified by the fact
that no solidification or other solubility threshold in the electrolyte 
is considered. In contrast, the potential jump  
at the cathode tends towards infinity due to the depletion of
active ions. This behaviour is corresponding to the concentration
overvoltages.}
\newtxt{Just as the overpotentials, the behaviour of the potential jumps at anode and cathode interface is similar.}\sout{According to Eq.\,\eqref{eq:2-31},} \newtxt{At the anode steady state is approached as no solidification or other solubility threshold in the electrolyte is considered. In contrast, overpotential and jump at the cathode tend towards infinity due to the depletion of active ions.}\sout{Equation\,\eqref{eq:2-32} points towards similar conclusions.}

%
%\begin{figure*}[h]
%    \centering
%    \begin{subfigure}[b]{0.475\textwidth}
%        \centering
%        \includegraphics[width=\textwidth]{chapters/figures/application/varyI/jumpAnode.pdf}
%        \caption[]%
%        {{\small}}  
%        \label{fig:caseAppPotAn}
%    \end{subfigure}
%    \hfill
%    \begin{subfigure}[b]{0.475\textwidth}  
%        \centering 
%        \includegraphics[width=\textwidth]{chapters/figures/application/varyI/jumpCathode.pdf}
%        \caption[]%
%        {{\small}}  
%        \label{fig:caseAppPotCat}
%    \end{subfigure}
%    \vskip\baselineskip
%    \begin{subfigure}[b]{0.475\textwidth}   
%        \centering 
%        \includegraphics[width=\textwidth]{chapters/figures/application/varyI/legend1.pdf}
%        {{\small}}
%    \end{subfigure}
%    \caption[]
%    {\small Comparison of the potential jumps at (a) anode and (b) cathode interfaces for different discharge current densities. The second axis in (b) describes a potential shift of +1V for better comparability with experimental results.} 
%    \label{fig:resApp_6}
%\end{figure*}

\begin{figure}[h]
  \centering
  \includegraphics[width=0.5\linewidth]{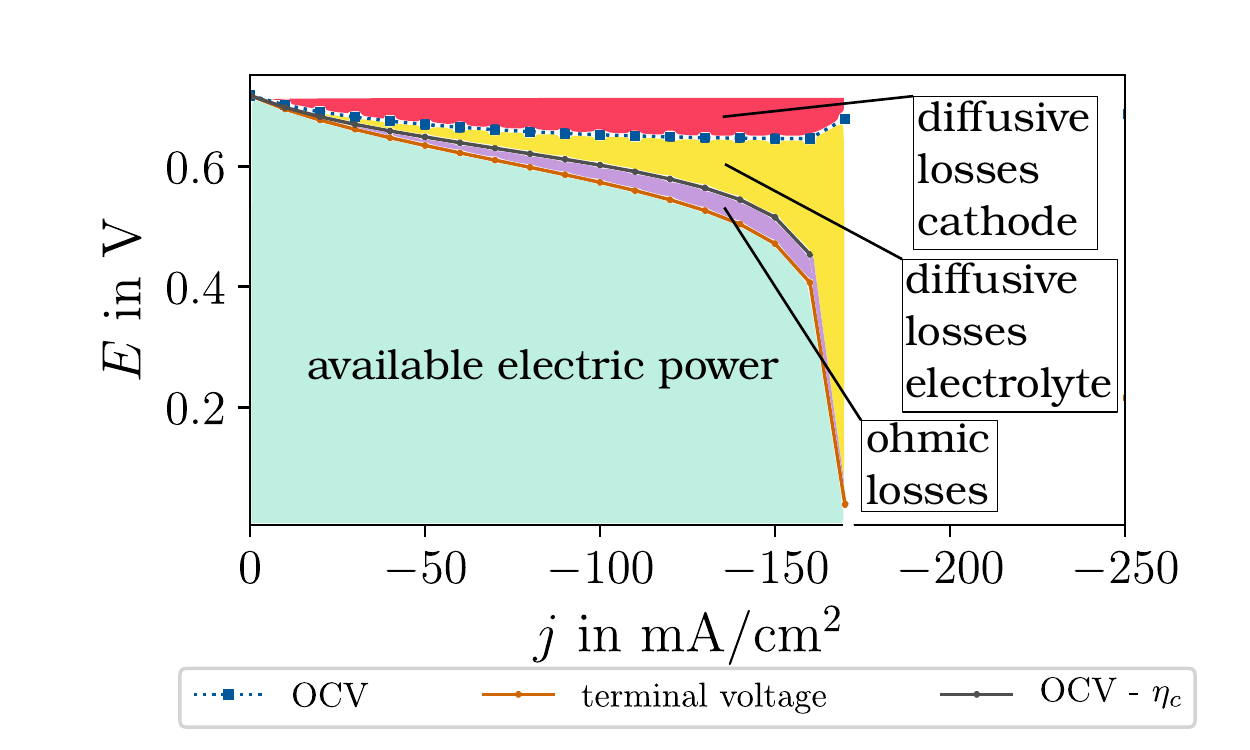}
  \caption{Polarisation curve of the Li|LiCl-KCl|Bi liquid metal battery.}
  \label{fig:caseAppPol}
\end{figure}

%
Polarisation curves are used, amongst others, for fuel cells
\cite{Heider2017} to analyse the available electric power\sout{ of a
cell}. Such a polarisation curve for the Li$\|$Bi LMB is generated and
shown in Fig.\,\ref{fig:caseAppPol}. In order to achieve realistic
potentials, a constant activity coefficient of Li in Bi of $4.2\cdot
10^{-5}$~m$^2$/s is assumed \cite{Cao2014}. Simulations are performed until
reaching a steady state and with a sufficiently thick cathode to avoid
any influence of (a lower) state of charge.

At first glance, the diagram shows precisely that there is a limiting
current density, of approximately $j=-165$mA/cm$^2$, which is a little higher than the previously calculated $-$130mA/cm$^2$. \newtxt{Since not only diffusion -- as considered in calculating the limiting current density -- contributes to mass transfer in the boundary layer, migrational species transport can lead to an in- or decrease in the limiting current}
\cite{Bagotskii2006}\newtxt{. In case of the reactant ion -- lithium in the
present case -- having a diffusive current density being considerably
different from the other ions, the influence of migration on the limiting current density is expected to be quite high} \cite{Newman2004}.\sout{Justifiable by the diffusive current density of the
lithium ion being considerably different from the one of chloride and
potassium, as shown in Fig.\,ref{fig:resCase3_2}, the influence of
the migrational current density on the limiting current density is
quite high.} The ratio $j_{\mathrm{lim}}/j_{\mathrm{lim,d}}=165/130 \approx 1.27$ indicates that migration enhances the limiting current by 27\%. Conversely, this also means that the thickness of the boundary that determines the limiting current is
reduced due to migrational species transport.

Overall, the three \newtxt{relevant} types of overpotentials \newtxt{-- as introduced in section} \ref{potDiffOver} -- can be identified. \sout{Firstly, concentration overpotentials in the cathode are caused by slow mass transport of Li in Bi (red). Secondly, the diffusive losses that express mass transfer and concentration overpotentials in the electrolyte will appear (yellow).}\newtxt{Of particular interest is the fact that the diffusive losses in the electrolyte increase largely }\sout{As they are increasing with }\newtxt{with an increasing current. This indicates }\sout{it is indicated} \newtxt{that the diffusive current density must also increase hereby. As the influence of the diffusive current density was discussed previously, this may lead to the assumption that the cell performance at higher current densities is more influenced by diffusion than at lower current.}\sout{exhibit the same behaviour and is further discussed in } \ref{chapter:appJMigDiff} \newtxt{further elaborates on the ratio of the different current density components.} \sout{Thirdly, ohmic losses that are based on the migrational current and the electric conductivity in all regions reduce the cell potential further (purple).}\newtxt{Moreover}, the area beneath the polarisation curve shows the
available electric power of the cell. 

After the point where $\eta_{\mathrm{c}} \approx \eta_{\Omega}$, which
is at approximately 30mA/cm$^2$, diffusive losses start to dominate
the overpotentials in the cell. Here, it must be noted that this is a
very small value compared to typical battery current densities, which
are in the order of 200-300mA/cm$^2$ \cite{Ning2015}. Further,
parallels to Fig.\,\ref{fig:caseAppOverpotAll} can be seen: \sout{Obviously,
for large discharge times and at typical current densities, ohmic
losses are always lower than concentration overpotentials.}\newtxt{concentration overpotentials can -- under certain conditions -- exceed the ohmic losses.}

Summing up, it has been shown that concentration overpotentials can
have an important influence on the cell performance: the corresponding
losses easily reach more than 50\% of the total losses of electric
power. This statement must, of course, be seen in context of the
limitations of the model. The simplifications concerning the ion
activity, which were introduced by Temkin's model, seem to be
justified. Molten salt experiments by Vallet gave similar
concentration overpotentials. On the other hand, the obtained limiting
current of 165mA/cm$^2$ seems to be extremely small. In various Li||Bi
cell experiments, ten times larger current densities were observed
\cite{Shimotake1969,Swinkels1971,Cairns1973,Temnogorova1979,Ning2015}. It
is therefore obvious that convection must always be present in the
electrolyte of real batteries and may be triggered by a number
of effects, likely in combination of several causes. While it is
impossible to identify them with certainty, some can be ruled out.
The Tayler instability cannot appear in small cells and thin layers
\cite{Weber2014,Herreman2015} and the metal pad roll instability would
need much larger Li||Bi cells \cite{Weber2017} to arise. While the
critical Rayleigh numbers of internally heated convection
\cite{Shen2015,Personnettaz2018,Kollner2017,Personnettaz2022} are not
exceeded for the 5\,mm thin electrolyte and the moderate current
densities in the paper at hand, internally heated convection is likely
to occur for the higher current densities and thicker electrolyte
layers used in experiments. Depending on the experimental setup,
electro-vortex flows \cite{Ashour2018,Herreman2019}, localised
heating/cooling \cite{Ashour2018}, solutal convection in the
electrolyte, convection driven by viscous coupling from the cathode or
anode \cite{Personnettaz2022}, and Marangoni convection
\cite{Kollner2017} are additional possible means to intensify mass
transfer.

\section{Conclusion and Outlook}
Concentration gradients and overpotentials in molten salt electrolytes
have been well-known for more than forty years. However, when modelling
liquid metal batteries, the latter have always been ignored -- with
the notable exception of the work of Newhouse
\cite{Newhouse2014}. With this motivation, mass transport in a
Li$\|$Bi concentration cell with a LiCl-KCl electrolyte is studied.

For this purpose, an OpenFOAM model relying on the parent-child mesh
approach that is suitable for electrochemical simulations of LMBs is
developed. The solver is able to calculate potential and species
distribution in a full LMB where all three regions -- anode,
electrolyte and cathode -- can be coupled. Further, modelling 
interfacial processes without resolving the electrical double layer
microscopically or using experimental values is successfully realised
using potential jumps and cyclic boundary conditions at both internal
interfaces. The solver is thoroughly \sout{validated}\newtxt{verified} using four exemplary test cases \newtxt{and validated through a comparative study with experimental results}.

Modelling a realistic Li$\|$Bi LMB with a mixed cation electrolyte leads to three main
conclusions. Firstly, it is confirmed that concentration gradients in
the electrolyte are indeed important. Without convection, mass
transfer overpotentials can reach or even exceed the ohmic
losses. While the latter appear instantaneous, concentration 
gradients develop rather slowly on a scale of several
minutes. Reducing the cell potential by several tens of millivolts and
leading to solidification of the electrolyte in the worst case,
concentration gradients can by no means be ignored.

A second important finding concerns diffusive mass transport in the
electrolyte. Depending on the relation of the diffusion coefficients
between the electrolyte's species, the overall diffusive current can
be directed in the same, or the opposite direction of the migrational
current. Thus, it can ``artificially'' increase or reduce the ohmic
overpotential. This means at the same time that computing the ohmic
cell losses by the full cell current -- instead of only the
migrational current -- will always be an estimate.

Finally, it is shown that the limiting current is difficult to predict
for real cells. The diffusive limiting current density can simply be
calculated, but underestimates the ``true'' value by 30\% \newtxt{in our case} as it does
not account for migration. The fact that short circuit current densities were
obtained in Li$\|$Bi cell experiments, which exceed the simulated
value by a factor of ten, is a clear evidence that convection must be
present in the molten salt. The source and intensity of this flow
shall therefore be subject of further investigation.

With this motivation, it appears recommendable to implement the
Navier-Stokes equations into the present model and perform battery simulations
with fluid motion. \newtxt{The fact that electrochemical transport is now successfully integrated in a finite volume solver is highly beneficial and a particularly good starting point for developing a solver that accounts for fluid flow as well.} Likewise it could be important to consider phase change of the electrolyte dependent on the composition\sout{ as well}. Apart from enhancing the capabilities of the numerical solver, accurate
values for diffusion coefficients as well as activities are
fundamental to obtain reliable numerical results. Finally, the
presented model can -- with small changes -- be applied to many other
electrochemical devices. It is particularly well suited for modelling
cells with sharp interfaces between electrode and electrolyte \newtxt{as well as with very thin double layers}\sout{ -- such as high voltage and high concentration applications}. 
%
%-----------------------------------------------------------------%
%%% Local Variables:
%%% mode: latex
%%% TeX-master: "../paper"
%%% TeX-parse-self: t
%%% TeX-auto-save: t
%%% TeX-PDF-mode: t
%%% eval: (auto-fill-mode 1)
%%% eval: (flyspell-mode 1)
%%% eval: (reftex-mode 1)
%%% ispell-dictionary: "british"
%%% End:

%%%%%%%%%%%%%%%%%%%%%
\section*{Acknowledgements}
This project has received funding from the European Union’s Horizon
2020 research and innovation program under grant agreement No.
963599 and in frame of the
Helmholtz - RSF Joint Research Group ``Magnetohydrodynamic
instabilities: Crucial relevance for large scale liquid metal
batteries and the sun-climate connection'', contract No.~HRSF-0044 and
RSF-18-41-06201. We would like to thank K.~E.~Einarsrud, G.~Mutschke,
M.~Huang, S.~Beale and G.~Natarajan for fruitful discussions about the electrochemical simulations. 
%%%%%%%%%%%%%%%%%%%%%
\section*{Data availability statement}
The solver and validation test case of Vallet et al. are openly available at
\url{https://doi.org/10.14278/rodare.2313}.

\appendix
\section{Species transport equation}
\label{chapter:appSpeciesEqn}
Solving the Poisson equation Eq.\,\eqref{eq:2-18} gives results for the desired variables. But, using those values can lead to inaccurate results for the species transport. To avoid this, the potential gradient in Eq.\,\eqref{eq:3-2} must be replaced. In the following it is shown, why $\varphi$ or $\nabla \varphi$ cannot be taken directly, but require special care, when solving the species
transport equation. To start, the most simple test case of an electrolyte with two ions\sout{ (binary)} and corresponding equal diffusion coefficients $D_- = D_+ = D$ and $\nu_- = - \nu_+$ is considered. Due to electroneutrality $c_- = c_+ = c$ follows. With this, Eq.\,\eqref{eq:2-17} reduces to Eq.\,\eqref{eq:2-5} and $\nabla \varphi = -\mathbf{j} / \sigma$. Having a closer look at the migration term $\nabla \cdot \left( \Psi c_i \right)$ in Eq.\,\eqref{eq:3-1}, the following can be found:
\begin{equation}
\label{eq:3-3}
\nabla \cdot \left( \Psi c \right) = \nabla \cdot \left( \pm \nu_+ \frac{D F}{R T} \frac{\mathbf{j}}{\sigma} c \right).
\end{equation}
By simplifying Eq.\,\eqref{eq:2-16} to $\sigma_{\mathrm{salt}} = 2 \cdot F^2 D c \nu_+^2 / (RT)$ this leads to
\begin{equation}
\label{eq:3-4}
\nabla \cdot \left( \pm \frac{\mathbf{j}}{2F\nu_+} \right) \Leftrightarrow \pm \frac{1}{2F\nu_+} \left(\nabla \cdot \mathbf{j} \right) \pm \mathbf{j}\cdot \nabla \left(\frac{1}{2F\nu_+} \right).
\end{equation}
Due to conservation of charge, the first term is zero. The second term is zero as well. So for the most simple case,
\begin{equation}
\label{eq:3-5}
\nabla \cdot \left( \Psi c_i \right) = 0
\end{equation}
must be satisfied. Now a closer look to a single cell is necessary. To meet the divergence-free condition, $c_i \nabla \varphi$ needs to be equal in the cell centre c and at the cell boundary b. % as shown in Fig.\,\ref{fig:cellElectrolyte} . 
%
%\begin{figure}[h]
%  \includegraphics[width=\linewidth]{chapters/cell.pdf}
%  \caption{Simulation domain}
%  \label{fig:cellElectrolyte}
%\end{figure}
%
Simply, this means that
$\left. c_{\mathrm{b}} \frac{\partial\varphi}{\partial
x} \right|_{\mathrm{b}}
= \left. c_{\mathrm{c}} \frac{\partial\varphi}{\partial
x} \right|_{\mathrm{c}}$ must be true, otherwise numerically generated
divergence will occur. Due to the given
boundary conditions, $c_{\mathrm{b}} \ne c_{\mathrm{c}}$ is
valid. Therefore $\left. \frac{\partial\varphi}{\partial
x} \right|_{\mathrm{b}} \ne \left. \frac{\partial\varphi}{\partial
x} \right|_{\mathrm{c}}$ needs to be satisfied, too. Within OpenFOAM,
gradient fields are computed using the Gauss integral, which means
that the face values of the potential are integrated to obtain the
cell-centred potential gradient \cite{Greenshields2015}. However, 
this means also that the gradient of the potential is never computed 
on the boundary, but simply assumed to be the same as in the cell 
centre. Using such a gradient field directly would therefore lead 
to considerable calculation errors \newtxt{on the boundary itself}.
%
%-----------------------------------------------------------------%
%
\section{Current density composition}
\label{chapter:appJMigDiff}
\begin{figure*}[h]
    \centering
    \begin{subfigure}[b]{0.475\textwidth}
        \centering
        \includegraphics[width=\textwidth]{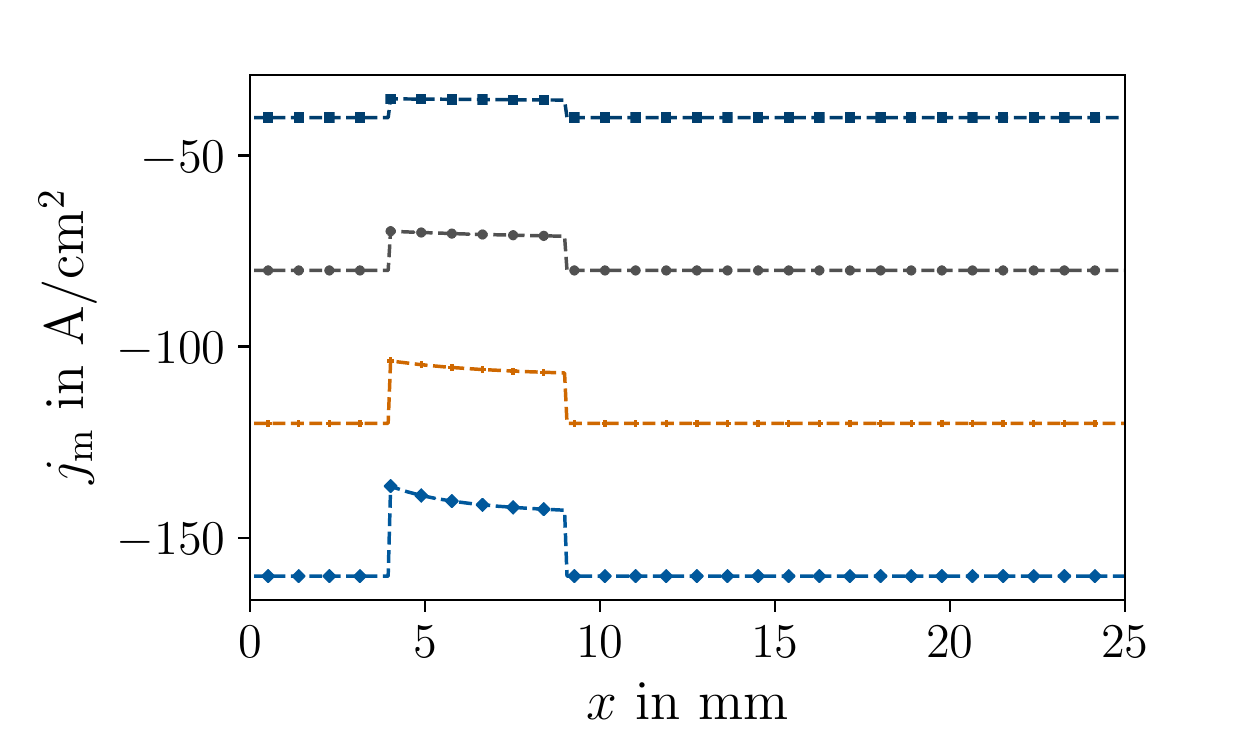}
        \caption[]%
        {{\small}}  
        \label{fig:caseAppPotJMig}
    \end{subfigure}
    \hfill
    \begin{subfigure}[b]{0.475\textwidth}  
        \centering 
        \includegraphics[width=\textwidth]{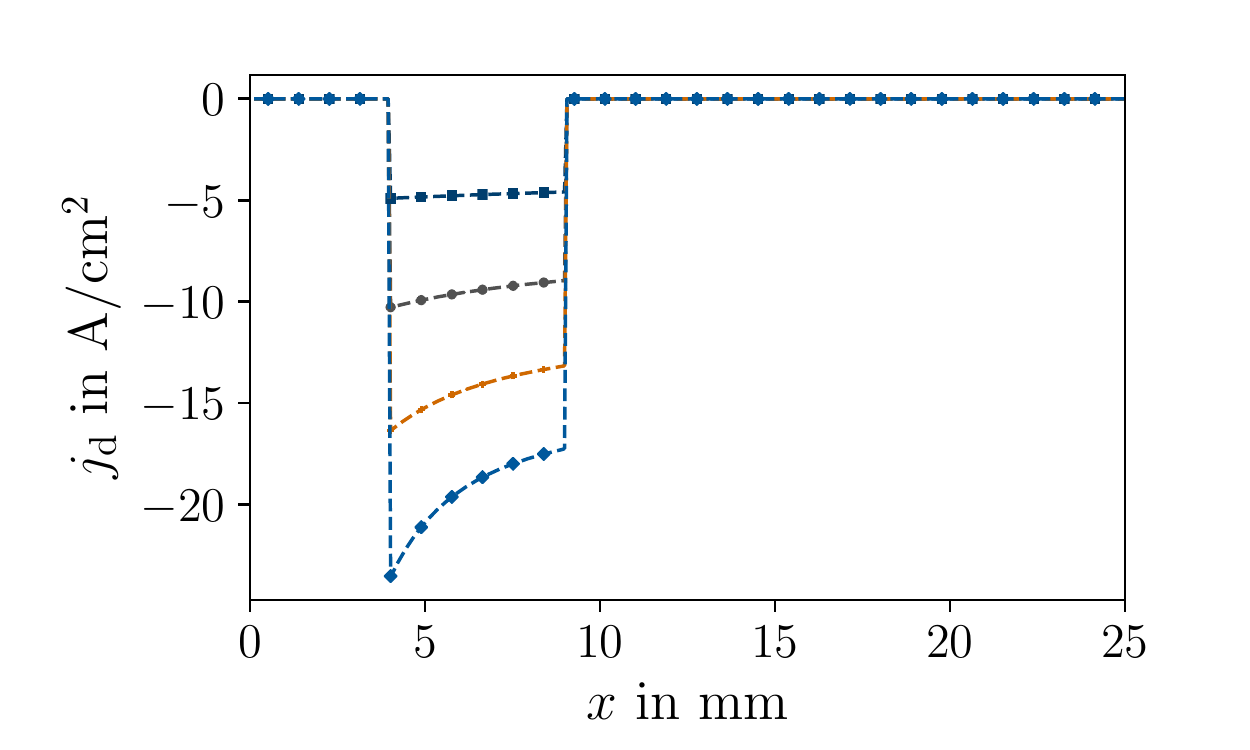}
        \caption[]%
        {{\small}}  
        \label{fig:caseAppPotJDiff}
    \end{subfigure}
    \vskip\baselineskip
    \begin{subfigure}[b]{0.475\textwidth}  
        \centering 
        \includegraphics[width=\textwidth]{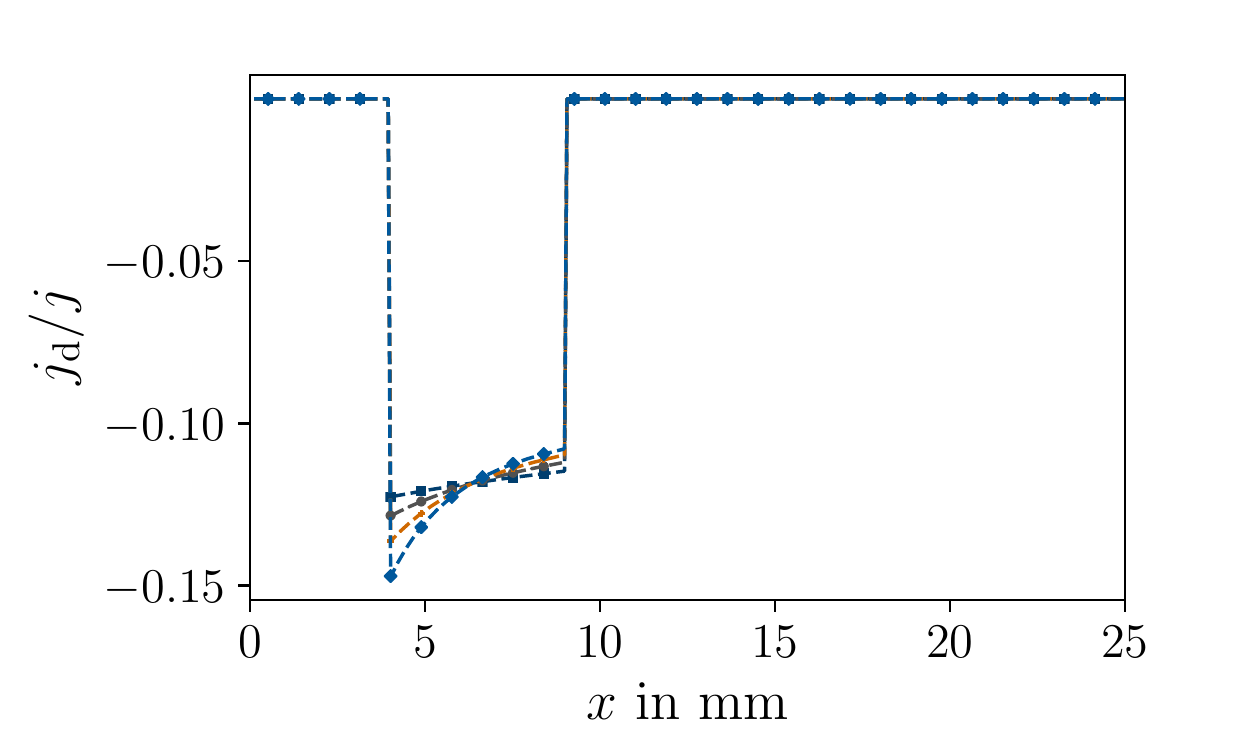}
        \caption[]%
        {{\small}}  
        \label{fig:caseAppPotJDiffRatio}
    \end{subfigure}
    \vskip\baselineskip
    \begin{subfigure}[b]{0.85\textwidth}   
        \centering 
        \includegraphics[width=\textwidth]{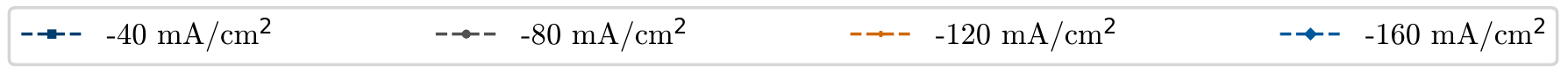}
        {{\small}}
    \end{subfigure}
    \caption[]
    {\small Distribution of (a) migrational and (b) diffusive current density as well as (c) the ratio $j_{\mathrm{d}}/j$ over the cell domain at steady state species distribution for different discharge current densities. } 
    \label{fig:resApp_7}
\end{figure*}

Within the electrolyte, the current density consists -- as shown in Eq.\,\eqref{eq:2-17} -- of a diffusive and migrational part. In order to examine the particular composition of the applied current density, some specific points below the detected limiting current are considered. The results are shown in Fig.\,\ref{fig:resApp_7}.

First, it is noticeable that the \newtxt{absolute value of the migrational current}\sout{is reduced} in the electrolyte \newtxt{is lower} compared to the electrodes. \sout{The ``missing'' part}\newtxt{This difference} is compensated by the diffusional current \newtxt{so that charge conservation is ensured. With increasing cell current, the diffusional part increases concurrently.}
Since this has direct influences on the charge
transport in the electrolyte, it can be deduced that more lossless
charge transfer is present that could enhance the cell performance of
the battery. Nevertheless, the ratio
$j_{\mathrm{d}}/j$ does not vary significantly for
the investigated currents, which means that the general influence of
the diffusive current is almost independent of the applied current.

%%% Local Variables:
%%% mode: latex
%%% TeX-master: "../paper"
%%% TeX-parse-self: t
%%% TeX-auto-save: t
%%% TeX-PDF-mode: t
%%% eval: (auto-fill-mode 1)
%%% eval: (flyspell-mode 1)
%%% eval: (reftex-mode 1)
%%% ispell-dictionary: "british"
%%% End:

%%%%%%%%%%%%%%%%%%%%%
\clearpage
\newpage
\bibliography{literature}
%%%%%%%%%%%%%%%%%%%%%
\newpage
\renewcommand*{\thesection}{Supplementary material S\arabic{section}}
\renewcommand*{\thefigure}{S\arabic{section}.\arabic{figure}}
\pagenumbering{roman}
\setcounter{page}{1}

\setcounter{section}{0}
%\section{\colorbox{yellow}{Comparison and verification with COMSOL results}}
\section{Comparison and verification with COMSOL results}
\label{chapter:appValidationComsol}
%\counter
%
\subsection{Test case 1}
\label{appCase1}
\begin{figure*}[h]
    \centering
    \begin{subfigure}[b]{0.475\textwidth}
        \centering
        \includegraphics[width=\textwidth]{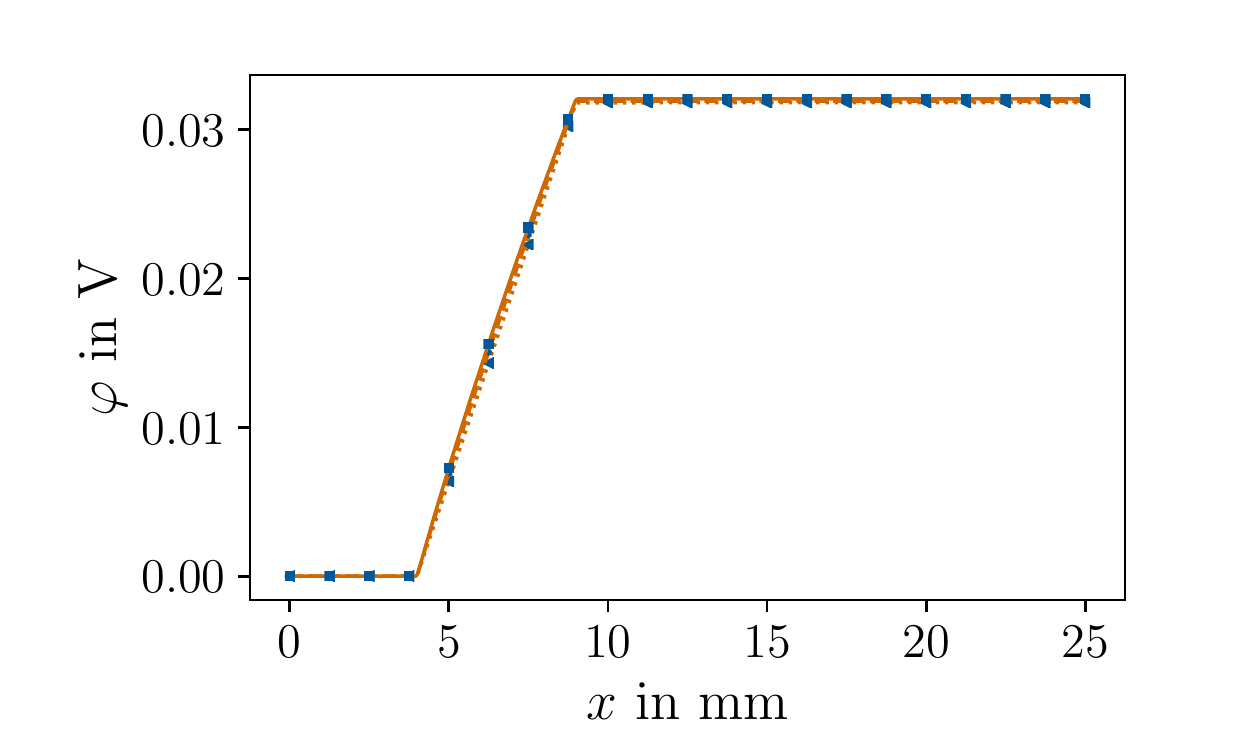}
        \caption[]%
        {{\small}} 
        \label{fig:case1Pot}
    \end{subfigure}
    \hfill
    \begin{subfigure}[b]{0.475\textwidth}  
        \centering 
        \includegraphics[width=\textwidth]{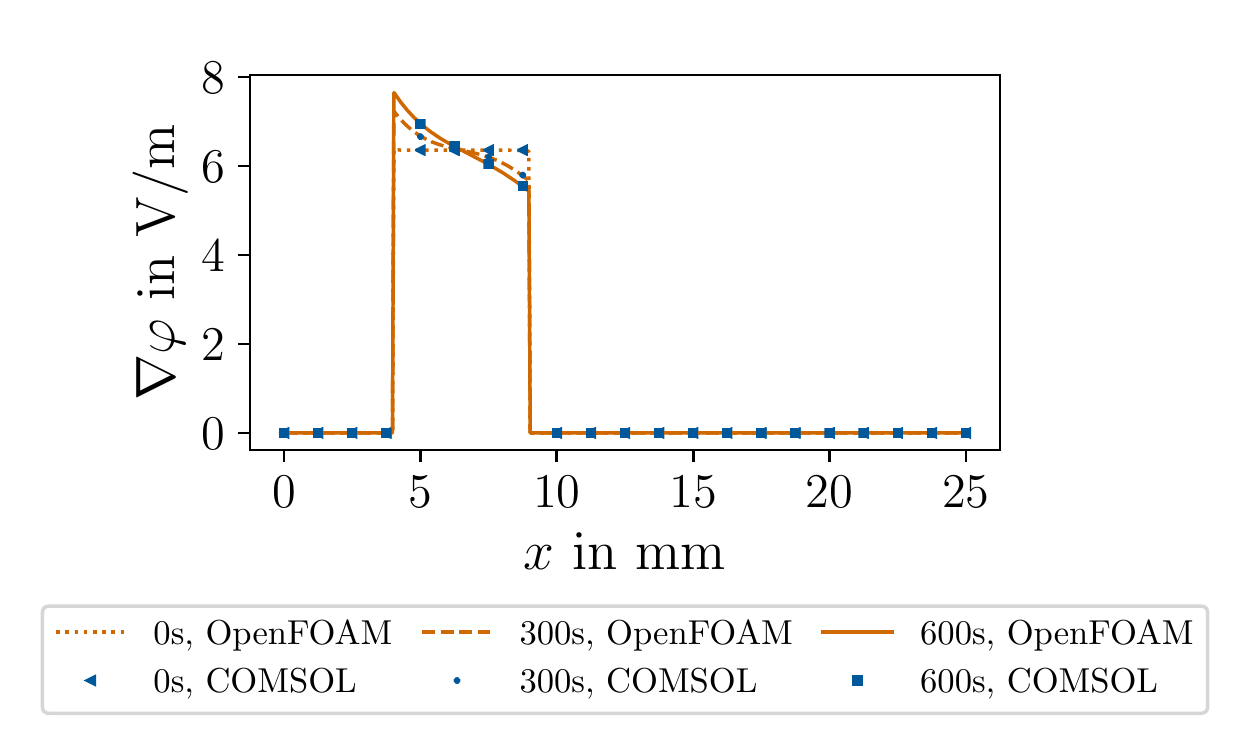}
        \caption[]%
        {{\small}} 
        \label{fig:case1gradPot}
    \end{subfigure}
    \vskip\baselineskip
    \begin{subfigure}[b]{0.475\textwidth}   
        \centering 
        \includegraphics[width=\textwidth]{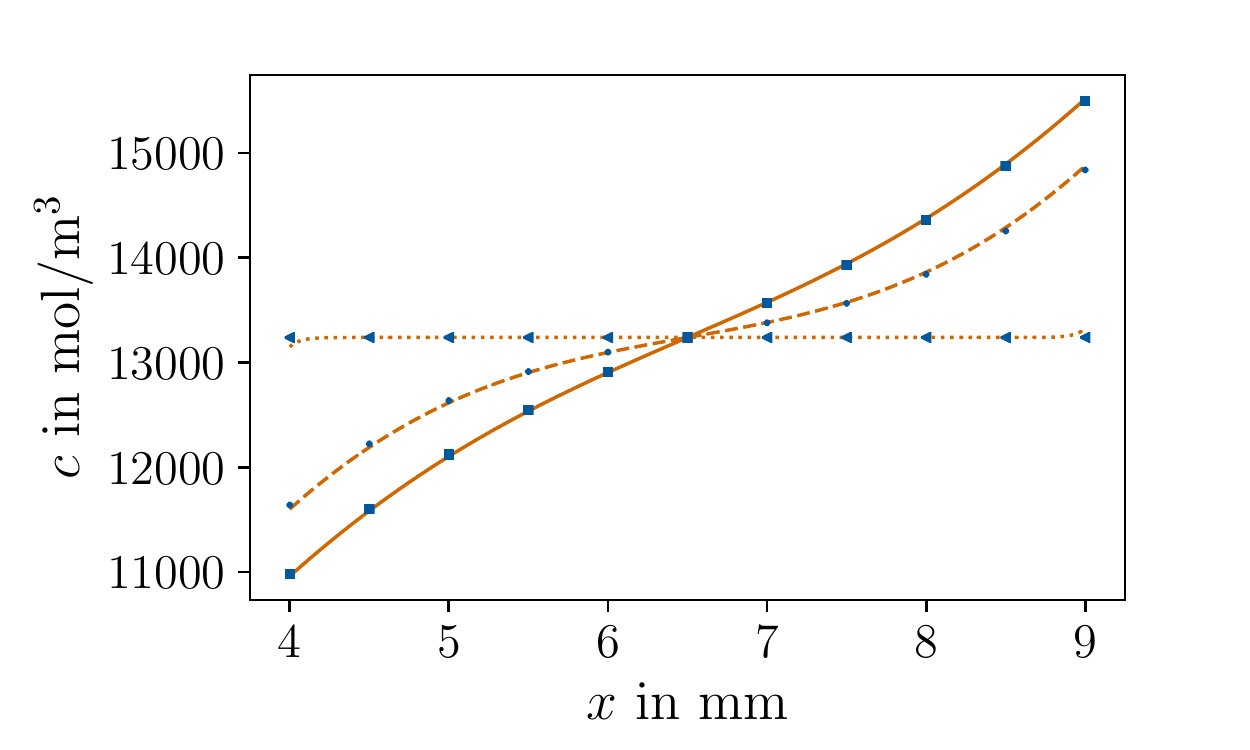}
        \caption[]%
        {{\small}} 
        \label{fig:case1Spec}
    \end{subfigure}
    \hfill
    \begin{subfigure}[b]{0.475\textwidth}   
        \centering 
        \includegraphics[width=\textwidth]{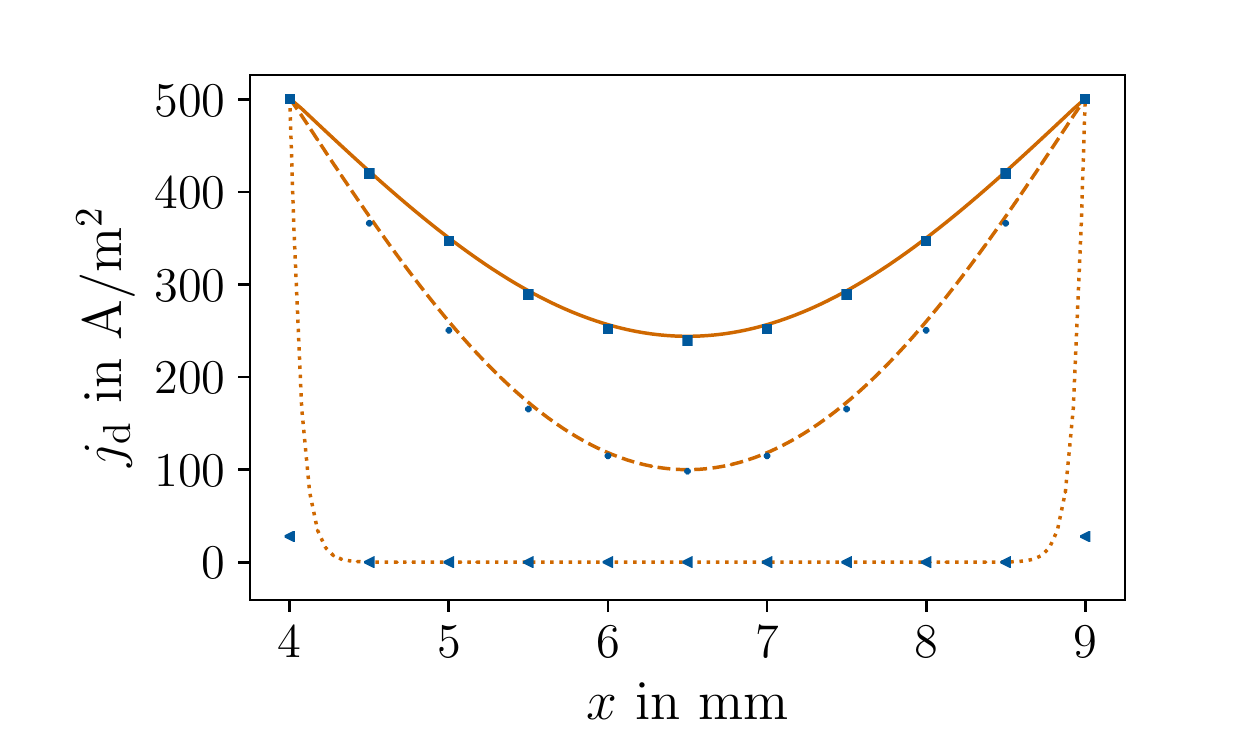}
        \caption[]%
        {{\small}} 
        \label{fig:case1gradSpec}
    \end{subfigure}
    \vskip\baselineskip
    \begin{subfigure}[b]{0.475\textwidth}   
        \centering 
        \includegraphics[width=\textwidth]{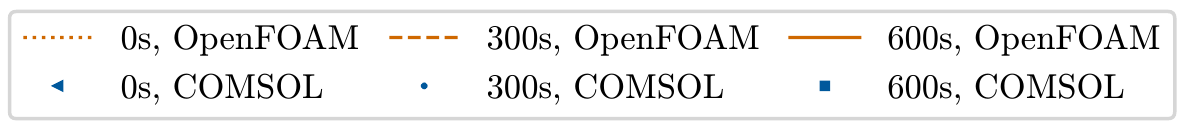}
        {{\small}}
    \end{subfigure}
    \caption{\small Test case 1; results for OpenFOAM and COMSOL simulations: (a) potential distribution (b) potential gradient in the whole cell, (c) species distribution and (d) diffusive current density of the Li$^+$ ion in the electrolyte.} 
    \label{fig:resCase1}
\end{figure*}

\sout{The simulation}\newtxt{Simulation} results related to test case 1 are shown in Fig.\,\ref{fig:resCase1}.
\newtxt{Due to $D_{\mathrm{Cl}^-} = D_{\mathrm{Li}^+}$,} the species distribution and species concentration gradient for the Li$^+$ and Cl$^-$ ions are equal, since the electroneutrality condition Eq.\,\eqref{eq:2-14} is preserved within a tolerance of machine accuracy. Further, point symmetrical behaviour of the species distribution -- as Eq.\,\eqref{eq:2-22} and Eq.\,\eqref{eq:2-23} as well as Fig.\,\ref{fig:case1Spec} show -- is associated with equal concentration gradients. Pursuing, the former indicates that the corresponding electric conductivity following Eq.\,\eqref{eq:2-17} needs to be symmetric as well. Furthermore, a symmetric trend in the electrolyte is also observable for the potential gradient in Fig.\,\ref{fig:case1gradPot}, as the latter is linearly dependent on the species gradient (see Eq.\eqref{eq:2-17}). 
\newtxt{Besides that, the diffusive current density of the Li$^+$ and Cl$^-$ ions differs only in terms of a reversed sign. Hence, it can be concluded that the simulation satisfies the theory by confirming that the diffusive current density is zero.} 

\newtxt{In terms of verification,} it can be seen that the solutions of OpenFOAM and COMSOL are in good agreement and match perfectly with each other in every case except of
the species gradient at $t_0$. \newtxt{Initial OpenFOAM results are analytically confirmed, whereas}\sout{In contrast,} the initial COMSOL results are not reliable. 
\subsection{Test case 2}
\label{appCase2}
\begin{figure*}[h]
    \centering
    \begin{subfigure}[b]{0.475\textwidth}
        \centering
        \includegraphics[width=\textwidth]{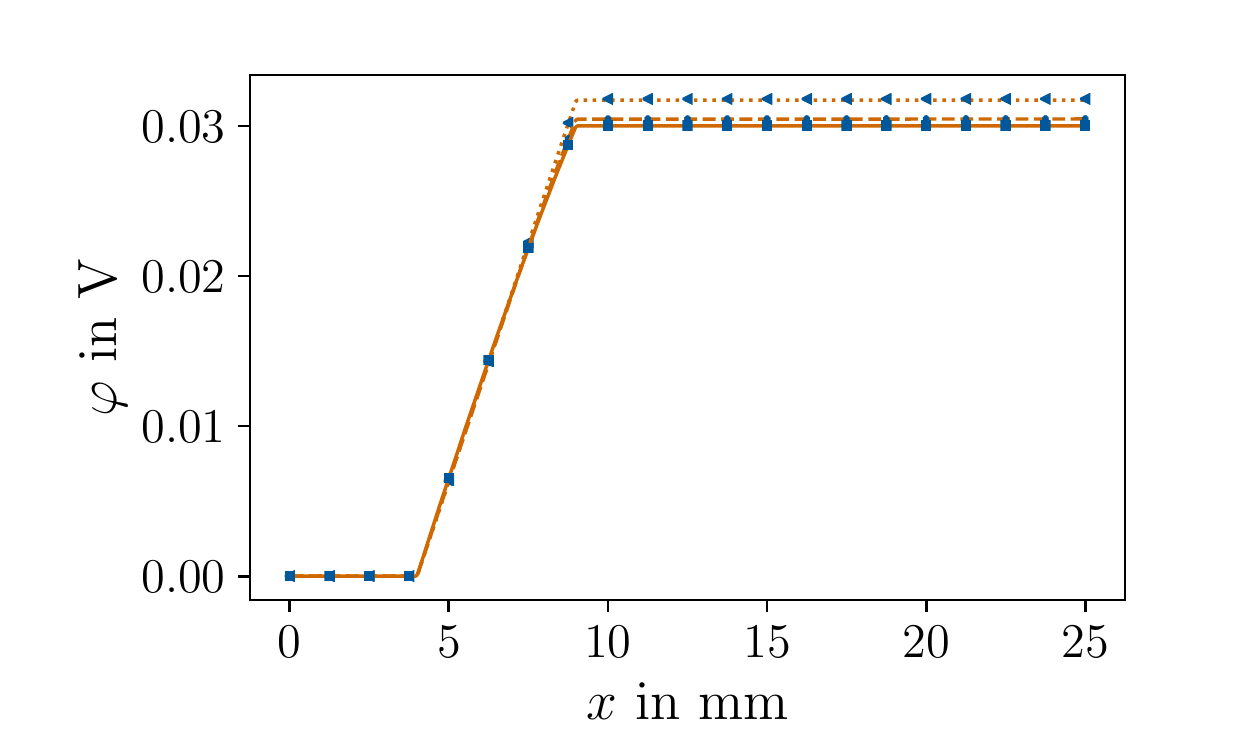}
        \caption[]%
        {{\small}}  
        \label{fig:case2Pot}
    \end{subfigure}
    \hfill
    \begin{subfigure}[b]{0.475\textwidth}  
        \centering 
        \includegraphics[width=\textwidth]{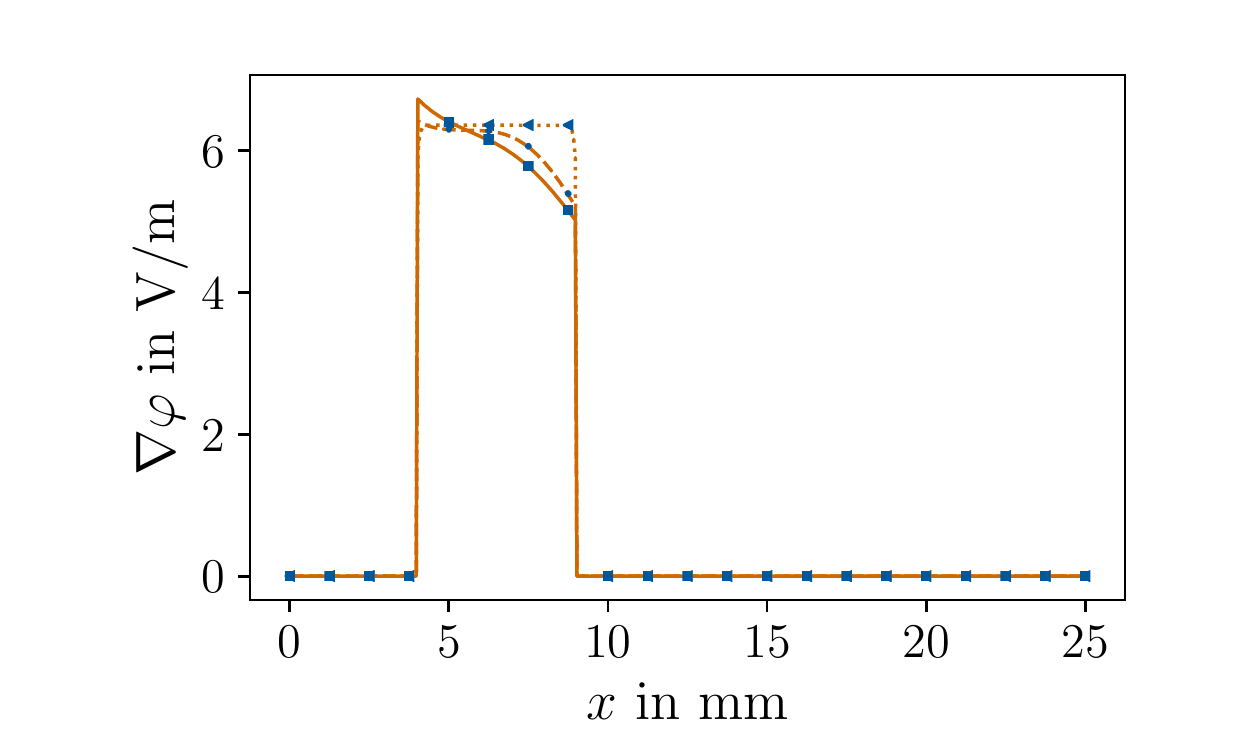}
        \caption[]%
        {{\small}} 
        \label{fig:case2gradPot}
    \end{subfigure}
    \vskip\baselineskip
    \begin{subfigure}[b]{0.475\textwidth}   
        \centering 
        \includegraphics[width=\textwidth]{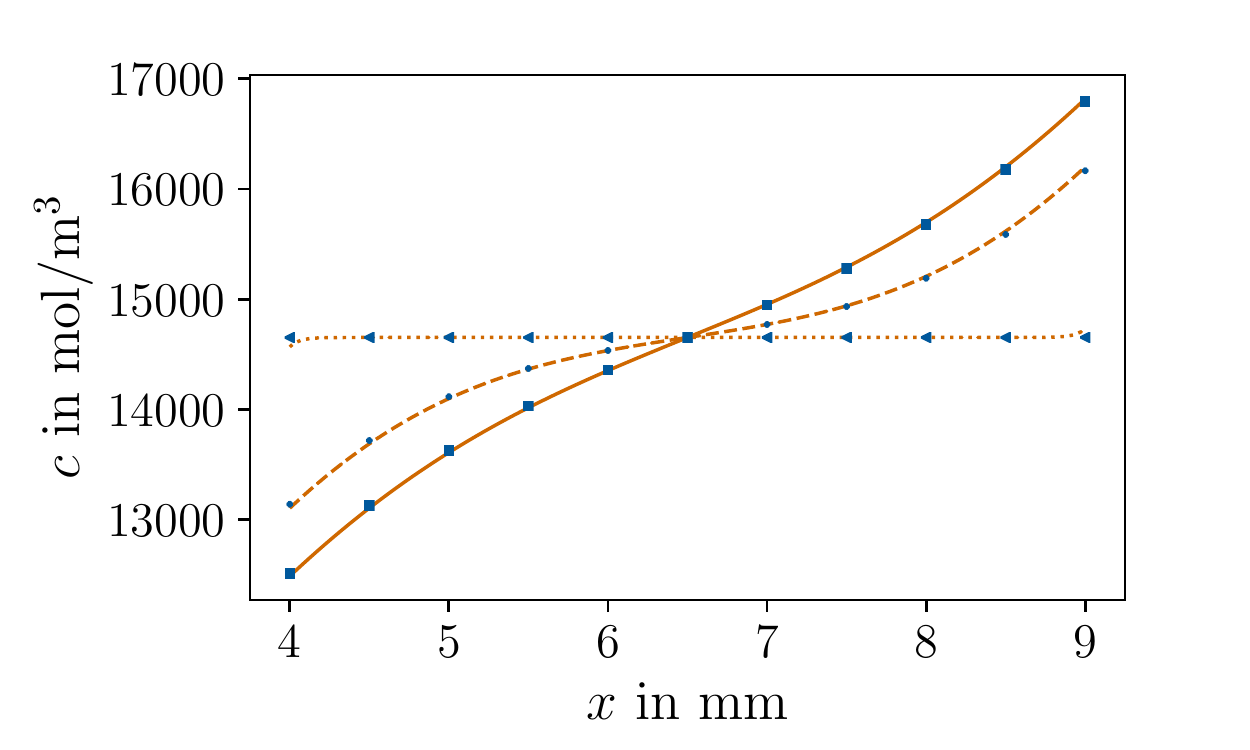}
        \caption[]%
        {{\small}}
        \label{fig:case2Spec}
    \end{subfigure}
    \hfill
    \begin{subfigure}[b]{0.475\textwidth}   
        \centering 
        \includegraphics[width=\textwidth]{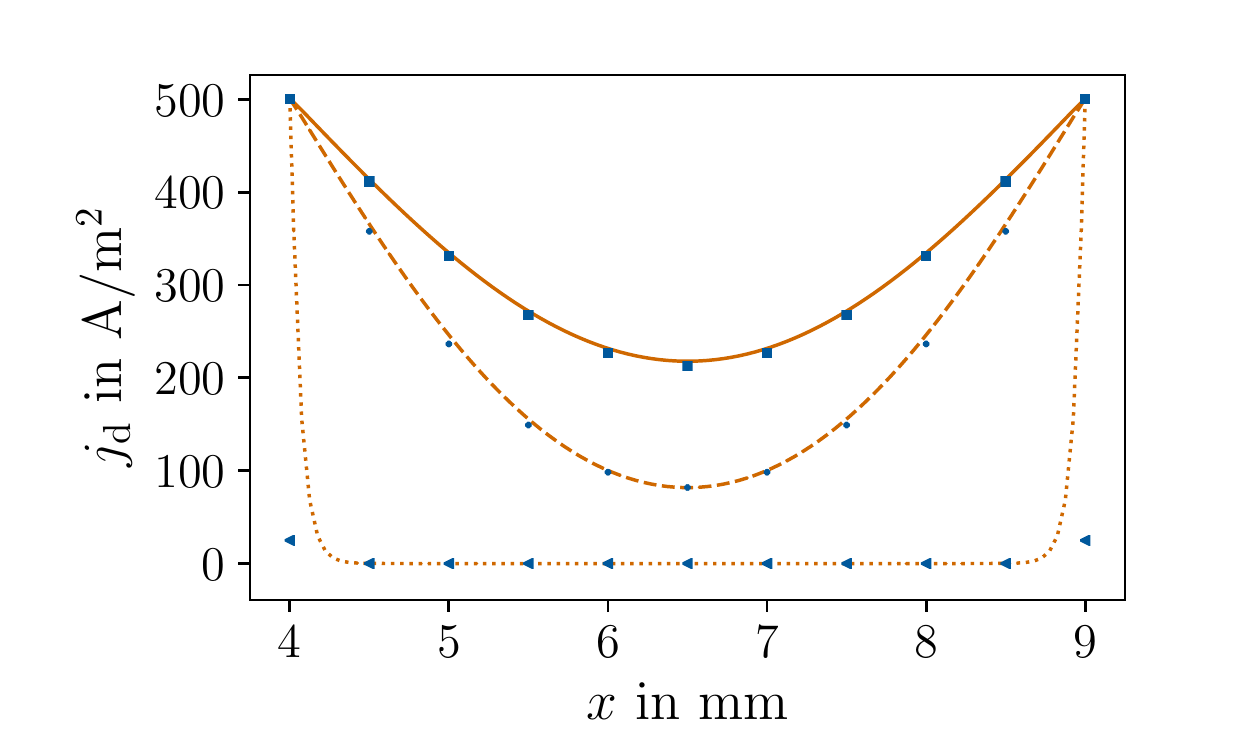}
        \caption[]%
        {{\small}}  
        \label{fig:case2gradSpec}
    \end{subfigure}
    \vskip\baselineskip
    \begin{subfigure}[b]{0.475\textwidth}   
        \centering 
        \includegraphics[width=\textwidth]{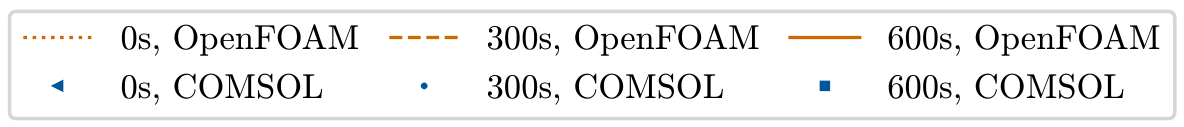}
        {{\small}}
    \end{subfigure}
    \caption[]
    {\small Test case 2; results for OpenFOAM and COMSOL simulations: (a) potential distribution (b) potential gradient in the whole cell, (c) species distribution and (d) diffusive current density of the Li$^+$ ion in the electrolyte.} 
    \label{fig:resCase2}
\end{figure*}

As well as in test case 1, validity can be confirmed since the results  of COMSOL and OpenFOAM shown in Fig.\,\ref{fig:resCase2} match perfectly except for the diffusive current density in COMSOL at the initial time step.

Species distribution and gradient, diffusive current density and electric conductivity exhibit the same behaviour as outlined in the previous section. However, the potential gradient in the electrolyte is no longer symmetric. Here, it can be concluded that the diffusive current density has a significant influence on the potential gradient. \newtxt{Additionally and in contrast to test case 1, the potential changes over time.}
As the slope of the potential curve flattens in the electrolyte region over time, the electric conductivity there seems to increase. This may indicate that the electric loss in the electrolyte
\begin{equation}
\label{eq:4-5}
U = \frac{\mathbf{j} l}{\sigma} 
\end{equation}
with $l$ being the resistor length, decreases over time. But, the electric conductivity is constant on average during the whole simulation. With this, it is made clear that the diffusive current density influences the electric losses significantly.
\subsection{Test case 3}
\label{appCase3}
\begin{figure*}[h]
    \centering
    \begin{subfigure}[b]{0.475\textwidth}
        \centering
        \includegraphics[width=\textwidth]{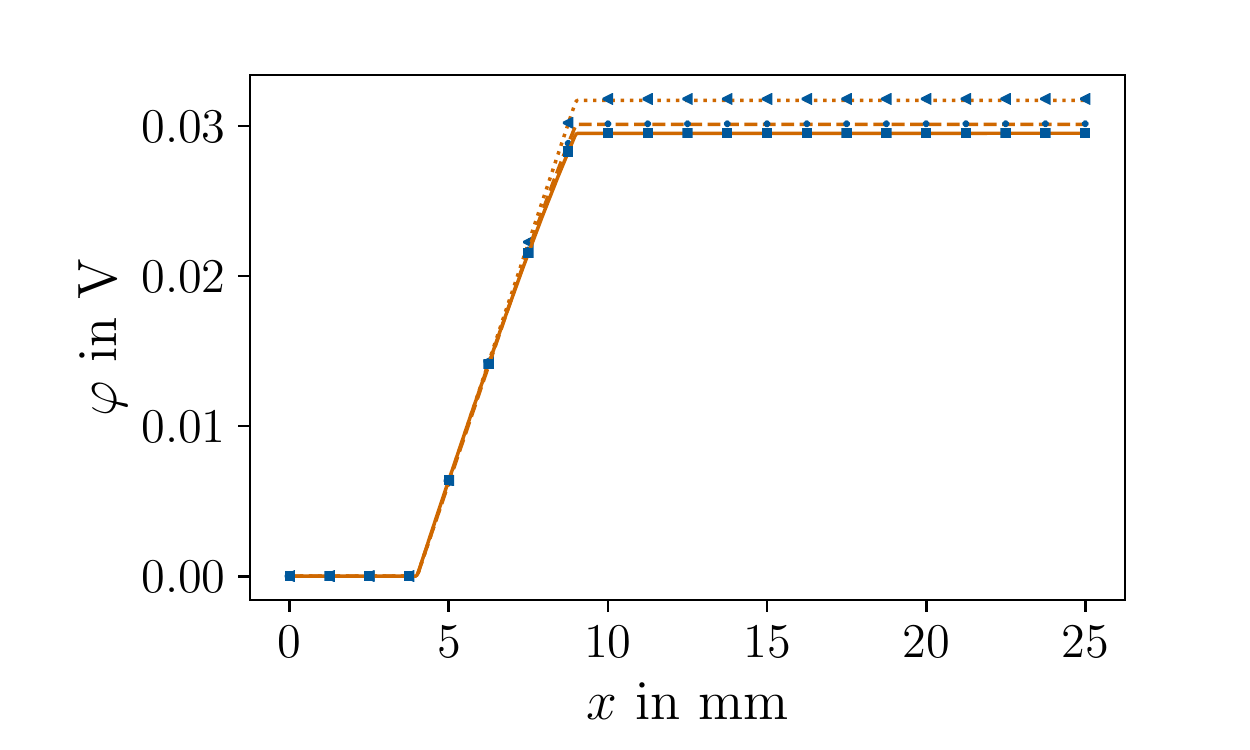}
        \caption[]%
        {{\small}}  
        \label{fig:case3Pot}
    \end{subfigure}
    \hfill
    \begin{subfigure}[b]{0.475\textwidth}  
        \centering 
        \includegraphics[width=\textwidth]{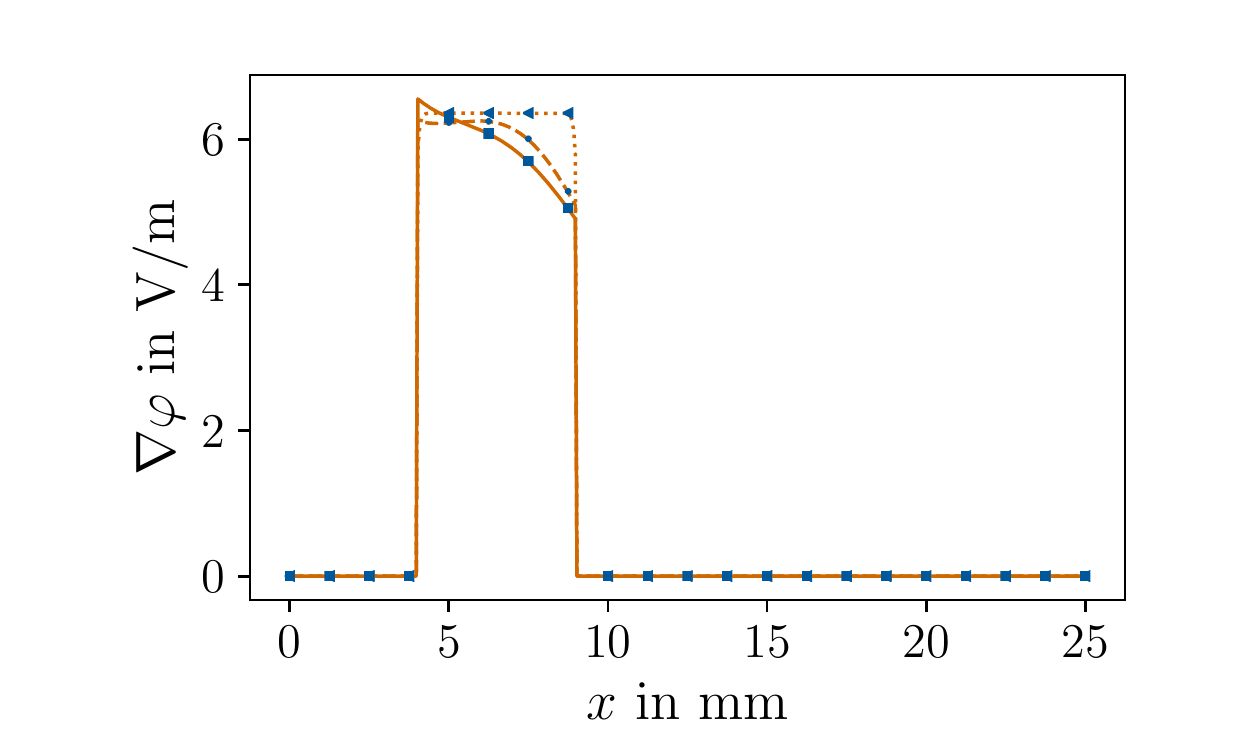}
        \caption[]%
        {{\small}}  
        \label{fig:case3gradPot}
    \end{subfigure}
    \vskip\baselineskip
    \begin{subfigure}[b]{0.475\textwidth}   
        \centering 
        \includegraphics[width=\textwidth]{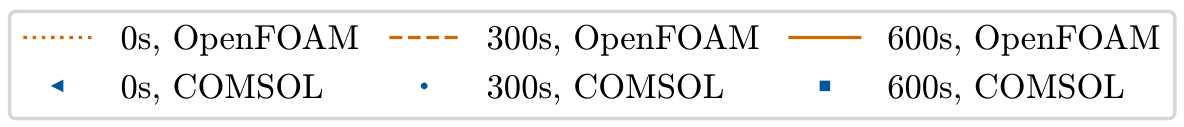}
        {{\small}}
    \end{subfigure}
    \caption[]
    {\small Test case 3; results for OpenFOAM and COMSOL simulations: (a) potential distribution (b) potential gradient in the whole cell.} 
    \label{fig:resCase3_1}
\end{figure*}

\begin{figure*}[htpb]
    \centering
    \begin{subfigure}[b]{0.475\textwidth}   
        \centering 
        \includegraphics[width=\textwidth]{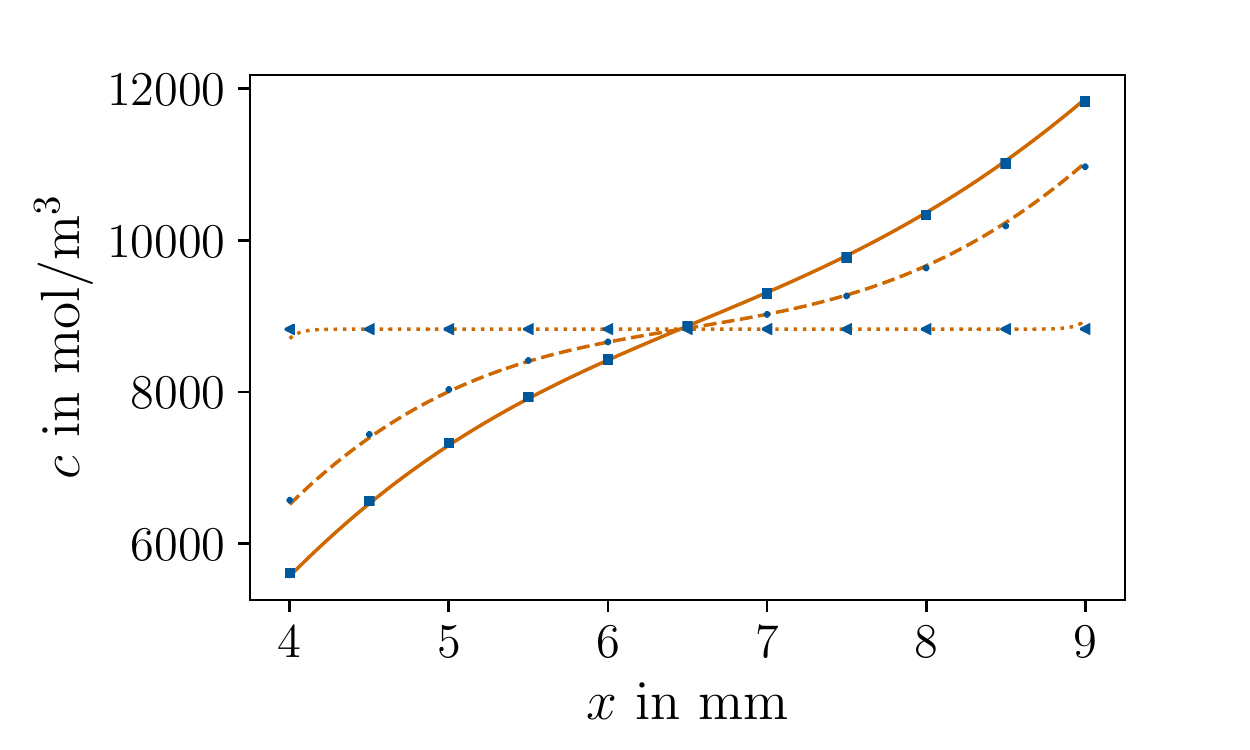}
        \caption[]%
        {{\small}} 
        \label{fig:case3SpecLi}
    \end{subfigure}
    \hfill
    \begin{subfigure}[b]{0.495\textwidth}   
        \centering 
        \includegraphics[width=\textwidth]{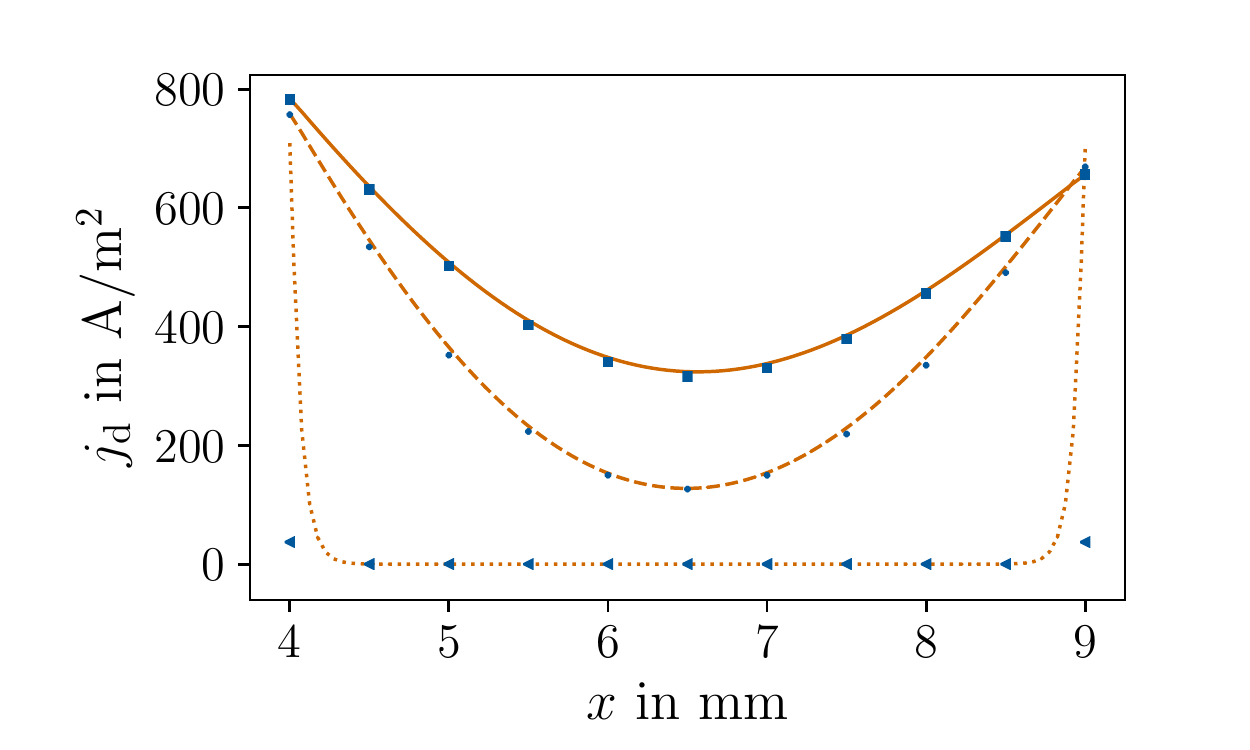}
        \caption[]%
        {{\small}}
        \label{fig:case3gradSpecLi}
    \end{subfigure}
    \vskip\baselineskip
    \begin{subfigure}[b]{0.495\textwidth}   
        \centering 
        \includegraphics[width=\textwidth]{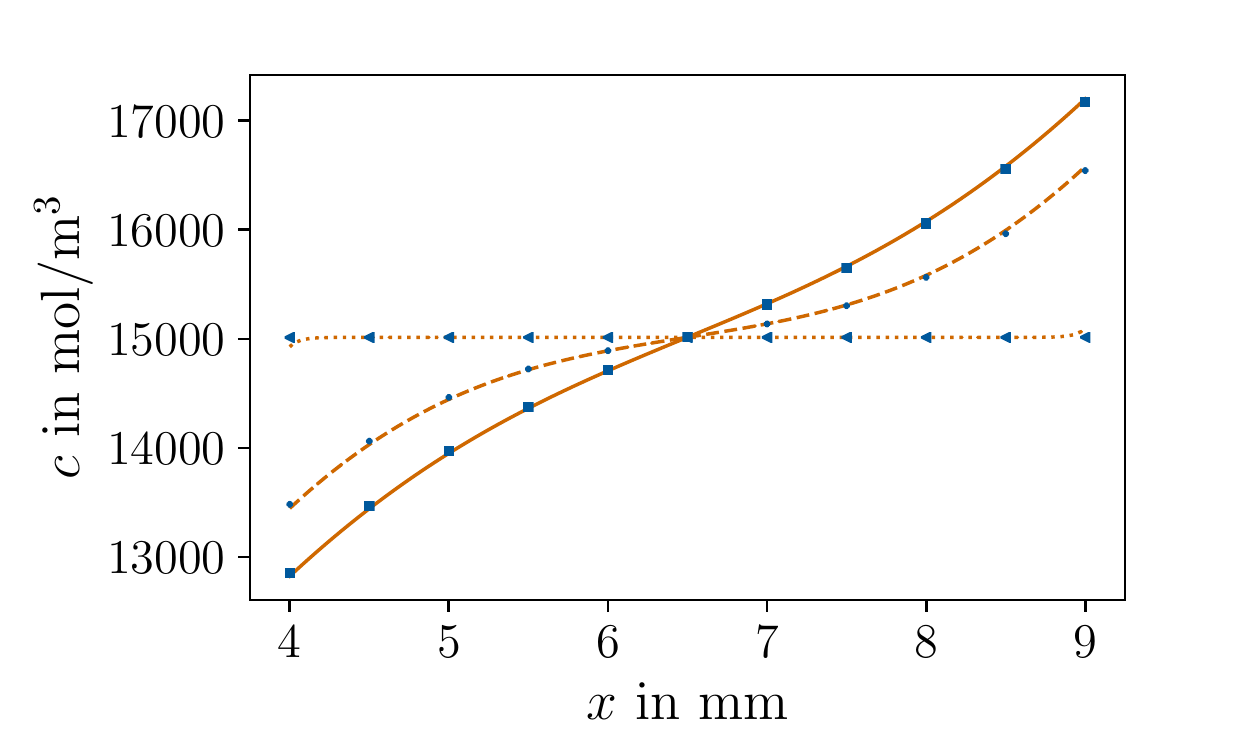}
        \caption[]%
        {{\small}} 
        \label{fig:case3SpecCl}
    \end{subfigure}
    \hfill
    \begin{subfigure}[b]{0.495\textwidth}   
        \centering 
        \includegraphics[width=\textwidth]{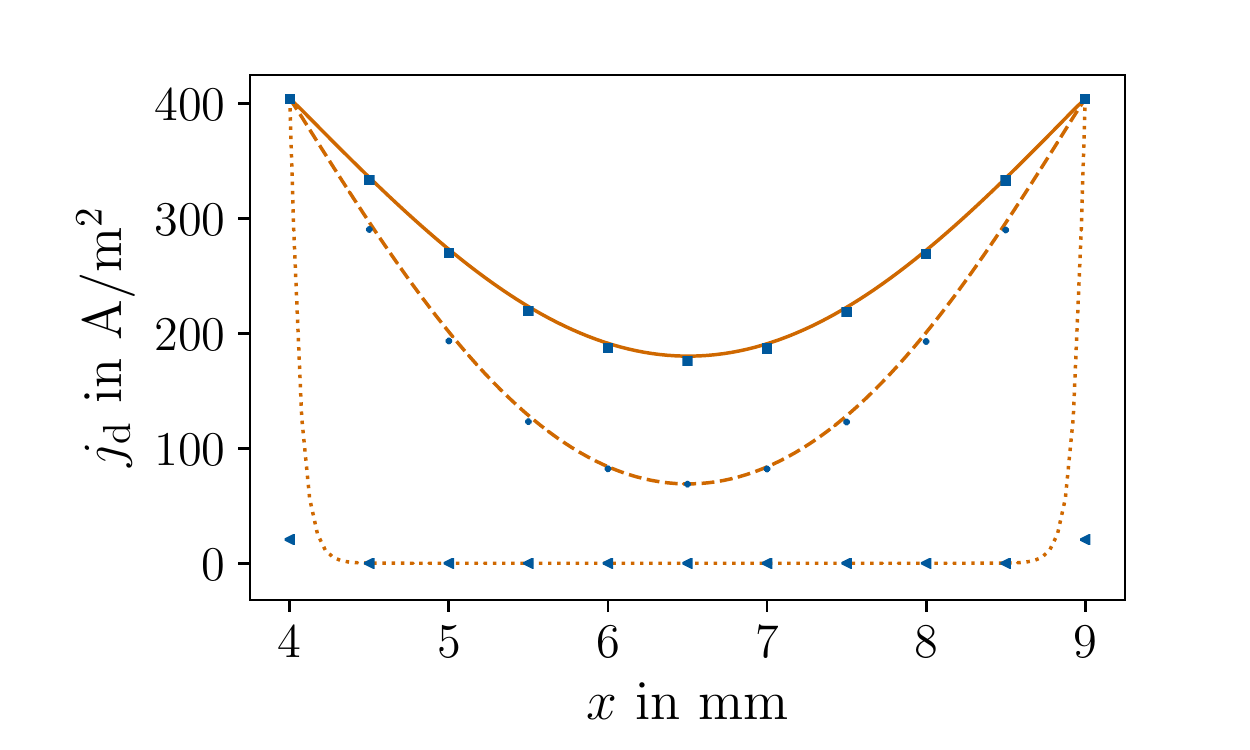}
        \caption[]%
        {{\small}} 
        \label{fig:case3gradSpecCl}
    \end{subfigure}
    \vskip\baselineskip
    \begin{subfigure}[b]{0.495\textwidth}   
        \centering 
        \includegraphics[width=\textwidth]{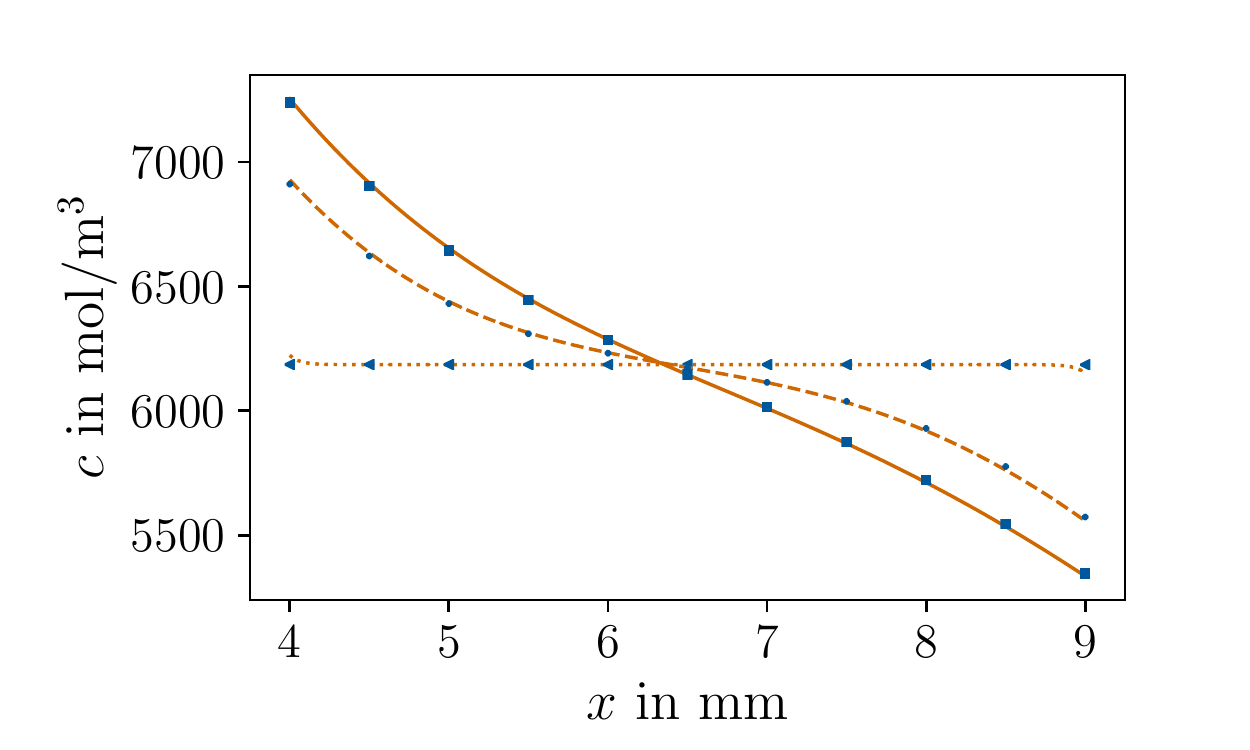}
        \caption[]%
        {{\small}} 
        \label{fig:case3SpecK}
    \end{subfigure}
    \hfill
    \begin{subfigure}[b]{0.495\textwidth}   
        \centering 
        \includegraphics[width=\textwidth]{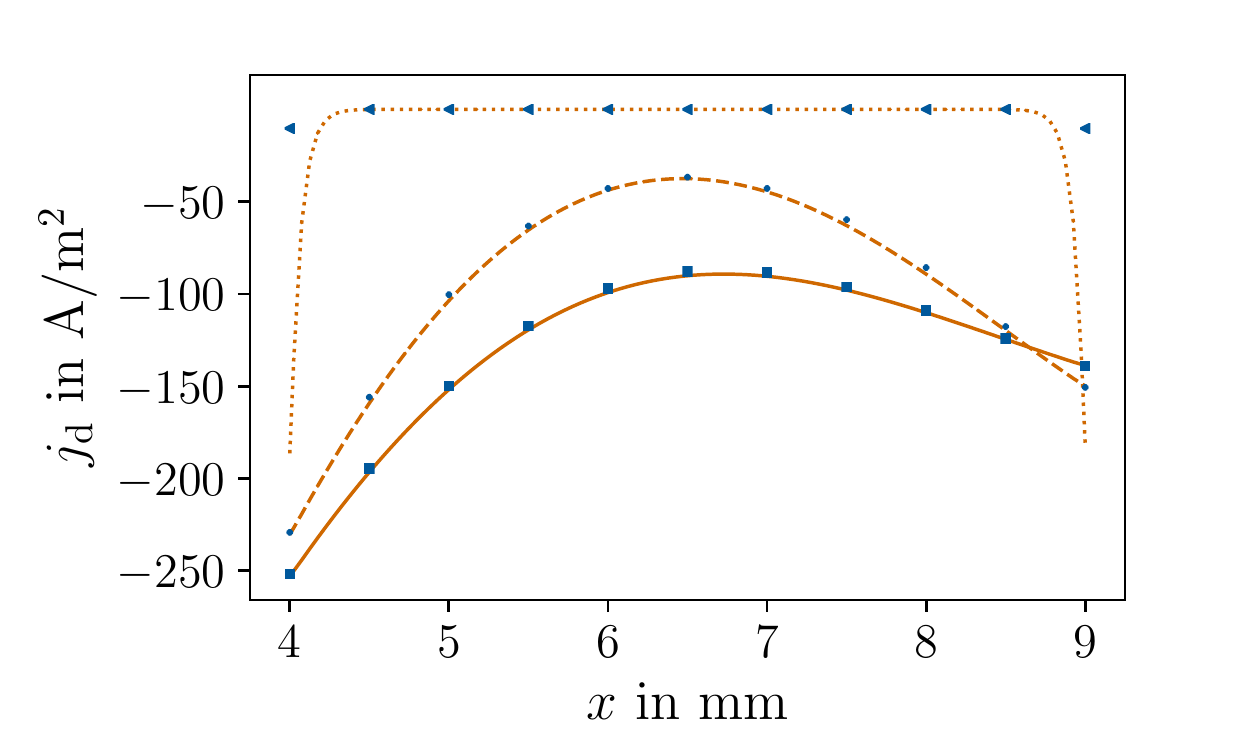}
        \caption[]%
        {{\small}}  
        \label{fig:case3gradSpecK}
    \end{subfigure}
    \vskip\baselineskip
    \begin{subfigure}[b]{0.475\textwidth}   
        \centering 
        \includegraphics[width=\textwidth]{c3_legend2.pdf}
        {{\small}}
    \end{subfigure}
    \caption[]
    {\small Test case 3; results for OpenFOAM and COMSOL simulations: species distribution (left) and diffusive current density (right) of the (a)\&(b) Li$^+$ ion, (c)\&(d) Cl$^-$ ion and (e)\&(f) K$^+$ ion.} 
    \label{fig:resCase3_2}
\end{figure*}

\newtxt{Comparing OpenFOAM to COMSOL results for test case 3} -- as presented in Fig.\,\ref{fig:resCase3_1} and Fig.\,\ref{fig:resCase3_2} shows validity of the former.

In terms of the electric loss identifiably by the temporal development of the potential distribution, the conclusions drawn in the previous case can be confirmed as non-random.
In contrast to the previous binary test cases, it can be seen in Fig.\,\ref{fig:resCase3_2} that the species distributions and gradients -- here expressed as diffusive current density -- for a ternary electrolyte are not equal any more. 
Nevertheless, the electric conductivity $\sigma$ is constant on average over time.
For the Cl$^-$ ion, being the common ion in both salts, the species distribution is symmetric, while this is not the case for the Li$^+$ and K$^+$ ions. This can be explained by the differing diffusion coefficients.
Moreover, an asymmetric potential gradient is -- as in test case 2 -- observable within the electrolyte. This leads to the logical consequence that the reason for the asymmetry must lie in the diffusive part of the material balance for the species transport
Eq.\,\eqref{eq:2-19}. 
\subsection{Test case 4}
\label{appCase4}
\begin{figure*}[h]
    \centering
    \begin{subfigure}[b]{0.495\textwidth}
        \centering
        \includegraphics[width=\textwidth]{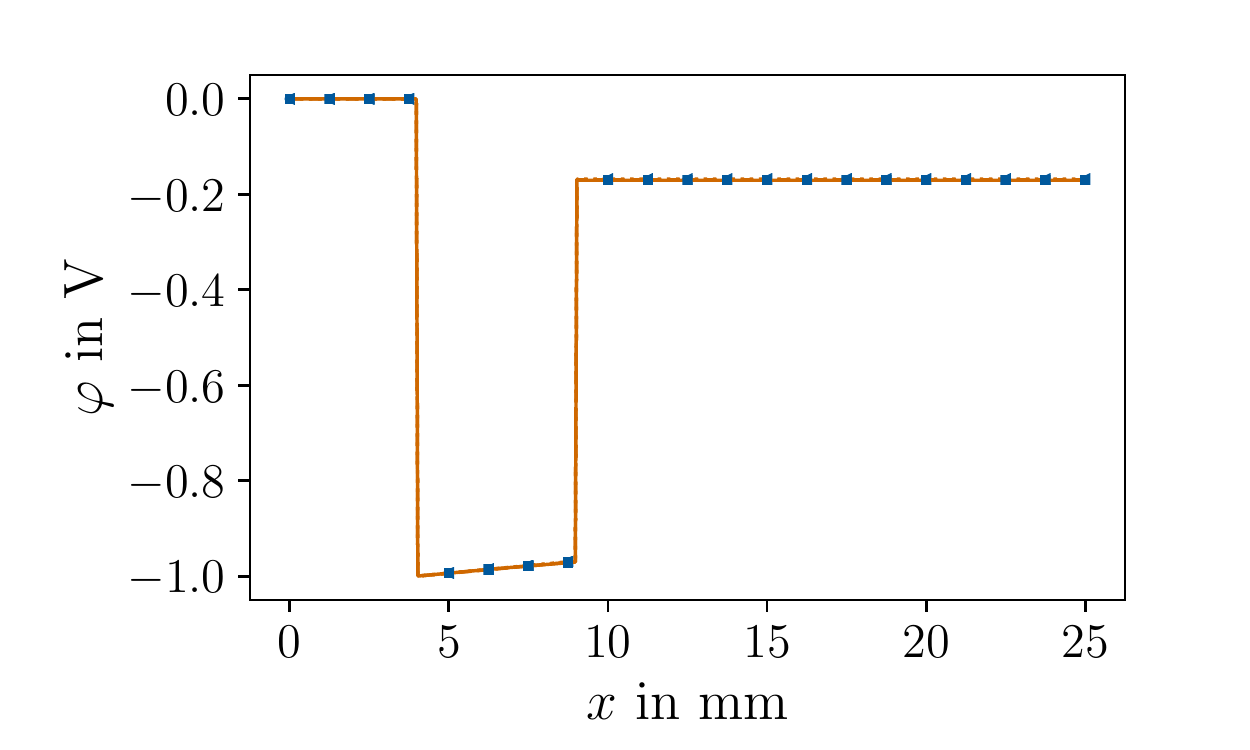}
        \caption[]%
        {{\small}}  
        \label{fig:case4Pot}
    \end{subfigure}
    \hfill
    \begin{subfigure}[b]{0.495\textwidth}  
        \centering 
        \includegraphics[width=\textwidth]{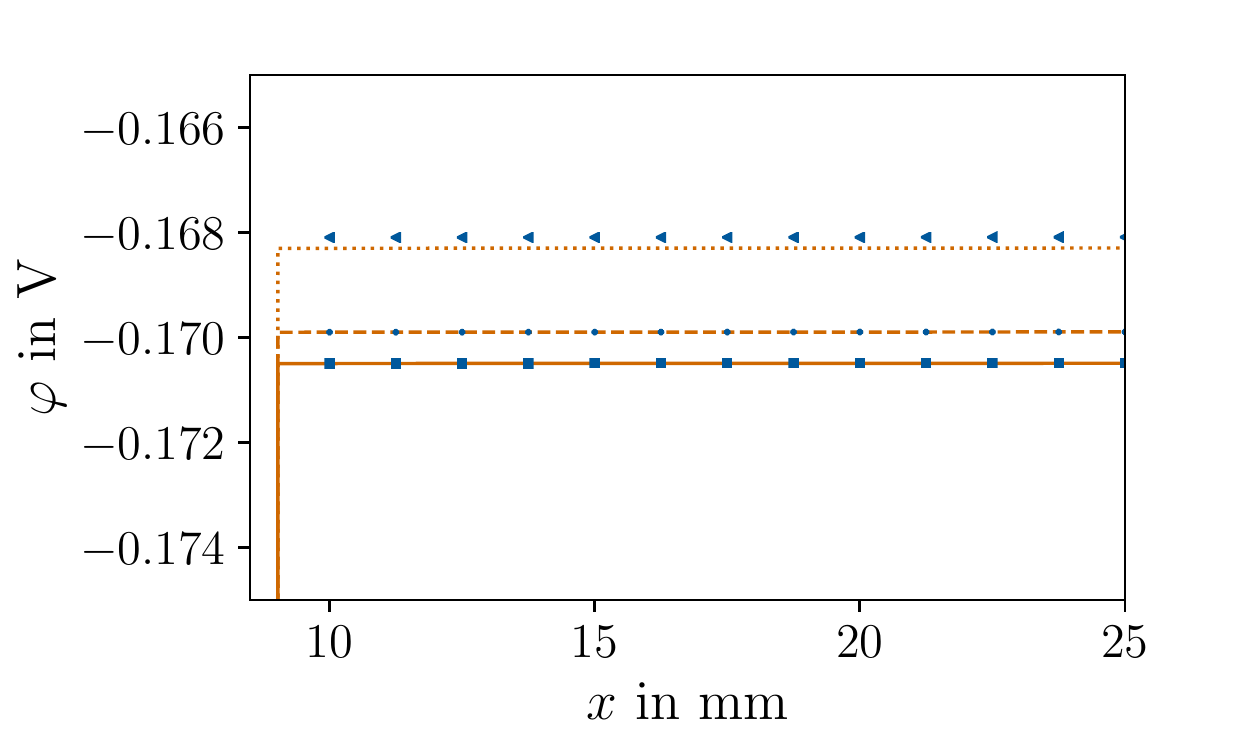}
        \caption[]%
        {{\small}}  
        \label{fig:case4PotAn}
    \end{subfigure}
    \vskip\baselineskip
    \begin{subfigure}[b]{0.475\textwidth}   
        \centering 
        \includegraphics[width=\textwidth]{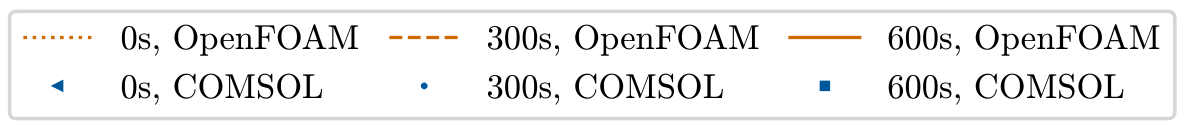}
        {{\small}}
    \end{subfigure}
    \caption[]
    {\small Test case 4; results for OpenFOAM and COMSOL simulations: potential distribution in the (a) whole cell and (b) anode.} 
    \label{fig:resCase4_1}
\end{figure*}

In Fig.\,\ref{fig:resCase4_1} the potential distribution in the cell, where the two potential jumps are clearly visible, is shown. Since $\nabla \varphi_{\mathrm{case}4} \approx \nabla \varphi_{\mathrm{case}3}$, the potential gradient is not presented again and it is referred to Fig.\,\ref{fig:case3gradPot}.
As in the previous \sout{validation}\newtxt{verification} test cases, the results of OpenFOAM and COMSOL match very well except for the initial time step.
%
%--------------------------------------------
%\section{\colorbox{yellow}{Grid study}}
\section{Grid study}
\label{chapter:appGrid}
%\counter

\begin{figure*}[h]
    \centering
    \begin{subfigure}[b]{0.475\textwidth}
        \centering
        \includegraphics[width=\textwidth]{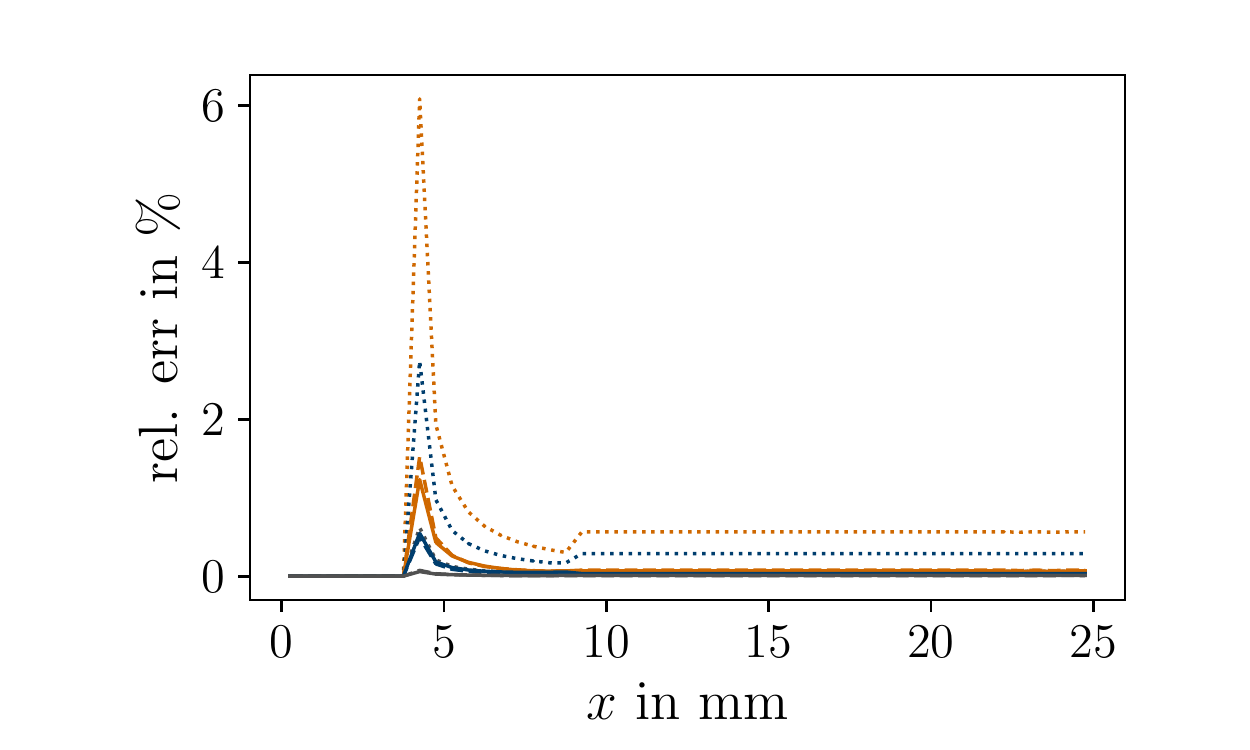}
        \caption[]%
        {{\small}}  
        \label{fig:gridCase3Pot}
    \end{subfigure}
    \hfill
    \begin{subfigure}[b]{0.475\textwidth}  
        \centering 
        \includegraphics[width=\textwidth]{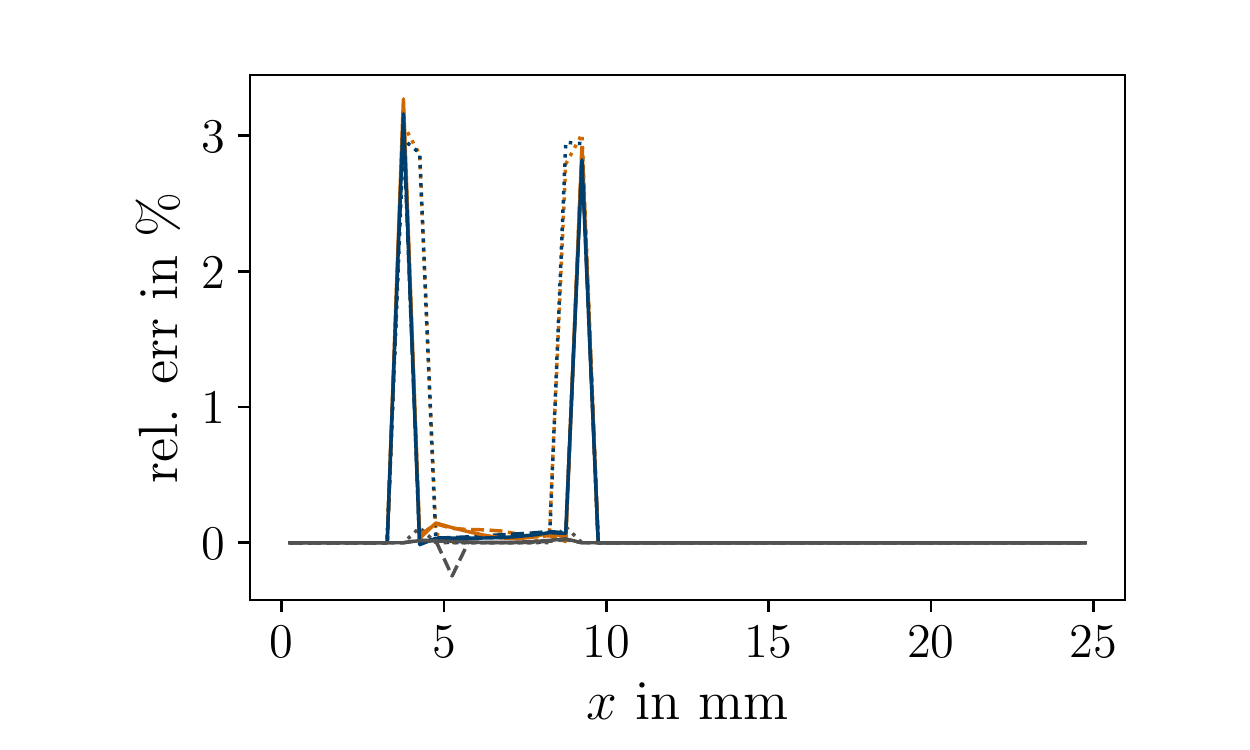}
        \caption[]%
        {{\small}}  
        \label{fig:gridCase3gradPot}
    \end{subfigure}
    \vskip\baselineskip
    \begin{subfigure}[b]{0.475\textwidth}   
        \centering 
        \includegraphics[width=\textwidth]{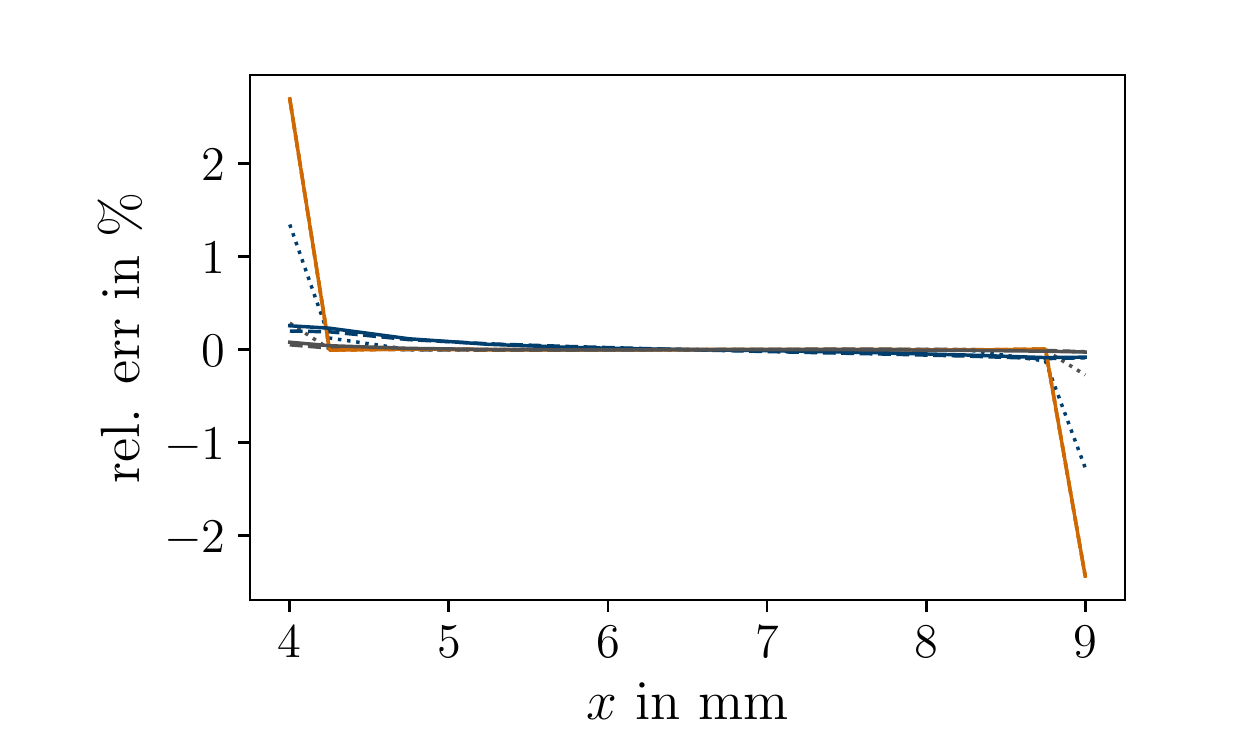}
        \caption[]%
        {{\small}}  
        \label{fig:gridCase3Spec}
    \end{subfigure}
    \hfill
    \begin{subfigure}[b]{0.475\textwidth}   
        \centering 
        \includegraphics[width=\textwidth]{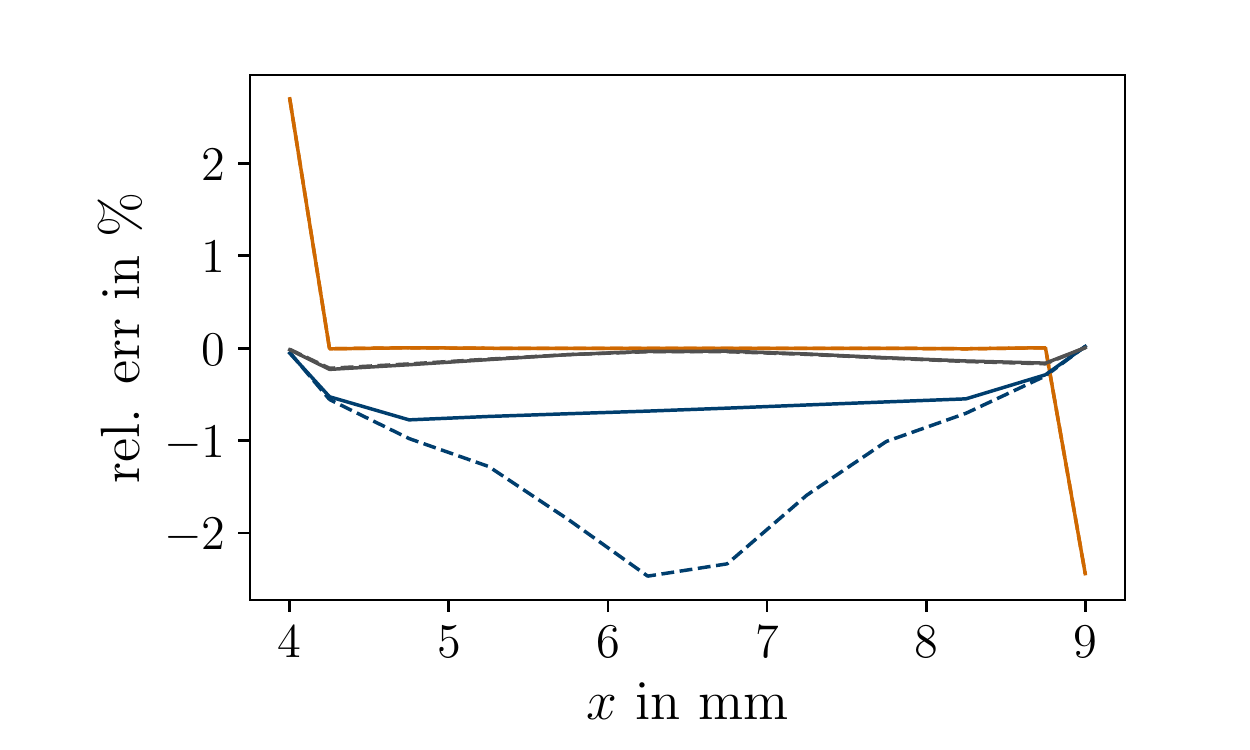}
        \caption[]%
        {{\small}}  
        \label{fig:gridCase3gradSpec}
    \end{subfigure}
    \vskip\baselineskip
    \begin{subfigure}[b]{0.475\textwidth}   
        \centering 
        \includegraphics[width=\textwidth]{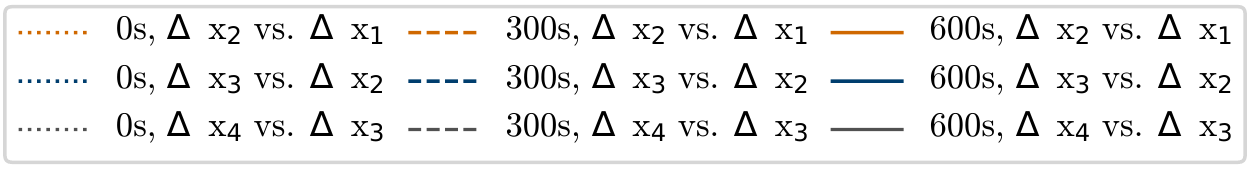}
        {{\small}}
    \end{subfigure}
    \caption[]
    {\small Grid study with OpenFOAM using simulation test case 3: relative error in the (a) potential distribution, (b) potential gradient, (c) species distribution of Li$^+$ and the (d) species gradient of Li$^+$.} 
    \label{fig:gridCase3}
\end{figure*}

%
A grid study for\sout{validation / verification} test case 3 is performed, to show that the results do not contain grid dependent errors. All simulations presented in section \ref{validation} and \ref{chapter:appValidationComsol} were done with a uniform grid spacing of $\Delta$ x$_4 = 5 \cdot 10^{-5}$m. Here, $\Delta$ x$_1 = 5 \cdot 10^{-4}$m, $\Delta$ x$_2 = 2.5 \cdot 10^{-4}$m and $\Delta$ x$_3 = 1 \cdot 10^{-4}$m. The relative error between the larger grid spacing to the smaller grid spacing is calculated for potential, potential gradient, species distribution and species gradient of Li$^+$ and the results are shown in Fig.\,\ref{fig:gridCase3}.
It can be seen clearly that the error shows peaks at the interfaces between electrolyte and electrode which are decreasing, when the number of cells is increased.\sout{This applies for potential, potential gradient, concentration and concentration gradient.} When the finest grid is used, the error is less than 1\%, which is small enough to claim that the solutions, shown in section \ref{validation} and \ref{chapter:appValidationComsol} are grid independent. 
%
%-----------------------------------------------------------------%
%
%\section{\colorbox{yellow}{Electroneutrality study}}
\section{Electroneutrality study}
\label{chapter:appElectro}
\begin{figure}[h]
  \centering
  \includegraphics[width=0.5\linewidth]{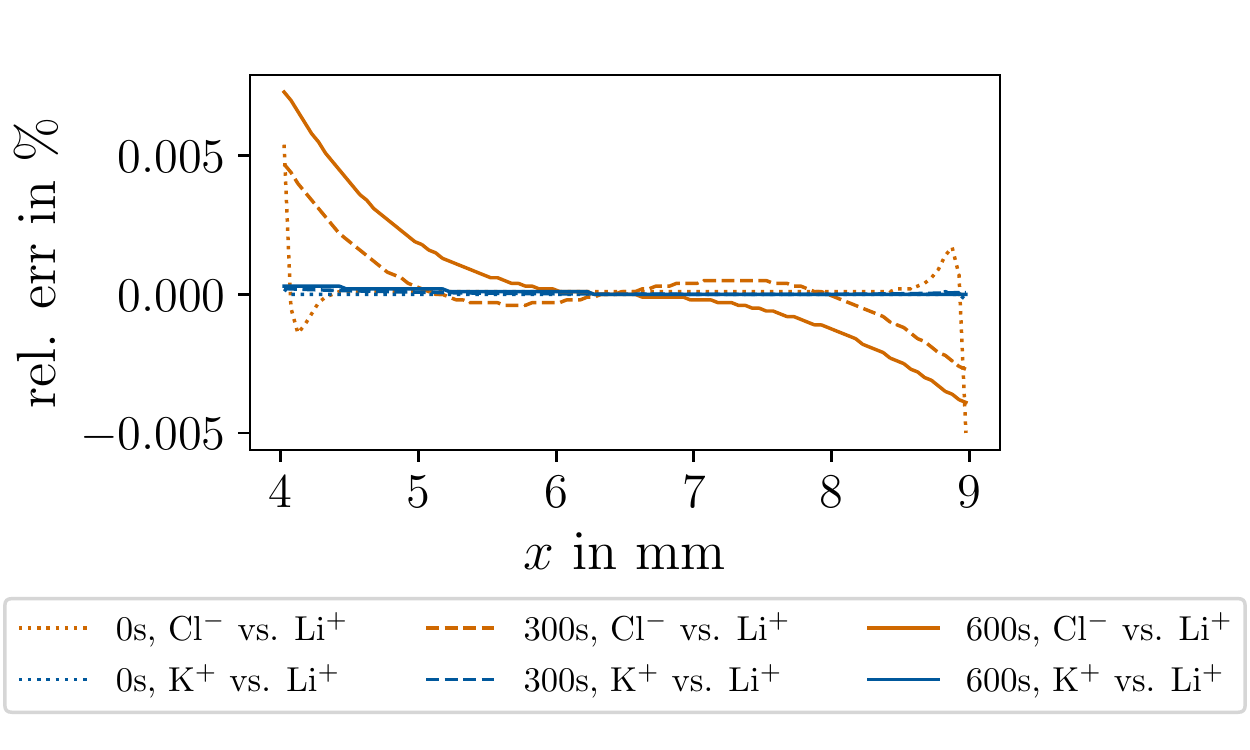}
  \caption{Electroneutrality study in OpenFOAM using simulation test case 3: relative error in the electric conductivity. Calculating Li$^+$ from electroneutrality was taken as reference.}
  \label{fig:nucCase3sigma}
\end{figure}

\begin{figure*}[h]
    \centering
    \begin{subfigure}[b]{0.475\textwidth}   
        \centering 
        \includegraphics[width=\textwidth]{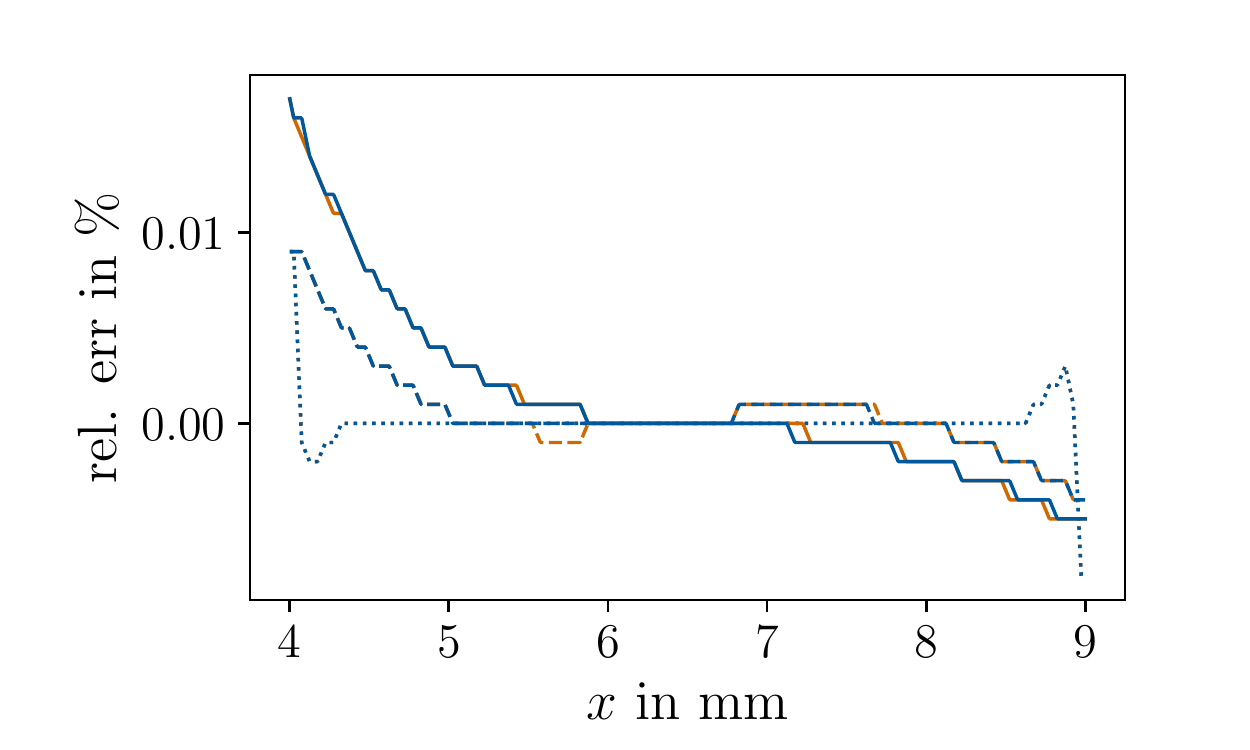}
        \caption[]%
        {{\small}}  
        \label{fig:nucCase3Spec}
    \end{subfigure}
    \hfill
    \begin{subfigure}[b]{0.475\textwidth}   
        \centering 
        \includegraphics[width=\textwidth]{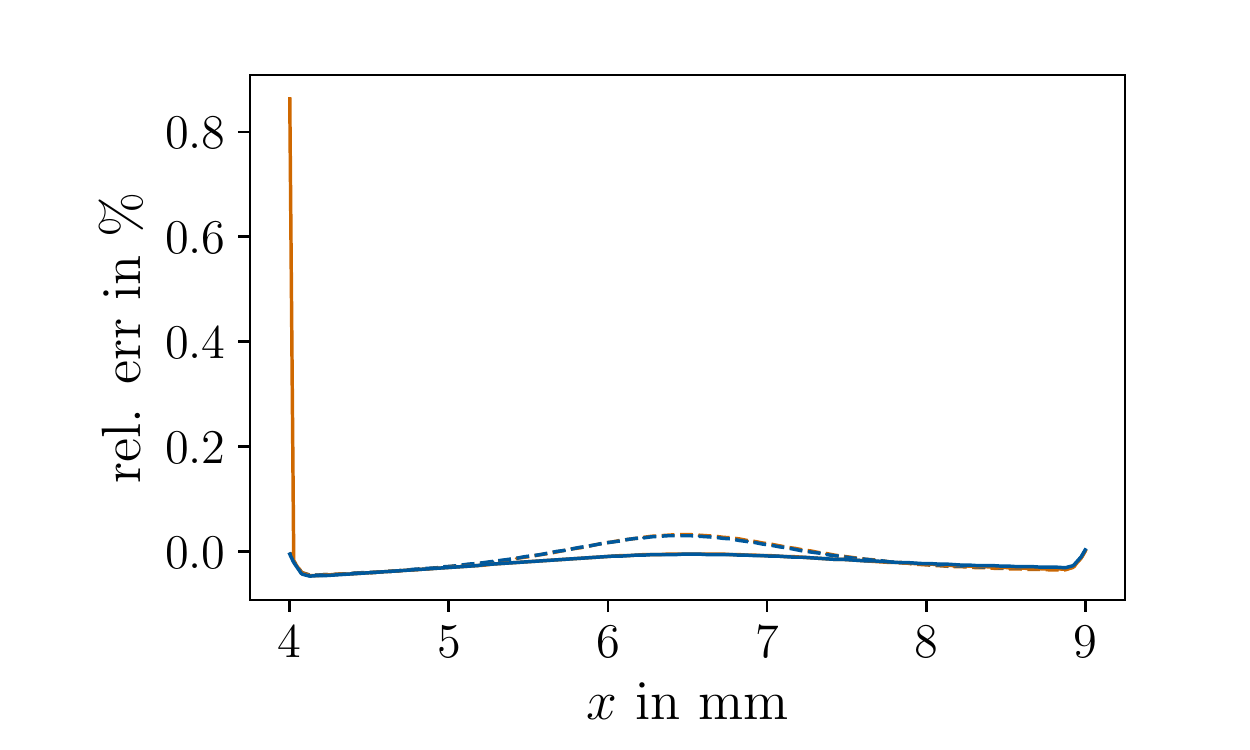}
        \caption[]%
        {{\small}}  
        \label{fig:nucCase3gradSpec}
    \end{subfigure}
    \vskip\baselineskip
    \begin{subfigure}[b]{0.475\textwidth}   
        \centering 
        \includegraphics[width=\textwidth]{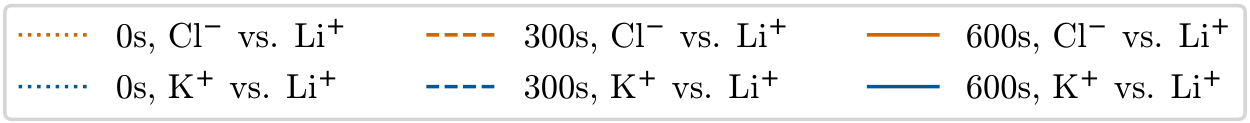}
        {{\small}}
    \end{subfigure}
    \caption[]
    {\small Electroneutrality study in OpenFOAM using simulation test case 3: relative error in the (a) species distribution of Li$^+$ and the (b) species gradient of Li$^+$. Calculating Li$^+$ from electroneutrality was taken as reference.} 
    \label{fig:nucCase3}
\end{figure*}

%\counter
%
As it was mentioned in section \ref{eqnElec}, the solver is able to
differentiate, if a species is active or passive. The results should
be the same no matter which species in the electrolyte is calculated
from electroneutrality. So, it is investigated, if there are any
differences in the solution of test case 3 when either Li$^+$, Cl$^-$ or K$^+$ are calculated from electroneutrality. The former is taken as reference case.\\
Fig.\,\ref{fig:nucCase3sigma} shows that the error in the electric conductivity is less than 0.01\%.
Further, it can be seen in Fig.\,\ref{fig:nucCase3} that the species distribution and the species gradient in the electrolyte are almost the same for all three simulations. This means that the solver is robust and it is not necessary to add additional boundary conditions for the active species as it is needed in COMSOL.

%%% Local Variables:
%%% mode: latex
%%% TeX-master: "../paper"
%%% TeX-parse-self: t
%%% TeX-auto-save: t
%%% TeX-PDF-mode: t
%%% eval: (auto-fill-mode 1)
%%% eval: (flyspell-mode 1)
%%% eval: (reftex-mode 1)
%%% ispell-dictionary: "british"
%%% End:

%%%%%%%%%%%%%%%%%%%%%
\end{document}